\documentclass[12pt]{article}
\usepackage[cmtip,arrow]{xy}\usepackage{pb-diagram,pb-xy}%

\input xy
\xyoption{all}
\input xy
\xyoption{all}
\usepackage{heck}
\usepackage{cite}
\usepackage{graphicx}
\usepackage{makeidx}
\usepackage{multicol}
\usepackage{geometry}
\usepackage{amsfonts}
\usepackage{mathrsfs}
\usepackage{amssymb}
\usepackage{amsmath}%

\providecommand{\U}[1]{\protect\rule{.1in}{.1in}}
\numberwithin{equation}{section}

\newcommand{\bea}{\begin{eqnarray}}
\newcommand{\eea}{\end{eqnarray}}
\newcommand{\be}{\begin{equation}}
\newcommand{\ee}{\end{equation}}

\newcommand{\bem}{\begin{pmatrix}}
\newcommand{\eem}{\end{pmatrix}}

\def\U{\Upsilon}

\def\cb{{\cal B}}
\def\cc{{\cal C}}

\def\cn{{\cal N}}

\def\cq{{\cal Q}}

\def\cs{{\cal S}}

\def \N {{\mathcal N}}
\def \Z {{\mathbb Z}}
\def \C {{\mathbb C}}
\def \R {{\mathbb R}}

\xyoption{arc}

\begin{document}

\bibliographystyle{utphys}

\date{March, 2011}


\institution{SISSA}{\centerline{${}^{1}$Scuola Internazionale Superiore di Studi Avanzati, via Bonomea 265, I-34100 Trieste, ITALY}}

\institution{HarvardU}{\centerline{${}^{2}$Jefferson Physical Laboratory, Harvard University, Cambridge, MA 02138, USA}}%

\title{Classification of Complete $\cn=2$ Supersymmetric Theories in $4$ Dimensions}
%

\authors{Sergio Cecotti\worksat{\SISSA}\footnote{e-mail: {\tt cecotti@sissa.it}} and Cumrun Vafa\worksat{\HarvardU}\footnote{e-mail: {\tt vafa@physics.harvard.edu}}}%

\abstract{We define the notion of a complete $\cn=2$ supersymmetric theory in 4 dimensions
as a UV complete theory for which all the BPS central charges can be arbitrarily varied as we vary its Coulomb branch parameters, masses, and coupling constants.  We classify all such theories whose BPS spectrum can be obtained via
a quiver diagram.  This is done using the 4d/2d correspondence and by showing that such complete $\cn=2$ theories map to quivers of finite mutation type.
The list of such theories is given by the (generalized) Gaiotto theories consisting of two 5-branes wrapping Riemann surfaces
with punctures, as well as 11 additional exceptional cases, which we identify.}

\maketitle

\tableofcontents

\pagebreak

\section{Introduction}

Supersymmetric gauge theories with high enough number of supersymmetries are relatively rigid. 
For example $\cn=4$ supersymmetric theories in 4 dimensions are completely classified by the choice of the gauge group.
However, the ones with lower number of supersymmetries are more flexible.  In particular $\cn=1$ theories
in 4 dimensions are far from being classified.  An interesting intermediate case in four dimensions arises for
$\cn=2$ theories, which are in some ways partially rigid, but still not rigid enough to be trivially classified.
A large class of these theories are constructed as gauge theories with matter field representations, consistent with asymptotic freedom.  On the other hand it is known that there are additional $\cn=2$ theories,
that can be obtained from string theory, but which are not easily obtained from gauge theories.  These include $\cn=2$ theories with exceptional symmetry groups obtained from 3-brane probes of F-theory, as well as ones which arise from singularities of Calabi-Yau compactifications of type II strings.  It is thus natural to ask to what extent we can classify all UV complete $\cn=2$ theories in
4 dimensions.

A similar question arises in 2 dimensional theories with $\cn=2$ supersymmetry.  In that case a program for
their classification was initiated in \cite{Cecotti:1993rm} based on their BPS soliton/kink spectra.  For example it was shown that
a theory with two vacua can have only 1 or 2 solitons connecting the two, and these theories
were identified with cubic LG theories and $\C\mathbb{P}^1$ sigma models respectively.  The data of the 2d kinks
are universal, except that as one changes the parameters of the theory, there could be jumps
in the number of BPS states, which are easily computable.  This computable change of data
of the BPS kinks in 2d will be called a `mutation'.
 Four dimensional
theories with $\cn=2$ also have an interesting set of BPS states, which in a sense characterize
the theory.  Moreover for typical such theories, there is an associated supersymmetric
quantum mechanical quiver (with 4 supercharges), whose ground states correspond to such BPS states.
 It was proposed in \cite{cnv} that the classification problem for $\cn=2$ theories in $2d$ and  $4d$ 
are linked.   The basic idea is that $\cn=2$ theories in $4d$ can be engineered in terms
of type II string theories.  And the type II theories have an associated 2d worldsheet theory with $\cn=2$ supersymmetry,
 which has, in addition to 2d Liouville field, a massive $\cn=2$ theory (for fixed value
of Liouville field) with central charge $\hat c \leq 2$.  Moreover the BPS quiver of the 4d theory was mapped to the vacua
and soliton data of the 2d theory.  In particular the nodes of the 4d BPS quiver were mapped
to vacua of the 2d theory, and the bifundamentals of the quiver, were mapped to solitons
connecting the pairs of 2d vacua.   Moreover the mutation of the 2d quiver gets mapped to the analogs
of Seiberg-like dualities for the supersymmetrical quantum mechanics which gives the number
of solitons in different chambers of the $4d$ theory.
 Even though this 4d/2d correspondence was not proven in general, it was checked
in a number of non-trivial cases and in this paper we continue to assume this holds generally and use
it to classify 4d theories with $\cn=2$ supersymmetry.  

Classification of 2d theories with $\cn=2$ supersymmetry with $\hat c \leq 2$ is already very non-trivial.  However, we can
refine our classification, by asking
if a natural subclass can be defined from the 4d point of view that can be effectively classified using this correspondence.
In this paper we find that there is one natural condition from 4d perspective that can be defined and be used to classify
in this way:  We define the notion of `complete' $\cn=2$ supersymmetric theories.  If we have
a $U(1)^r$ gauge symmetry at a generic point on Coulomb branch, and a rank $f$ flavor symmetry, the BPS
lattice is $2r+f$ dimensional, corresponding to (electric, magnetic, flavor)  charges.  The
maximal allowed deformation we would imagine in this case is $2r+f$ complex dimensional, corresponding to arbitrary local variations
of the central charges of the BPS lattice.   This could come from $r$ Coulomb branch parameters,
$f$ masses, and $r$ coupling constants of the $U(1)^r$ theory.   We call an $\cn=2$ theory complete
if all the BPS central charges can be deformed in this way, and thus in particular its moduli space is $2r+f$ dimensional.
 Note, however, that not all $\cn=2$ thoeries are complete.  For example
for an $SU(r+1)$ gauge group, we have $r$ Coulomb branch parameters, but only 1 coupling constant, and not
$r$ independent ones. On the other hand, the product of $SU(2)$ theories with asymptotically free matter representation
is `complete' in the above sense, because each $SU(2)$ can have its own coupling constant.
 We will argue that this criteria for `completeness' maps to 2d theories with $\hat c
\leq 1$.  Moreover, the corresponding BPS quivers have a finite number of elements in the mutation orbit.
In other words, they are of {\it finite mutation type}.  Since the quivers of finite mutation types have been classified mathematically \cite{fomin,derksen,felikson},
we can identify the corresponding theories. 

The quivers of finite mutation type turn out to come in two types:
They are either associated to a Riemann surface with punctures (with extra data
at the punctures), or they belong to one of the 11 exceptional cases.
The ones associated to Riemann surfaces get mapped to (generalized) Gaiotto theories
with two five branes wrapping the corresponding Riemann surfaces.   The 2d version
of them correspond to Landau-Ginzburg theories whose fields live on Riemmann
surface, with a superpotential with specified poles.
  Nine of the eleven
exceptional cases correspond to type IIB on certain local Calabi-Yau singularities
(three of them can also be viewed as an M5 brane wrapping a specific singular curve).
These again map to 2d Landau-Ginzburg theories with $\hat c=1$ and their deformations,
as well as the exceptional minimal $\cn=2$ LG models.
The last two correspond to a massive deformation of the genus 2 Gaiotto theory without
punctures, and a certain limit of it.  The 2d version of these last two theories
is not known.
It is remarkable that all complete $\cn=2$ gauge theories that admit a quiver realization
for their BPS states are classifiable, and even more surprisingly identifiable!
This gives further motivation for an even more complete classification of
$\cn=2$ theories by relaxing the completeness criteria.

The organization of the remainder of this paper is as follows:  In section 2 we discuss the general notion
of quivers relevant for finding the BPS states of 4d, $\cn=2$ theories.  In section 3
we give a definition of complete $\cn=2$ theories.  In section 4 we review the 4d/2d
correspondence advanced in \cite{cnv}.  In section 5 we discuss why the complete
$\cn=2$ theories map to quivers of finite mutation type and review the mathematical classification
of quivers of finite mutation type.  In section 6 we identify the class corresponding
to Riemann surfaces with punctures.  In section 7 we identify the exceptional ones.
In section 8 we identify the conformal subset. In section 9 we discuss some physical properties of gauging certain $\cn=2$ subsystems. Finally
in section 10 we present our conclusions.  Appendices A and B deal with certain technical computations.

\section{BPS Quivers}\label{sec:BPSquivers}

Quivers have been studied in the context of supersymmetric gauge theories
in two different ways.  In one context one uses them to describe gauge
theories with products of $U(N_i)$, one factor group per node, with bifundamental
matter being captured by links between nodes.  In another
approach, one uses quiver to describe BPS states of supersymmetric gauge
theories.  In this context \cite{douglasmoore,denef} one is considering a supersymmetric quantum
mechanical system, again with the $U(N_i)$ gauge groups at the nodes and
bifundamental matter.  In this latter sense, each node corresponds
to an elementary BPS state and one considers all
possible ranks $N_i$ for the gauge groups. The normalizable zero modes
for the quantum mechanics signify BPS bound states with the quantum numbers of $N_i$ copies of each elementary state.  It is this second
sense of quivers that would be of interest in the present paper.
We shall call the quivers interpreted in this sense the \emph{BPS quivers}.

\subsection{Generalities of quivers}\label{genQuivers}

Consider an $\cn=2$ theory in 4 dimensions which at a generic point on the Coulomb
branch has an abelian rank $r$ gauge symmetry $U(1)^r$.  In addition we assume
the theory has a rank $f$ flavor symmetry group  given by $U(1)^f$ for generic
values of mass deformations.  Then the total rank $D$ of the charges determining
the BPS mass of the $\cn=2$ theory is given by
$$D=2r+f$$
given by $r$ electric, $r$ magnetic and $f$ flavor charges.  The set of
BPS states should thus include at least $2r+f$ states.
We say an $\cn=2$ supersymmetric gauge theory admits a BPS quiver, if the following
conditions are satisfied: 

1) There are $2r+f$ BPS hypermultiplets with charges $\alpha_i\in \Gamma^{2r+f}$ with spin
0, with their $\cn=2$ central charge lying on the same half plane, and such that all the BPS states are given by a $positive$ linear combination of them, up to an overall conjugation.  
In other words, if there is a BPS particle of charge $\beta$, then
$$\beta =\pm \sum_{i=1}^{2r+f} n_i \,\alpha_i$$
where $n_i$ are positive integers.

2) There is a quiver supersymmetric quantum mechanics with 4 supercharges,
and $2r+f$ nodes,
with unitary gauge groups on each node, such that as we vary the ranks
of the unitary group, the ground states of the theory
are in 1--1 correspondence with the BPS states.  Moreover the nodes
are in 1--1 correspondence with the BPS states with charges $\alpha_i$, and
the ground states corresponding to the supersymmetric quantum mechanics with
gauge group $\prod U(n_i)$ corresponds to state(s) with charge $\beta=\sum_i
n_i \,\alpha_i$.

3) The number of bi-fundamental between the nodes $i,j$ is given by
the electro-magnetic skew--symmetric inner product $\alpha_i \cdot \alpha_j$.

4) As we change the parameters of the theory, and in particular when
one of the central charges $Z(\alpha_i)$ is about to exit
the same half plane as the other  $\alpha$'s, we replace
the corresponding BPS generator $\alpha_i$ with the conjugate state
with charge $-\alpha_i$.  Furthermore we replace all the
other BPS states with charge $\alpha_j$ which have positive inner product $n_{ij}$ with $\alpha_i$
with other BPS generators having charge
$$\alpha_j^\prime= \alpha_j+n_{ij}\, \alpha_i$$
leading to a new quiver which is {\it mutated} (see sect.\,\ref{sec:quiversoffinite} for more details on mutation).
As we will discuss below this is a Seiberg-like duality for quivers.

Given a BPS quiver, we can read off $r,f$ as follows:  Consider the skew-symmetric
matrix which we can read off from the quiver links, that is $B_{ij}=\alpha_i\cdot \alpha_j$.
The rank of $B$
is $2r$ while $f$ is the corank of $B$, \textit{i.e.}\! $f=D-2r$.

\subsection{Examples of BPS quivers}

Let us give examples of BPS quivers.  Consider for example type IIA in the
presence of $A_{n-1}$ singularity.  We model this by $\C^2/\Z_n$.
As it is well known \cite{douglasmoore}, if we consider BPS
states for this geometry we end up with the affine $A_{n-1}$ quiver,
corresponding to a supersymmetric quantum mechanical problem with 8
supercharges.  The bound states of this theory include states that correspond to the
roots of $SU(n)$.   These are the BPS states which complete the
$U(1)^{n-1}$ vector bosons to an $SU(n)$ vector multiplet.  These BPS
states correspond to D2 branes wrapped over the 2-cycles of this geometry.
Other examples, more relevant for this paper, are the local
Calabi-Yau threefolds. For example consider type IIA in the geometry
of $\C^3/\Z_3$.  Then the corresponding BPS states are given by the quiver
consisting of 3 nodes with three directed arrows (see Fig.\eqref{threearrows}):
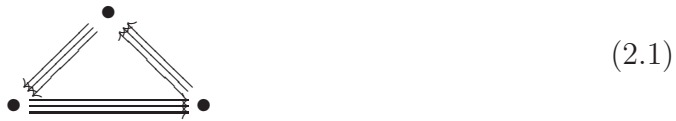
\begin{equation}\label{threearrows}
 \begin{gathered}
  \xymatrix{& \bullet \ar@<-0.45ex>[ld]\ar[ld]\ar@<0.45ex>[ld]& \\
\bullet \ar@<-0.45ex>[rr]\ar[rr]\ar@<0.45ex>[rr] && \bullet \ar@<-0.45ex>[ul]\ar[ul]\ar@<0.45ex>[ul]}
 \end{gathered}
\end{equation}

This theory corresponds to a supersymmetric quantum mechanical problem
with 4 supercharges (the same number as $\cn=1$ in 4d) which captures the
BPS states of the $\cn=2$ theory in 4d.  The presence of three nodes
reflects the fact that this theory can have bound states of D0, D2 and D4
branes, and for each of them there is only one allowed topological class.
Each node corresponds to a linear combination of these three charges. 
Note that, for generic ranks at each node, the number of incoming and outgoing
arrows at each node are not equal.  Of course this is not a problem for
the quantum mechanical system (unlike the 4d case, where the same quiver
would lead to an anomalous gauge theory unless the rank of the three nodes
are the same).  In addition to the quiver, this theory also has a
superpotential.  In principle for each closed oriented loop we can introduce a term
in the superpotential, and this theory indeed does have a superpotential
of the form
$$W=\epsilon_{ijk}\epsilon^{IJK} \,\mathrm{Tr}(A_{I}^iA_J^jA_K^k)$$
Where the $A_I^i$ label the $3\times 3$ bifundamental matter.
In addition the supersymmetric ground states of the quantum mechanics
depend on the choice of the FI parameters for each node, which depend
on the choice of moduli. Moreover as we change the moduli sometimes
the BPS quiver undergoes Seiberg-like dualities, known as mutations.
In this way, one of the nodes is replaced by a dual node (corresponding
to reversing the charge of that node), reversing the
direction of the arrows to that node, replacing the corresponding bifundamentals from the
node,
$q_j,\tilde q_i \rightarrow \tilde Q_j Q_i$, and adding to the new dual theory all the
meson fields which pass through the node $M_{ij}=\tilde q_i q_j$.   In addition one needs to
add, a term to the superpotential given by 
$$\delta W=Q_iM_{ij}\tilde Q_j.$$
That this Seiberg-like duality should take place is natural in the context of string theory.
Indeed applying T-duality to  D0-branes, and replacing them by spacetime filling D3-branes, leads to the same quiver.
Moreover in the cases where the resulting 4d theory is non-anomalous, the Seiberg-like
duality for BPS quiver, becomes T-dual to the standard Seiberg duality. 
 
The ground states of the new quiver may be different from that of the old one,
related to it by a suitable wall-crossing formula, as in \cite{ks1,Gaiotto:2008cd,Dimofte:2009bv,Dimofte:2009tm,Cecotti:2009uf}.

There is another general fact which follows from the geometry of the
D-branes.  As we noted, each node of the quiver corresponds to a BPS state,
which one can imagine as a brane wrapped over a cycle.  If we have two
nodes, corresponding to two different BPS states, clearly there will be
bifundamental strings at the intersections of the branes.  Thus we expect
the net number of bifundamentals between two nodes to be given by the
inner product of the corresponding classes.

So far we have given examples of simple quivers which arise from orbifolding.
However it is known that many other $\cn=2$ theories in 4d also have a BPS quiver.
For example it is known that the BPS quiver for the pure $SU(2)$ gauge theory
is given by the affine Dynkin diagram $\hat{A_1}$ \cite{denef}.  In fact this can simply be deduced
by the condition that one is looking for a basis of the BPS states which can
generate all the others by $positive$ linear combinations (up to overall
conjugation).  Inside the curve of marginal stability, we know that there are
only two BPS states, given by a monopole with (electric, magnetic) charge
given by 
$$\alpha_0=(0,1)$$ 
and a dyon with inner product two with the monopole, given by
$$\alpha_1=(2,-1)$$
Note that the electro-magnetic inner product given by
$$(e_1,m_1)\cdot (e_2,m_2)=e_1m_2-m_1e_2$$
in this case yields
$$\alpha_1 \cdot \alpha_0=2$$
Thus we obtain the quiver of the $SU(2)$ theory as given by the (oriented) affine Dynkin diagram  $\hat{A_1}$.  In the math literature the quiver corresponding to the affne $\widehat{A}_1$ Dynkin diagram with both arrows in the same direction is called the \emph{Kronecker quiver}:
\vglue 12pt
\begin{equation}\label{dinKhatA1}
 \begin{gathered}
\xymatrix{\alpha_0 && \alpha_1\ar@/^1.6pc/[ll]\ar@/_1.6pc/[ll]} 
 \end{gathered}
\end{equation}\vglue 12pt

The two nodes of the quiver have FI-terms.  The $U(1)$ part of the D-term for this quantum
mechanical problem will involve
$$(|q_1|^2+|q_2|^2+(f_0-f_1))^2$$
where $q_i$ denote the two bifundamentals, and $f_i$ denote the FI D-term
for each of the two nodes.  It is clear that for one sign of the FI term there
is no supersymmetric ground state.  This means that the only ground state arises when one of the
two nodes has zero rank, and so we will not have any $q_i$ fields.  As we
change the sign of FI-term we cross the curve of marginal stability, and now
we can have a bound state.  

The ground states of this theory have been studied by mathematicians \cite{kac,king,gabriel,MR2197389,auslander} in relation with the representations of quivers. See refs.\cite{douglas1,douglas2,fiol1,fiol2,denef} for discussions in the physical literature.  For this case it was shown that
the only allowed representations will have charges given by
\begin{equation}\label{A1spect}\alpha_0+ n(\alpha_0+\alpha_1)\ \ \mathrm{or} \ \ \alpha_0+\alpha_1.\end{equation}
The first series corresponds to dyons in the weak coupling region and the latter
correspond to the massive W boson \cite{fiol2,denef}.
Physically this result is obtained by analyzing the $D$--term equation \cite{douglas1,douglas2,fiol1,fiol2,denef}; we shall review the argument in a more general context in \S.\,\ref{sec:reptheory}.

Encouraged by this example, and assuming there is a BPS quiver description,
we can come up with a unique possibility for each matter representation of
$SU(2)$.  For example consider adding a quark in the fundamental representation.  Let us
consider the regime given by large quark mass.  In this limit the massive field
decouples without affecting the bound state structure for the pure $SU(2)$.
So we would still have the light degrees of freedom captured by the 
$\hat{A_1}$.  On the other hand we have in addition two massive fields which
should now be read off from the quiver as well.  These two have electric/magnetic charges given by
$(1,0),(-1,0)$.   In addition they both carry a charge $+1$ under
the additional $U(1)$ flavor symmetry. We need to add one of these two to
generate all the fields in terms of them.  We note that since $\alpha_0+\alpha_1=(2,0)$,
adding the $\alpha_2=(-1,0)$ as a new node for the quiver, would allow us to obtain the $(1,0)$
state using positive combination of the three nodes.  Thus we end up with the proposed
node charges for this theory given by 
$$(0,1),(2,-1),(-1,0)$$
leading to the quiver
\begin{equation}
 \begin{gathered}
  \xymatrix{\alpha_1\ar@<-0.9ex>[dd]\ar@<-0.1ex>[dd] & \alpha_2\ar[l]\\
&\\
\alpha_0\ar[uur] &}
 \end{gathered}
\end{equation}

We will later present evidence that this quiver correctly
reproduces the BPS states for $SU(2)$ with one fundamental field.
If we consider the matter representation of spin $j$, we get the
same quiver except with $2j$ lines connecting the extra node
with the first two nodes.  This is because the additional node
needed to generate all the BPS states is simply given by $(-2j,0)$.
In particular for the $\cn=2^*$ model, corresponding to mass deformations of the $SU(2)$
$\cn=4$ theory, we obtain:
\begin{equation}\label{markovqui}
 \begin{gathered}
  \xymatrix{\alpha_1\ar@<-0.8ex>[dd]\ar@<-0.2ex>[dd] & \alpha_2\ar@<-0.3ex>[l]\ar@<0.3ex>[l]\\
&\\
\alpha_0\ar@<-0.3ex>[uur]\ar@<0.3ex>[uur] &}
 \end{gathered}\qquad \text{(\textit{a.k.a.}\! Markov quiver).}
\end{equation}

Similarly for $N_f$ fundamentals, by the same decoupling
argument applied to $N_f$ very massive quarks, we get the quiver
obtained by adding $N_f$ nodes each of which is connected to the original
two nodes in the same way (\textit{i.e.}\! by single arrows making oriented triangles together with the $SU(2)$ double arrow):
\begin{equation}\label{SU(2)Nfquiver}
 \begin{gathered}
  \xymatrix{\alpha_1\ar@<-0.8ex>[dd]\ar@<-0.2ex>[dd] & \alpha_2\ar[l]\\
& \alpha_3\ar[lu]\ar@{..}[d]\\
\alpha_0\ar[uur]\ar[ur]\ar[r] & \alpha_{N_f+1}\ar[luu]}
 \end{gathered}
\end{equation}

We expect that, with generic enough superpotential for the
quiver, the resulting ground states are universal and insensitive to the precise
choice of the superpotential.   Moreover changing the FI--terms may result
in wall--crossing phenomena, but should not be necessary to specify the $\cn=2$
theories if we are to study them up to moduli deformation.
Since the BPS quiver captures the BPS degeneracies, it is natural
to ask if the quiver completely captures the corresponding $\cn=2$ theory.
The results of this paper seem to suggest that indeed BPS quivers faithfully represent the theory.  

The characterization of $4d$, $\cn=2$ theories using quivers is very powerful.
This shifts the classification of $\cn=2$ theories to classification of
allowed BPS quivers up to mutations.  
Above we have seen examples of $\cn=2$ theories for which
there is a quiver description.  Note that whether an $\cn=2$ theory
admits a quiver description may and in fact does depend
on which point on its moduli space we are considering.  An example of this
is the $\cn=2^*$ theory, say for the $SU(2)$ gauge group.  As we
have indicated, for sufficiently large mass for the adjoint matter,
there is a quiver description.  However, if the mass is turned off
we obtain an $\cn=4$ gauge theory.  It is easy to see that for this value
of moduli the $\cn=2^*$ theory cannot admit a quiver realization.
The reason is that we would need to come up with three BPS states
(since $r=1,f=1$) whose positive span contains all the BPS charges.
On the other hand we know that the BPS states of $\cn=4$ are given by one hypermultiplet and one vector multiplet (in the $\cn=2$ counting)
for each relatively prime $p,q$
with electromagnetic charge $(p,q)$.  Clearly this cannot be given by the positive
span of three vectors which are in the same half-plane.  In fact
quite generally if we consider the phase of the central charge of $\cn=2$
BPS states, the condition that they be spanned by a finite number of BPS states
implies that the phases of BPS central charges do not
form an everywhere dense subset of the circle, which is not the case for this theory.
Thus we have learned that there are some $\cn=2$ theories which
have BPS quivers in some region of the moduli but not at all points on the moduli.
 
From this example one may be tempted to conclude that all the $\cn=2$ theories
have at least some points on their moduli for which there is a BPS quiver
description.  However, this turns out not to be the case.  In fact
all the Gaiotto theories of rank 2 with $g>2$ and with no punctures are believed
to be of this type \cite{Gaiotto:2009hg,gaiottopriv}. These theories admit no mass deformation,
and in some sense are the analog of the $\cn=2^*$ at $m=0$ which are permanently
stuck there.
The case of $g=2$ with no punctures is different.  In one duality
frame, that theory corresponds to an $SU(2)^3$ theory with two
half-hypermultiplets in $(\mathbf{2},\mathbf{2},\mathbf{2})$.  The two half-hypermultiplets, form
one full hypermultiplet and that can receive a mass (though
its IR Seiberg-Witten geometry, unlike the $m=0$ point which is
given by Gaiotto curve, is unknown).
It is natural to conjecture that all the $\cn=2$ theories whose
BPS phases do not form a dense subspace of the circle admit a BPS quiver
description (of course as discussed this is a necessary condition).

\subsection{BPS spectra and representation theory}\label{sec:reptheory}

The BPS spectrum of an $\cn=2$ may also be understood in terms of the representation theory of the associated quiver $Q$ \cite{douglasmoore,douglas1,douglas2,fiol1,fiol2,denef}. A representation associates a vector space $V_i$ to each node $i$ of $Q$ and a linear map $V_i\xrightarrow{\phi_a} V_j$ to each arrow $i\xrightarrow{a} j$. We write $d_i=\dim V_i$ ($i=1\dots, D$) for the dimension vector of the representation; in terms of quiver quantum mechanics, $d_i$ corresponds to the rank $n_i$ of the gauge group at the $i$--th node.  

As a first example, consider 
the BPS spectrum of the $ADE$ Argyres--Douglas theories determined\footnote{\ See \cite{Bkell} {\textbf{Corollary 1.7}} for an equivalent mathematical statement.} in \cite{cnv,Shapere:1999xr}. The quiver $Q_\mathfrak{g}$ of these theories is simply the Dynkin diagram of the associated Lie algebra $\mathfrak{g}\in ADE$ with some orientation of the edges (all orientations being equivalent up to mutation \cite{fominIV}), so that the charge lattice gets identified with the root lattice of $\mathfrak{g}$,
$\Gamma\simeq \sum_i \Z\, \alpha_i$. The $ADE$ Argyres--Douglas theories have two\footnote{\ In fact many such chambers corresponding to different orientations of the Dynkin graph. These chambers have the same spectrum but differ for the BPS phase order \cite{cnv}. See also appendix \ref{app:strongcoupling}.} special\footnote{\ For $\mathrm{rank}\,\mathfrak{g}>2$ there are other BPS chambers as well. The BPS spectrum is always finite.} chambers, \textbf{(S)} and \textbf{(W)}, having a finite BPS spectrum consisting, respectively, of
\begin{description}
\item[(S)] one BPS hypermultiplet for each simple root with charge vector $\alpha_i$;
\item[(W)] one BPS hypermultiplet for each positive root of $\mathfrak{g}$ with charge vector the same positive root $\sum_i n_i\,\alpha_i$, ($n_i\geq 0$).
\end{description}

This result may be understood in terms of the Gabriel theorem \cite{gabriel,MR2197389,auslander} which puts the above Argyres--Douglas models in one--to--one correspondence with the quivers having finitely  many non--isomorphic indecomposable representations. The Gabriel map sends the representation of a Dynkin quiver with dimension vector  $d_i$ into the element of the root lattice $\sum_i d_i\, \alpha_i\in \Gamma_\mathfrak{g}$. Under this map, the simple representations correspond to the simple roots $\alpha_i$, and the indecomposable representations to the positive roots. 

Gabriel theorem has being generalized to arbitrary quivers by Kac \cite{kac}. So the charge lattice may be always identified with the root lattice of some Lie algebra, and stable BPS states are mapped to positive roots under this identification. Real positive roots correspond to \emph{rigid} indecomposable representation (no continuous moduli) so they are naturally related to BPS hypermultiplets; imaginary positive roots have moduli so, in general, they correspond to higher spin BPS multiplets.
Which positive roots actually correspond to stable BPS particles depends on the particular chamber. Concretely, 
given a quiver $Q$ we consider the central charge function $Z(\cdot)$ which associates to a representation $R$, having dimension vector $d_i(R)$, the complex number $Z(R)=\sum_i d_i(R)\, Z_i$, where $\arg Z_i\in [0,\pi[$. We say that a representation $R$ is stable (with respect to the given $Z(\cdot)$) if \cite{Bkell} 
\begin{equation}\label{stabcond}
\arg Z(S)<\arg Z(R)
\end{equation}
for all proper subrepresentations $S$ of $R$ (this condition is called $\Pi$--stability in \cite{douglas1,douglas2}). Physically, this is the requirement that the BPS state of charge vector $\sum_i d_i(R)\,\alpha_i$ cannot decay into states having charge $\sum_i d_i(S)\,\alpha_i$ because there is no phase space.

Notice that simple representations, associated to the simple roots $\alpha_i$, correspond to BPS hypermultiplets which are stable in all chambers. The existence of such a spanning set of universally stable hypermultiplets is a necessary condition for the $\cn=2$ theory to admit a quiver in the present sense. 

As anticipated above, this representation--theoretical stability condition may be understood from the quiver quantum mechanics viewpoint as a consequence of the $D$--term equation in presence of  FI terms which depend on the given central charges $Z_j=m_j\, e^{i\theta_j}$. Without changing the chamber, we may assume that the $\arg Z_i$'s are all very close together. Then, if $\arg Z(R) =\alpha$,
\begin{equation}\begin{split}
Z(S)/Z(R) &= \frac{\sum_j d_j(S)\, m_j\, e^{i(\theta_j-\alpha) }}{|Z(R)|} \approx \frac{1}{|Z(R)|} \Big(\sum_j d_j(S)\, m_j +i \sum_j d_j(S)\, m_j\,(\theta_j-\alpha)\Big)\\
&= r_1+\frac{i}{|Z(R)|}\sum_j d_j(S)\,\vartheta_j\end{split}
\end{equation}
where $r_1$ is real positive and $\vartheta_j=m_j(\theta_j-\alpha)$. Thus the stability condition
\eqref{stabcond} is equivalent to the condition that 
\begin{equation}\label{stabcon2}
\sum_i d_j(S)\, \vartheta_j <0
\end{equation} 
for all proper subrepresentations $S$ of $R$ (this condition is called $\vartheta$--stability \cite{king}).
A 
 theorem by King (\textbf{Proposition 6.5} of \cite{king}) states that an indecomposable representation $R$ is $\vartheta$--stable if and only if it satisfies the equation
\begin{equation}
\sum_{t(\alpha)=j}\Phi_\alpha^\dagger\Phi_\alpha-\sum_{h(\alpha)=j}\Phi_\alpha\Phi^\dagger_\alpha= \vartheta_j\, \boldsymbol{1},
\end{equation} 
 which is the $D$--term equation in presence of the FI terms $\vartheta_j$.
 \medskip

After the $ADE$ Argyres--Douglas models, the next simplest instances  are the $\cn=2$ theories having a quiver $Q$ whose underlying graph is an affine  $\widehat{A}\widehat{D}\widehat{E}$ Dynkin diagram with arrows oriented in such a way that there are no oriented cycles. Up to equivalence, the affine quivers are
\begin{enumerate}
 \item $\widehat{A}(p,q)$, with $p\geq q\geq 1$, corresponding to the $\widehat{A}_{p+q-1}$ Dynkin diagram oriented in such a way that $p$ arrows point in the positive direction and $q$ in the negative one. We exclude $q=0$ since $\widehat{A}(p,0)\sim D_p$ and we get back a Argyres--Douglas model;
\item $\widehat{D}_r$, $\widehat{E}_6$, $\widehat{E}_7$, and $\widehat{E}_8$. In these cases, since the Dynkin diagram is a tree, all orientations are mutation equivalent.
\end{enumerate}
The charge lattice is identified with the root lattice $\Gamma_{\widehat{\mathfrak{g}}}$, and the only charge vectors which may possibly correspond to stable BPS states are:
\begin{itemize}
 \item \emph{real} positive roots $\Rightarrow$ BPS hypermultiplets;
\item the indivisible imaginary root $\delta$ $\Rightarrow$ BPS vector--multiplet.
\end{itemize}
In particular, in any BPS chamber, we have \emph{at most} one vector; indeed one of the result of the present paper is that affine $\cn=2$ theories correspond to a single $SU(2)$ SYM coupled to a vector--less $\cn=2$ system.

The simple roots are always stable. In fact, there exists a chamber, corresponding to the strong coupling regime, in which the \emph{only} states are those associated to the simple roots\footnote{\ For an argument along the lines of \cite{cnv}, see appendix \ref{app:strongcoupling}. }.
Indeed, we may number the nodes of an affine quiver, without oriented cycles, from $1$ to $D$ in such a way that each vertex $i$ is a source in the full subquiver of vertices $1,\cdots, i$ \cite{Bkell,kellrecrel}. In this numeration, if we have
\begin{equation}
\arg Z_1< \arg Z_2 < \cdots < \arg Z_D
\end{equation}
we see recursively that the indecomposable representations are just the simple roots. 

In the weak coupling regime the state associated to $\delta$, \textit{i.e.}\! the $W$--boson, is stable together with a tower of hypermultiplets corresponding to a certain subset of $\Delta_+^\mathrm{re}$.

We close this section by checking these predictions for $SU(2)$ $\cn=2$ SQCD with $N_f=0,1,2,3$ fundamental hypermultiplets \cite{Seiberg:1994rs,Seiberg:1994rs2,Gaiotto:2009hg}.
The case $N_f=0$, corresponding to the quiver $\widehat{A}(1,1)$, was already discussed around eqn.\eqref{dinKhatA1}. It is easy to check that the stable representations in the weak coupling chamber, namely $\delta$ and the real positive roots, correspond to the BPS states present in the physical spectrum \cite{Bkell,fiol2}.
\vglue 9pt

\noindent\underline{$N_f=1$}\vglue 12pt

Mutating\footnote{\ Detailed definitions of the quiver mutations are given in section \ref{sec:quiversoffinite}. } the $N_f=1$ quiver \eqref{SU(2)Nfquiver} at the hypermultiplet vertex (indicated by a curled arrow in the figure) we get the affine $\widehat{A}_2(2,1)$ quiver
\begin{equation}
\begin{gathered}
\xymatrix{\bullet \ar@<-0.9ex>[dd]\ar@<-0.3ex>[dd] & \bullet\ar[l]\\
& \ar@<-0.5ex>@{~>}[u]\\
\bullet \ar[uur] &}
\end{gathered}\quad\boldsymbol{\longrightarrow}\quad
\begin{gathered}
\xymatrix{\alpha_1\ar[r]\ar@<-0.3ex>[dd] &\alpha_2\ar[ddl]\\
&\\
\alpha_0 &}
\end{gathered}
\end{equation}

One has $2e\equiv \delta=\alpha_0+\alpha_1+\alpha_2$ while the flavor charge is proportional to
$f=\alpha_2-(\alpha_0+\alpha_1)$, so in terms of the usual charges $(e,m,f)$ the affine simple roots are
\begin{equation}
 \alpha_0=(0,1,-1), \qquad \alpha_1=(1,-1,0),\qquad \alpha_2=(1,0,1).
\end{equation}
which is the correct strong coupling spectrum.
The known weak coupling spectrum is also consistent with representation theory.

\vglue 12pt

\noindent\underline{$N_f=2$}\vglue 12pt

Mutating the $N_f=2$ quiver \eqref{SU(2)Nfquiver} at both hypermultiplet vertices  we get the affine $\widehat{A}_3(2,2)$ quiver
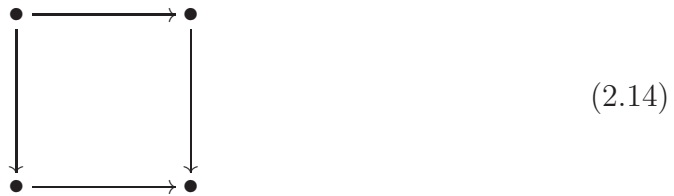
\begin{equation}
\begin{gathered}
\xymatrix{\bullet\ar[rr]\ar[dd]&& \bullet\ar[dd]\\
&&\\
\bullet\ar[rr] && \bullet}
\end{gathered}
\end{equation}
Again, the strong coupling BPS spectrum is given by four hypermultiplets of charges $\alpha_0, \alpha_1, \alpha_2,\alpha_3$. In the  weak coupling we have a vector multiplet of charge $\alpha_0+\alpha_1+\alpha_2+\alpha_3$ and a tower of BPS hypermultiplets whose charge vectors belong to $\Delta_+^\mathrm{re}(\widehat{A}_3)$.
\vglue 12pt

\noindent\underline{$N_f=3$}\vglue 12pt

Mutating the $N_f=3$ quiver one gets the $\widehat{D}_4$ {affine} quiver
  
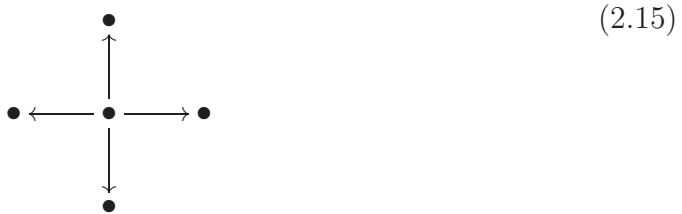
\begin{gather}\label{d4quiver}\xymatrix{
& \bullet & \\
\bullet & \bullet \ar[l]\ar[u]\ar[r]\ar[d] & \bullet\\
& \bullet &}
\end{gather}
Again, the strong coupling spectrum  consist of five hypermultiplets with charge vectors $\alpha_0,\alpha_1,\alpha_2,\alpha_3,\alpha_4$, while in the weak coupling we have one  BPS vector multiplet with charge vector
\begin{equation}\sum_{i\neq 1} \alpha_i +2\, \alpha_1,
\end{equation}
where $\alpha_1$ is the simple root associated to the central node in \eqref{d4quiver}, and the usual tower of dyons with charge vectors in $\Delta_+^\mathrm{re}(\widehat{D}_4)$.

\section{Definition of Complete $\cn=2$ theories}

In this section we motivate the definition of a special class of
$\cn=2$ theories which we will call `complete $\cn=2$ gauge theories'.
Consider an $\cn=2$ theory with $D=2r+f$ BPS charges. This in particular means
that we have $D$ central charges $Z_i\in \C$, with $i=1,...,D$ which appear in the BPS algebra.  It is natural to ask if they can be arbitrarily varied. 
In other words we are asking if the map from the moduli space ${\cal M}$ to D-dimensional complex plane, giveny by the central charges,
$$Z:\ {\cal M}\rightarrow \C^D$$ 
is at least locally onto.
 For
this to happen we need to have at least $D$ complex parameters in the moduli space ${\cal M}$
of the theory.  Quite generally we can identify $r$ complex parameters with
labelling the Coulomb branch, and $f$ parameters for varying the masses.
In addition there could be additional coupling constants.  In order
to vary the central charges independently, we need at least $r$ additional
parameters.  This suggests that if we can in addition vary the $r$ coupling constants
of the theory independently, then we have a complete $\cn=2$ theory.  Note that
this latter condition may not be possible in general.  For example, for
$SU(N)$ gauge theory we expect only one coupling constant but $r=N-1$ dimensional
Coulomb branch.  We can in principle formally deform the coupling
constants of the $U(1)$'s in the IR, but there is no guarantee that there is
a UV complete theory which allows this (in fact it follows from the results of this paper
that this is not possible).  Moreover, there are some $\cn=2$ theories
which do not even have a freedom to vary one coupling constant.  For example
the Minahan--Nemeschansky theories \cite{minahan1,minahan2} are of this type, where the coupling constant
is completely fixed by the masses and the point on the Coulomb branch.

On the other hand it is clear that an $\cn=2$ theory consisting of asymptotically
free matter spectrum with a gauge group $G=SU(2)^{\otimes r}$ is complete
in the above sense, because we have $r$ couplings, $r$ Coulomb branch parameters,
and one mass parameter for each matter representation.  In particular all the
rank 2 Gaiotto theories \cite{gaiotto} are complete in this sense.  
One can also ask if the dimension of ${\cal M}$ can be bigger than $D$.
This is in principle possible, because the coupling constants of a $U(1)^r$
theory is a symmetric complex $r\times r$ matrix, which has $(r^2+r)/2$ entries.
Nevertheless, the results of this paper imply that the dimension of ${\cal M}$ is at most D, which
gets saturated by complete theories.

The question we pose is the classification of all complete $\cn=2$ gauge theories
which admit a BPS quiver.  In order to accomplish this, we will use the 4d/2d correspondence of \cite{cnv} that we will review in the next section.

\section{4d-2d Correspondence Reviewed}\label{sec:2d4drev}
There has been a number of links between 4d $\cn=2$ theories and 2d QFT's.  In particular
two such correspondences were suggested in \cite{cnv}.  In this section
we review one of those conjectured correspondences, which proves
important for our applications.

This duality maps 4d theories with $\cn=2$ supersymmetry (with 8 supercharges) to 2d theories with $\cn=2$ (with 4 supercharges).   The specific case where
the map can be demonstrated explicitly is for $\cn=2$ theories in 4d which can be constructed in type II strings on local Calabi-Yau manifolds.  The idea is that the worldsheet of
the type II strings involves an $\cn=2$ superconformal theory, with $\hat c=3$.
Furthermore when the 4d $\cn=2$ theory can be decoupled from gravity,
one is discussing the geometry near a local singularity of Calabi-Yau.
In such a case, one can expect that the theory has a Liouville field,
and that the $\cn=2$ worldsheet theory decomposes to a mixed product of the Liouville field and an $\cn=2$ 2d QFT.  The accompanying $\cn=2$ QFT may be massive or conformal, which can be read off by freezing the value
of the Liouville field.  This worldsheet $\cn=2$ theory could be massive without contradiction as its coupling to Liouville can make it conformal.  Moreover,
since the central charge of the Liouville is $\hat c\geq 1$, this implies
that the central charge of the accompanying 2d theory is $\hat c\leq 2$.

An example of this is the following:  Consider again type IIA on the local
Calabi-Yau threefold given by $\C^3 / \Z_3$ or its blow ups, which is the total space of $O(-3)$ line bundle
over $\mathbb{P}^2$.  Then the worldsheet theory has a mirror Landau-Ginzburg
description given by \cite{Hori:2000kt,Hori:2000ck},
$$W=\exp(-Y_1)+\exp(-Y_2)+\exp(-Y_3)+ \exp(+Y_1+Y_2-3Y_3) \exp(-t)$$
where $Y_i$ are chiral $\C^*$ valued superfields, and $t$ denotes the complexified
Kahler class of $\mathbb{P}^2$.  We can treat an overall shift of $Y$ as a Liouville
field.  Fixing that, will yield a theory with one less field given by
$$W=\exp(-Y_3)\big[\exp(-Y'_1)+\exp(-Y_2')+1+ \exp(Y_1'+Y_2')\exp(-t)\big]=
\exp(-Y_3)\cdot W'(Y_1',Y_2')$$
where 
$$Y_1'=Y_1-Y_3, Y_2'=Y_2-Y_3$$
One recognize $W'(Y_1',Y_2')$ as the superpotential for massive
2d theory which is  the mirror of sigma model to $\mathbb{P}^2$ \cite{Cecotti:1993rm,Hori:2000kt}.

Similarly, if we consider the type IIA on a Calabi-Yau corresponding to $\C^2/\Z_2\times \Z_2$ or its blow up,
the total space of the $O(-2,-2)$ bundle over $\mathbb{P}^1\times \mathbb{P}^1$, similar
manipulations (see \cite{Hori:2000ck}) will yield a factor $W'$ given by
$$W'=\exp(-X_1)+\exp(X_1)\exp(-t_1)+ \exp(-X_2)+\exp(X_2)\exp(-t_2)+1$$
where the $t_i$ are the two complexified Kahler classes of the $\mathbb{P}^1$'s.
Again, one recognizes $W'$ as the mirror to the 2d sigma model on $\mathbb{P}^1\times \mathbb{P}^1$.
By taking a special limit (corresponding to taking one of the $\mathbb{P}^1$'s
much larger than the other) leads to geometric engineering of $\cn=2$ pure $SU(2)$
in 4 dimensions, leading to a 2d factor with superpotential (after an overall rescaling of
W')
$$W'\rightarrow \exp(-X_1)+\exp(+X_1)+X_2'^2+u$$
where one recognizes $W'=0$ as the SW curve for the pure $SU(2)$ theory.
This 2d factor is equivalent to the mirror of the sigma model on $\mathbb{P}^1$
(where the $X_2'$ part gives a trivial massive theory).

From these examples the general idea emerges that at least
for all the $\cn=2$ theories which can be engineered in type II strings,
we would obtain an accompanying 2d $\cn=2$ theory which is the factor
of the worldsheet theory.  However, there is more to this map.
The BPS quivers of the 4d theories naturally encode the soliton data
of the corresponding 2d theory.  The nodes of the 4d BPS quiver map to the
2d vacua, and the lines connecting them map to the soliton between them.\footnote{\ The extra
data of orientation of the arrows is also encoded in the 2d theory  in an implicit way, as we
discuss later in the context of examples.}  In particular we recognize the 4d BPS quiver
of the $\C^3/\Z_3$ model as encoding the three vacua of the $\mathbb{P}^2$ model
and the corresponding bifundamentals as mapping to the kinks connecting them, and similarly
that of the $\C^2/\Z_2\times \Z_2$, which maps to the 2d data of the $\mathbb{P}^1\times \mathbb{P}^1$
sigma model.   Another example is the theory
corresponding to $\cn=2$ theory for the pure $SU(2)$.
As we just saw the corresponding 2d theory corresponds to the sigma
model on $\mathbb{P}^1$.  This massive theory has two vacua and two solitons
between the two.  This is exactly the structure of the quiver
for the $SU(2)$ theory as we already discussed. 

The idea for this map is that there are canonical D-branes associated to LG vacua,
as discussed in \cite{Hori:2000ck}, corresponding to Lagrangian subspaces of LG.
These we can identify with the worldsheet description of the BPS states.
Moreover the intersection pairing between these Lagrangian cycles in 2d
was mapped in \cite{Hori:2000ck} to the number of kinks connecting the vacua.  On the
other hand the intersection of D-branes give bifundamental fields, thus
explaining this connection.

Based on many such examples it was suggested in \cite{cnv} that for every
$\cn=2$ theory in 4d, there is an associated 2d theory with $\cn=2$ supersymmetry.
Moreover it was proposed that the quiver of the 4d theory get mapped
to the vacua and kink structure of the 2d theory.  On the other hand
we know that not every 4d theory has a quiver description.  This
actually has a 2d counterpart:  Not every 2d theory has isolated
vacua and kinks between them.  Thus the 4d/2d correspondence is more
general than the map between their associated quivers.
In this paper we assume the validity of this correspondence and use
it to classify complete $\cn=2$ theories in 4d, which were defined in the previous section.

\section{Complete $\cn=2$ theories and quivers of finite mutation type}\label{sec:finitecomplete} 

In this section we argue that complete $\cn=2$ theories in 4d are mapped
to 2d theories with $\hat c\leq 1$ in the UV.  We will be interested
in the case where both theories admit a quiver, though we believe
the map is more general.  Furthermore we review the mathematical classification
of quivers of finite mutation type.

\subsection{Completeness and finiteness of mutation type}
The basic idea for showing the connection between completeness and finiteness of
mutation type for the quiver is very simple:
First we will assume that the 4d theory admits a BPS quiver.  In such a case
we are looking for theories whose dimension of moduli space is equal to the
number of nodes.  On the other hand, mapping this theory to 2d, and identifying
the nodes, with vacua, it means that we are looking for 2d theories which have
as many deformations as the number of massive vacua.  For 2d theories, with $\cn=2$
we know that, in the UV, the number of allowed deformations is given by the number of operators
with dimension less than or equal to 1, \textit{i.e.}\! relevant or marginal operators.
On the other hand there are as many chiral fields as the vacua, with the
highest chiral field having dimension $\hat c$.  Since the dimension
of deformations is equal to the number of vacua, this means all chiral
fields can be used to deform it, including the one with maximal dimension.  But given the bound on the allowable
deformations, this implies that $\hat c\leq 1$.

On the other hand we can ask the question of what kinds of quivers are allowed
for 2d theories with $\hat c\leq 1$.  We argue that these must have a finite
mutation type.  In other words, there cannot be infinitely many mutation
orbits of the quiver.  Indeed, as noted before, the mutation of the quiver
maps to wall crossing for the 2d BPS states.  But since we have as many
parameters to vary as the number of vacua, we can use this freedom
to induce arbitrary wall crossings for the 2d theory.  On the other hand
each wall crossing leads to a mutation of the quiver.  Thus arbitrary mutations
of the quiver are physically realized.  Moreover since we have enough
parameters we can decouple as many vacua as we wish.  In particular we can
decouple all vacua except for any fixed pair.  In this way we end up with
a theory with only two vacua with some kinks between them.  It is known 
\cite{Cecotti:1993rm} that the number of kinks between them is less than or equal to 2
for the theory to exist.  This implies that no matter what quiver mutations
we consider, the number of links between any pair cannot grow more than 2 for complete
$\cn=2$ theories.  This in particular implies that the quivers of complete
$\cn=2$ theories should be finite in number (otherwise this number would grow
at least for a pair of vacua).  

It turns out that the quivers of finite mutation type have been classified
by mathematicians \cite{fomin,derksen,felikson}.  From what we have said above, we need to further
restrict to quivers where there is no more than two links between any pairs
of nodes.  This turns out to be automatically true for all quivers of finite
mutation type with more
than two nodes and so we do not need to further impose this condition.

On the other hand for quivers with two nodes, we need to restrict to ones
with less than three links.  

Of course it is not clear that all the quivers of finite mutation type
(apart from the restriction for the two node case) do arise for some complete
$\cn=2$ gauge theory.  We have only shown that complete gauge theories
lead to finite mutation type quivers.  Nevertheless we show this is also
sufficient and identify each finite mutation type quiver with
a unique $\cn=2$ theory in 4d.  Before doing so, in the next subsection
we review the mathematical result for classification of quivers
of finite mutation type.

\subsection{Quivers of finite mutation type}\label{sec:quiversoffinite}

The class of quivers of interest in $\cn=2$ theories are the `\emph{$2$--acyclic}', namely the ones without loops (arrows which start and end in the same node) and no arrows with opposite orientations between the same two nodes.  Physically
this is because a loop corresponds to an adjoint matter which can be given mass and thus disappear from consideration
of BPS spectrum.  For the same reason only the net number of bi-fundamentals between pairs of nodes enter the discussion because the others can
be paired up by superpotential mass terms and disappear from the study of ground states of the SQM.   In
this paper when we discuss quivers we restrict to this class.
 Specifying such a quiver $Q$ with $D$ nodes is equivalent to giving  an integral $D\times D$ skew--symmetric matrix $B$  (called the \emph{exchange matrix}) whose $(i,j)$
entry is equal to the number of arrows from the $i$--th node to the $j$--th one (a negative number meaning arrows pointing in the opposite direction $j\rightarrow i$). 

A mutation of such a quiver $Q$ is given by a composition of elementary mutations. There is an elementary mutation for each vertex of $Q$. The elementary mutation at the $k$--th vertex, $\mu_k$, has the following effect on the quiver \cite{MR1887642,MR2004457,fominIV} (for reviews see \cite{MR2383126,MR2132323,cluster-intro}):
\begin{enumerate}
 \item It inverts the direction of all arrows going in/out the $k$--th vertex;
\item each triangle having $k$ as a vertex gets mutated as in the following figure
\begin{center}
 \begin{tabular}{|c|c||c|c|}\hline
$Q$ & $\mu_k(Q)$ & $Q$ & $\mu_k(Q)$\\\hline
 $\begin{diagram}
  \node{i} \arrow[2]{e,t}{r}\arrow{se,b}{s} \node{} \node{j}\\
\node{}\node{k}\arrow{ne,b}{t}
 \end{diagram}$
 &$
   \begin{diagram}
  \node{i} \arrow[2]{e,t}{r+st} \node{} \node{j}\arrow{sw,b}{t}\\
\node{}\node{k}\arrow{nw,b}{s}
 \end{diagram}
$ & $\begin{diagram}
  \node{i} \arrow[2]{e,t}{r}\arrow{se,b}{s} \node{} \node{j}\arrow{sw,b}{t}\\
\node{}\node{k}
 \end{diagram}$
 &$
   \begin{diagram}
  \node{i} \arrow[2]{e,t}{r} \node{} \node{j}\\
\node{}\node{k}\arrow{nw,b}{s}\arrow{ne,b}{t}
 \end{diagram}
$
\\\hline
$
 \begin{diagram}
  \node{i} \arrow[2]{e,t}{r} \node{} \node{j}\arrow{sw,b}{t}\\
\node{}\node{k}\arrow{nw,b}{s}
 \end{diagram}
$
& 
$
  \begin{diagram}
  \node{i} \arrow[2]{e,t}{r-st}\arrow{se,b}{s}\node{} \node{j}\\
\node{}\node{k}\arrow{ne,b}{t}
 \end{diagram}
$ & 
$
 \begin{diagram}
  \node{i} \arrow[2]{e,t}{r} \node{} \node{j}\\
\node{}\node{k}\arrow{nw,b}{s}\arrow{ne,b}{t}
 \end{diagram}
$
& 
$
  \begin{diagram}
  \node{i} \arrow[2]{e,t}{r}\arrow{se,b}{s}\node{} \node{j}\arrow{sw,b}{t}\\
\node{}\node{k}
 \end{diagram}
$
\\\hline
\end{tabular}\label{boxquivemut}
\end{center}
where $r$, $s$, $t$ are non-negative integers, and an arrow $i\xrightarrow{l} j$ with $l\geq 0$ means that $l$ arrows go from $i$ to $j$ while an arrow $i\xrightarrow{l} j$ with $l\leq 0$ means $|l|$ arrows going in the opposite direction.
 \end{enumerate}

In terms of the exchange matrix $B_{ij}$ the mutation $\mu_k$ reads \cite{fominIV,MR1887642,cluster-intro}
\begin{equation}
 \mu_k(B_{ij})=\begin{cases}
                - B_{ij} & \text{if } i=k\ \text{or }j=k;\\
B_{ij}+ \mathrm{sign}(B_{ik})\,\max\{B_{ik}B_{kj},0\} &\text{otherwise.}
               \end{cases}
\end{equation}
The definition implies that $\mu_k$ is an involution:
\begin{equation}\label{muinv}
 (\mu_k)^2=\text{identity}.
\end{equation} 
From the box we see that the mutation $\mu_k$ is particularly simple when the node $k$ is either a sink (all arrows incoming) or a source (all arrows outgoing). In these cases, $\mu_k$ just inverts the orientation of the arrows trough the $k$--th node.

Two quivers are said to be in the \emph{same mutation--class} (or mutation--equivalent) if one can be transformed into the other by a finite sequence 
of such elementary mutations.
A quiver is said to be \emph{mutation--finite} if its mutation--class contains only finitely many distinct quivers.

There is a Java applet due to B. Keller \cite{kellerapp} which implements the quiver mutations and computes the mutation--class of a quiver up to sink/source equivalence (\textit{i.e.}\! two quivers are identified if they differ by a mutation at a sink/source).
\medskip

\begin{figure}

 \begin{align*}
 &\widehat{\widehat{E}}_6\colon &&\begin{gathered}
\xymatrix{ & & \bullet\ar@<-0.35ex>[dd]\ar@<0.35ex>[dd] &\bullet\ar[l]& \bullet\ar[l]\\
\bullet \ar[r] &\bullet\ar[ur] & &\\
&& \bullet \ar[lu]\ar[ruu]\ar[r] & \bullet\ar[uul] \ar[r] &\bullet}
                            \end{gathered}\\
&\widehat{\widehat{E}}_7\colon &&\begin{gathered}
\xymatrix{ & \bullet\ar@<-0.35ex>[dd]\ar@<0.35ex>[dd] &\bullet\ar[l]& \bullet\ar[l]&\bullet\ar[l]\\
\bullet\ar[ur] & & &\\
& \bullet \ar[lu]\ar[ruu]\ar[r] & \bullet\ar[uul] \ar[r] &\bullet\ar[r]&\bullet}
                            \end{gathered}\\
&\widehat{\widehat{E}}_8\colon &&\begin{gathered}
\xymatrix{ & \bullet\ar@<-0.35ex>[dd]\ar@<0.35ex>[dd] &\bullet\ar[l]& \bullet\ar[l]&\bullet\ar[l]& \bullet\ar[l] &\bullet\ar[l]\\
\bullet\ar[ur] & & & & &\\
& \bullet \ar[lu]\ar[ruu]\ar[r] & \bullet\ar[uul]\ar[r] &\bullet&&&}
                          \end{gathered}\\
&X_6\colon && \begin{gathered}
\xymatrix{& \bullet\ar[rd] && \bullet\ar@<0.4ex>[dr]\ar@<-0.4ex>[dr]  &\\
\bullet \ar@<0.4ex>[ur]\ar@<-0.4ex>[ur] && \bullet \ar[ll]\ar[ur]\ar[d] && \bullet\ar[ll]\\
&& \bullet &&}\end{gathered}\\
&X_7\colon && 
\begin{gathered} \xymatrix{& \bullet\ar[rd] & & \bullet\ar@<0.4ex>[dr]\ar@<-0.4ex>[dr] &\\
\bullet\ar@<0.4ex>[ur]\ar@<-0.4ex>[ur]&&\bullet\ar[ll]\ar[ur]\ar[rd] && \bullet\ar[ll]\\
& \bullet\ar[ru] && \bullet \ar@<0.4ex>[ll]\ar@<-0.4ex>[ll] &}\end{gathered}
 \end{align*}
\caption{\label{ellipticEs}The three elliptic $E$--type Dynkin diagrams oriented as to give finite mutation quivers, and the two Derksen--Owen quivers.}
\end{figure}
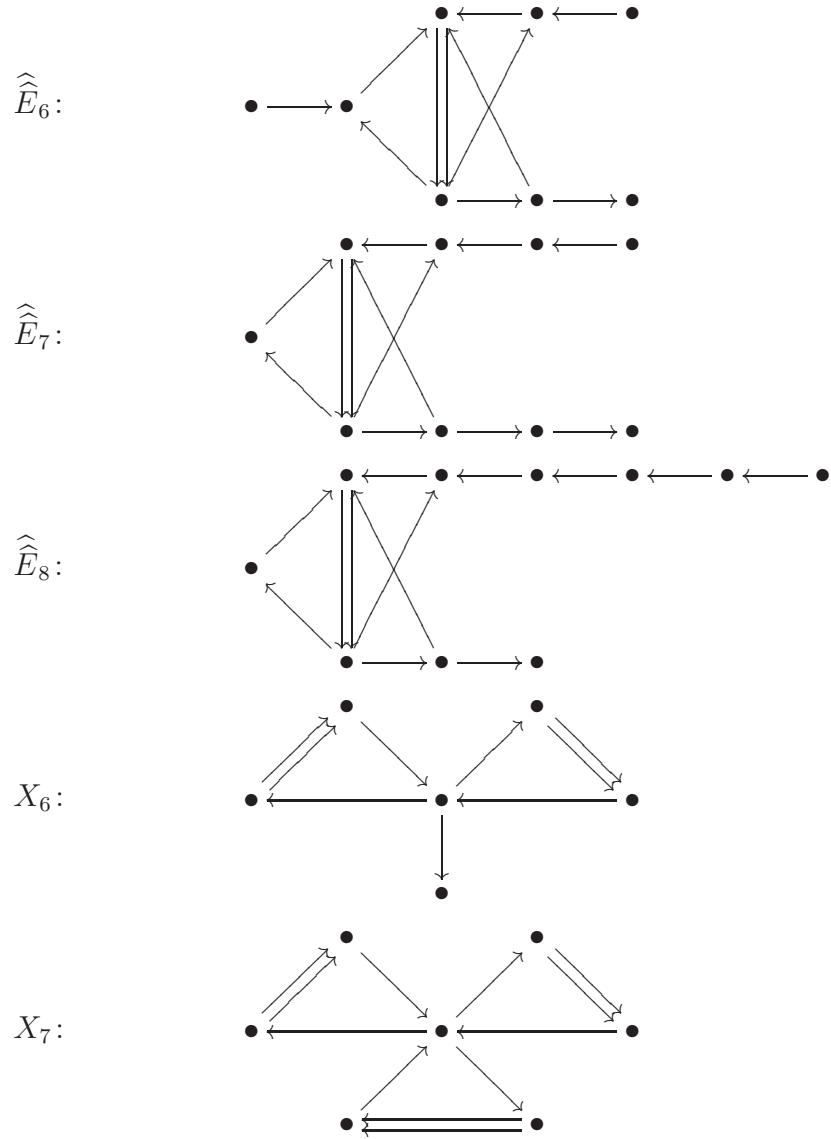

According to the Felikson--Shapiro--Tumarkin theorem \cite{felikson} the complete list of mutation--finite quivers is the following:
\begin{enumerate}
 \item quivers with at most two nodes;
\item quivers representing adjacency matrices of ideal triangulations of bordered surfaces with punctures and marked points on the boundaries \cite{fomin} (to be discussed in the next subsection);
\item the quivers mutation equivalent to the nine $E$--type Dynkin diagrams\footnote{\ In Saito's notation \cite{saito} the root system $\widehat{\widehat{E}}_r$ is written as $E_r^{(1,1)}$. }
\begin{align*}
  \text{finite:}&\qquad E_6, E_7, E_8\\
 \text{affine:}&\qquad \widehat{E}_6, \widehat{E}_7, \widehat{E}_8\\
 \text{elliptic:}&\qquad \widehat{\widehat{E}}_6, \widehat{\widehat{E}}_7, \widehat{\widehat{E}}_8,
\end{align*}
having rank $D$ equal to the sum of the subscript plus the number of hats.  The quivers associated to the
unhatted and single hatted $E$--theories are the usual Dynkin diagrams of the $E$--type, and different orientation
of the arrows give mutation equivalent quivers.
For $\widehat{\widehat{E}}_r$ the arrows are cyclicly oriented in all triangles (all such orientations are mutation equivalent) see figure \ref{ellipticEs};
\item the two Derksen--Owen mutation classes $X_7$ and $X_6$, (of rank $7$ and $6$, respectively) \cite{derksen}. There are five distinct quivers in the class of $X_6$, and just two in the one of $X_7$). See figure \ref{ellipticEs}.
\end{enumerate}
 
In particular, all finite--mutation quivers with more than 10 nodes arise from ideal triangulations of surfaces in the sense of \cite{fomin}. 

\medskip

In \cite{cnv} it was  shown that the only two--node quivers which correspond to sensible $4d$ $\cn=2$ theories are (orientations) of the Dynkin graphs of $A_1\times A_1$, $A_2$ and $\widehat{A}_1$. 
If $Q$ is a finite--mutation quiver with $D\geq 3$, all its mutation--equivalent quivers have at most \emph{double} arrows. The same is true for the three $D=2$ Dynkin quivers $A_1\times A_1$, $A_2$ and $\widehat{A}_1$. Then the property characterizing quivers corresponding to complete $\cn=2$ models is that in their mutation class there is no quiver with arrows of multiplicity $>2$. When in this paper we loosely refer to finite--mutation quivers, we mean those having this property.
It is remarkable that all such quivers correspond to meaningful $4d$ $\cn=2$ theories, in fact to complete ones in the present sense.

\subsubsection{Quivers from ideal triangulations of bordered surfaces}\label{sec:adjacencyquivers}

All but 11 mutation--finite classes arise from ideal triangulations of surfaces studied in ref.\cite{fomin}. Here we summarize the results of \cite{fomin} we need below. Let $\cc$ be an oriented surface of genus $g$ with $n$ punctures,
$b$ boundary components, and $c_i$ marked points on the $i$--th boundary component ($i=1,2,\dots, b$). By a compatible collection of arcs we mean a set of curves, identified up to isotopy, which end at the punctures or the marked points, do not intersect themselves or each other except at the end points, and cannot be contracted to a puncture or a boundary segment. Any maximal such compatible collection contains
\begin{equation}
D=6g-6+3n+\sum_i(c_i+3)
\end{equation}
arcs, and it is called an \emph{ideal triangulation} of $\cc$.
This definition allows for \emph{self--folded} triangles whose sides are not all distinct, see figure \eqref{selfoldedT}
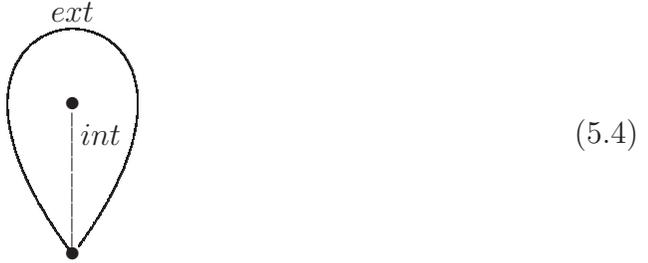
\begin{equation}\label{selfoldedT}
\begin{gathered}{\xy {(20,0)*+{\bullet}; (20,20)*+{\bullet} **\crv{}\POS?(0.8)*^+!L{int}};
{(20,0)*+{}; (20,0)*+{} **\crv{(5,20)&(20,40)&(35,20)}\POS?*_+!D{ext}}
\endxy}\end{gathered}
\end{equation}
%

Given an ideal triangulation we number the arcs as $1,2,\dots, D$, and define a skew--symmetric $D\times D$ integral matrix $B$ as follows \cite{fomin}: if $i$ and $j$ are not internal arcs of self--folded triangles (as is the arc $int$ in  figure \eqref{selfoldedT}) we set $B_{ij}$ to be the sum over all triangles $\triangle$ of which both arcs are sides of the weight $w^\triangle_{ij}$. $w^\triangle_{ij}$ is equal $+1$ (resp.\! $-1$) if the side $i$ of $\triangle$ follows (resp.\! precedes) the side $j$ in the counter-clockwise order. If $i$ is an internal arc of a self--folded triangle we set $B_{ij}\equiv B_{ext(i)j}$, where $ext(i)$ is the external arc of the self--folded triangle containing $i$ (see figure \eqref{selfoldedT}). The matrix $B$ is called the \emph{adjacency matrix} of the ideal triangulation.

The adjacency matrix $B$ defines a $2$--acyclic quiver as before. From the definition, one has
\begin{equation}\label{-1..2}
B_{ij}=-2,-1,0,1,2.
\end{equation}

One shows\cite{fomin} that two quivers, $Q_1$ and $Q_2$, representing adjacency matrices of two different ideal triangulations of the same surface $\cc$ are mutation equivalent. Moreover, any quiver which is mutation equivalent to the adjacency quiver of a surface is the adjacency quiver for some ideal triangulation of that surface. 
This, together with eqn.\eqref{-1..2}, implies that all adjacency quivers are of finite--mutation type.

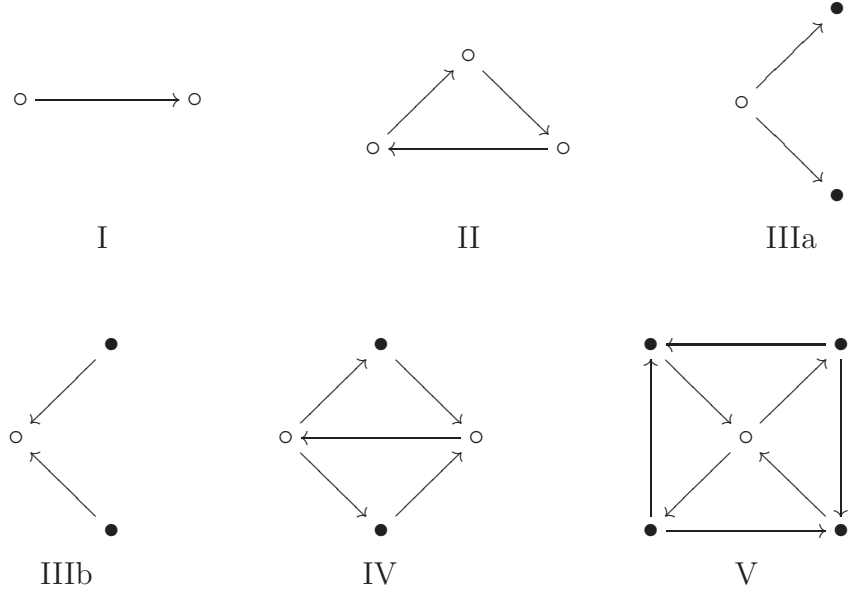
\begin{figure}
\begin{align*}
&\begin{gathered}
\xymatrix{\circ\ar[rr] &&\circ}
\end{gathered}
&&
\begin{gathered}\xymatrix{& \circ\ar[dr] & \\
\circ\ar[ur] & & \circ\ar[ll]}
\end{gathered}
&&
\begin{gathered}
\xymatrix{&\bullet\\
\circ\ar[ru]\ar[dr]\\
&\bullet}
\end{gathered} \\
&\hskip 1.3cm \mathrm{I} &&\hskip 1.4cm \mathrm{II} &&\hskip 0.6cm \mathrm{IIIa}
\end{align*}
\begin{align*}
&\begin{gathered}
\xymatrix{& \bullet\ar[dl]\\
\circ &\\
& \bullet\ar[ul]}
\end{gathered}
&&
\begin{gathered}
\xymatrix{& \bullet\ar[dr] & \\
\circ\ar[ur]\ar[dr] && \circ\ar[ll]\\
& \bullet\ar[ur] &}
\end{gathered}
&&
\begin{gathered}
\xymatrix{\bullet\ar[dr] & & \bullet\ar[ll]\ar[dd]\\
& \circ \ar[ur]\ar[dl] &\\
\bullet\ar[uu]\ar[rr] & & \bullet\ar[lu]}
\end{gathered}\\
&\hskip 0.6cm \mathrm{IIIb} &&\hskip 1.3cm \mathrm{IV} &&\hskip 1.4cm \mathrm{V}
\end{align*}
\caption{\label{figblocks} The quiver blocks of Type I--V \cite{fomin}.}
\end{figure}

A mutation invariant of the quiver is automatically a topological invariant of $\cc$. Since the rank of $B$ is invariant under mutation \cite{fominIII}, the corank of $B$ is a topological invariant equal to the number of punctures plus the number of boundary components with $c_i$ even \cite{fomin}
\begin{equation}\label{topologicaltheorem}
f=D-\mathrm{rank}\, B= n+\sum_{c_i\ \text{even}} 1.
\end{equation}
From the discussion in section \ref{genQuivers} we see that this topological invariant is equal to the number of flavor charges in the $\cn=2$ theory.
\medskip 

A quiver is the adjacency quiver of a bordered surface if and only if it can be decomposed into quiver blocks \cite{fomin}. There are five types of blocks (see figure \ref{figblocks}).  A quiver is an adjacency quiver of some bordered surface iff it can be obtained by gluing together a collection of blocks of types I, II, III, IV, and V by identifying together pairs of white nodes $\circ$. If the resulting quiver contains a pair of arrows connecting the same pair of vertices, but pointing in opposite directions, they must be removed.

White nodes represent arcs which are ordinary sides of triangles, and identifying pairs of them is equivalent to gluing the corresponding (generalized) triangles along that arc. 
More precisely, each block represents a piece of the triangulation \cite{fomin}:
\begin{itemize}
\item a block of type I represents a triangle with one side along the boundary of the surface $\cc$;
\item a block of type II represents a triangle with all three sides inner triangulation arcs;
\item a block type III represents a punctured $2$--gon\footnote{\ By an $n$--gon we mean a polygon with $n$ sides, that is a disk with $n$ marked points on the boundary.} with a side on the boundary;
\item a block of type IV represents a $2$--gon containing a folded triangle;
\item a block of type V represents a $1$--gon containing two folded triangles. 
\end{itemize}

Finally, if a quiver may be decomposed into blocks in a unique way, there is (topologically) precisely one surface $\cc$ whose triangulations correspond to the quivers of its mutation class; it is possible (but very rare) that two topologically distinct surfaces have the same class of adjacency quivers. The physical meaning of this non uniqueness will be discussed in the next section.

The two black nodes of a type III block are terminal nodes and in particular sink/sources. To avoid special cases in some of the statements below, it is convenient to adopt the following convention: whenever we have a quiver $Q$ with some type III blocks in its decomposition, we replace it by the physically equivalent quiver obtained by mutating $Q$ at one terminal node for each type III block. We call this sink/source equivalent quiver the \emph{normalized} quiver.

\subsection{Some basic features of mutation-finite quivers}\label{sec:mutfinvsSU(2)}

In this section we discuss some general features of mutation-finite quivers.
One basic features of mutation-finite quivers is that any full subquiver is also mutation finite.
We interpret this in the 4d language as saying that there is a choice of moduli which reduces the light degrees of freedom
of the theory to the corresponding subquiver.  This is the correct interpretation also from the
viewpoint of 4d/2d correspondence:  From the 2d perspective the nodes
correspond to 2d vacua and we can change the moduli of the 2d theory by taking
all the nodes outside the subquiver to have infinitely large value for the superpotential.
The inverse can also be done.  Namely one can start with a mutation-finite quiver and add
additional nodes
and arrows subject to maintaining mutation-finiteness.  This process should also be interpretable
physically as coupling a given physical theory to another one.    It is also interesting to ask if this
process would end, namely are there theories whose quivers are maximal and do not admit
any additional nodes, subject to mutation-finiteness.  The aim of this section is to analyze these questions.

As already noted, mutation-finite quivers have at most two arrows between any
pairs of nodes.  
The double arrows of a finite--mutation quiver have a simple physical interpretation. In section \ref{sec:BPSquivers}  we considered the example of $SU(2)$ SYM coupled to $N_f$ fundamental flavors. Its quiver,  see figure
 \eqref{SU(2)Nfquiver},
 has a double arrow subquiver $\xymatrix{\bullet\ar@<-0.4ex>[r]\ar@<0.4ex>[r]&\bullet}$
(\textit{a.k.a.}\! the Kronecker quiver), corresponding to the $SU(2)$ gauge sector, which is coupled by pairs of single arrows to each flavor node (which represents a fundamental hypermultiplet). The single arrows form together with the double one an oriented triangle, and stand for the gauge coupling of the SYM sector to the matter one. In section \ref{sec:BPSquivers} we saw how this particular arrangement of arrows precisely corresponds to the physics of the gauge couplings.

As already noted, a subquiver can be viewed as a subsector of the theory.  In particular
we can go to a point in moduli space where we have only the $SU(2)$ gauge theory degrees of freedom.  On the
other hand we could look at the couplings of the
Kronecker subquiver  $\xymatrix{\bullet\ar@<-0.4ex>[r]\ar@<0.4ex>[r]&\bullet}$ which represents a pair of dual electric/magnetic charges of an $SU(2)$ gauge sector, to the rest of the quiver and interpret
this as the coupling of the $SU(2)$ gauge sector to the rest of the system.  This can natually
be interpreted as saying that the rest of the quiver has an $SU(2)$ gauge symmetry which is
being gauged.  We now discuss some general aspects of such couplings.  

Let us ask then how the Kronecker quiver can be connected to the rest of the quiver.  It turns out
that generically quivers cannot have \emph{overlapping} Kronecker subquivers; more precisely, if a mutation--finite quiver $Q$ has a subquiver of the form\footnote{\ Here and below we use the following convention: Graphs with \emph{unoriented} edges stand for the \emph{full} family of quivers obtained by giving arbitrary orientations to the arrows. }
\begin{equation}
 \begin{gathered}
  \xymatrix{ & \bullet\ar@{-}@<-0.4ex>[rd]\ar@{-}@<0.4ex>[rd] &\\
\bullet \ar@{-}@<-0.4ex>[ur]\ar@{-}@<0.4ex>[ur] && \bullet}
 \end{gathered}
\end{equation}
then $Q$ must be the Markov quiver \eqref{markovqui}, and we have the $\cn=2^*$ theory \cite{derksen}. 
Thus other than this case, the Kronecker quivers are connected to the rest of the quiver only by single arrows.
Consider then another node connected to the Kronecker quiver.  It is either connected to both
nodes of the Kronecker quiver or just to one.
Note, however, that the following quivers
\begin{equation}
\begin{gathered}\xymatrix{\bullet \ar@{-}[r] &\bullet\ar@{-}@<-0.4ex>[r] \ar@{-}@<0.4ex>[r] &\bullet\ar@{-}[r]&\bullet}\end{gathered}\quad
\begin{gathered}\xymatrix{& \bullet &\\
\bullet\ar[ru] &&\bullet\ar@<-0.4ex>[ll]\ar@<0.4ex>[ll]\ar[lu]}
\end{gathered}\quad
\begin{gathered}\xymatrix{& \bullet \ar[dr]\ar[dl]&\\
\bullet &&\bullet\ar@<-0.4ex>[ll]\ar@<0.4ex>[ll]}\end{gathered}
\end{equation}
are not mutation--finite, and hence cannot appear as subquivers of finite--mutation quivers. Hence a Kronecker subquiver $\mathbf{Kr}$ of a quiver $Q$ which corresponds to a complete $\cn=2$ theory is attached to the rest of the quiver $Q$ through oriented triangles, so that, locally around the double--arrow, the quiver looks like that of
$SU(2)$ with $N_f$ flavors (see figure
 \eqref{SU(2)Nfquiver}), where $N_f$ is the number of oriented triangles in $Q$ which have $\mathbf{Kr}$ as a side.

The quiver \eqref{SU(2)Nfquiver} is not of mutation--finite type for $N_f\geq 5$; this corresponds to the fact that the corresponding gauge theory is not UV complete having a Landau pole. For $N_f=4$ the quiver \eqref{SU(2)Nfquiver} is of mutation--finite type, but no (connected) finite--mutation quiver may have it as a proper subquiver. Physically, this corresponds to the fact that $SU(2)$ with four flavor is conformal, and coupling extra matter makes the gauge beta function  positive, losing UV completeness.  Therefore\vglue 6pt

\textbf{Kronecker Coupling:} \textit{Let $Q$ be a quiver with a double--arrow describing a complete $\cn=2$ theory  which is not
pure $SU(2)$, $\cn=2^*$ $SU(2)$, or $SU(2)$ with $N_f=4$. Then, locally near the double--arrow, $Q$ has one of the following three subquivers}
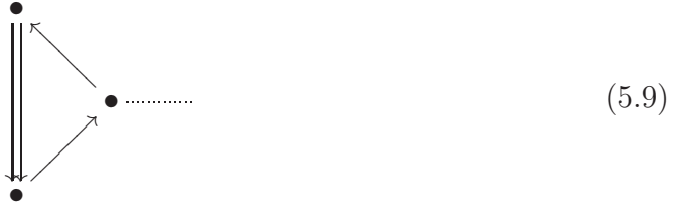
\begin{gather}
 \begin{gathered}\xymatrix{\bullet\ar@<-0.3ex>[dd]\ar@<0.3ex>[dd] &&\\
& \bullet\ar[lu] \ar@{..}[r]&\\
\bullet\ar[ur] &&}\end{gathered}
\end{gather}
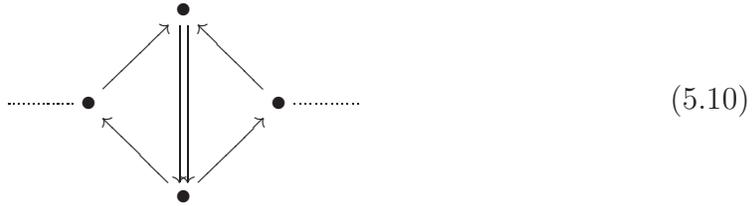
\begin{gather}
 \begin{gathered}\xymatrix{&&\bullet\ar@<-0.3ex>[dd]\ar@<0.3ex>[dd] &&\\
& \bullet\ar@{..}[l] \ar[ur] && \bullet\ar[lu] \ar@{..}[r]&\\
&&\bullet\ar[ur]\ar[ul] &&}\end{gathered}
\end{gather}
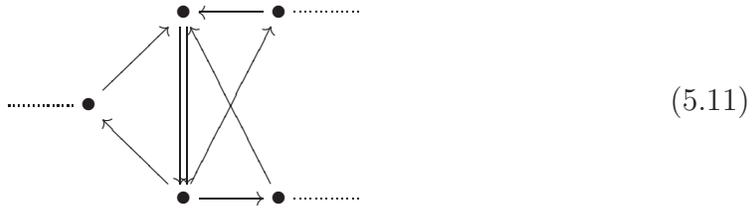
\begin{gather}
 \begin{gathered}\xymatrix{&&\bullet\ar@<-0.25ex>[dd]\ar@<0.25ex>[dd] &\bullet\ar[l]\ar@{..}[r]&\\
& \bullet\ar@{..}[l] \ar[ur] && &\\
&&\bullet\ar[uur]\ar[ul]\ar[r] &\bullet\ar[uul]\ar@{..}[r]&}\end{gathered}
\end{gather}
\textit{dashed lines standing for arrows connecting the subquiver to the rest of the quiver $Q$.}\vglue 6pt

The above situation is naturally interpreted as the coupling of the $SU(2)$ SYM represented by the Kronecker subquiver $\mathbf{Kr}$ to, respectively, one, two, or three $\cn=2$ systems represented by the subquivers $\xymatrix{\bullet\ar@{..}[r]&}$. The simplest instance is when these subquivers are just a node, $\bullet$, in which case we get $SU(2)$ SQCD with $N_f=1,2$ and $3$, respectively. 
We stress that, in general, the subquiver $\cn=2$ systems are coupled together also by other interactions, corresponding to the arrows connecting them in the full quiver $Q$. A specially simple case is when the elimination of the Kronecker subquiver $\mathbf{Kr}$ disconnects $Q$ into a maximal number of `elementary' components $\xymatrix{\bullet\ar@{..}[r]&}$.

The allowed subquivers $\xymatrix{\bullet\ar@{..}[r]&}$ are severely restricted by the mutation--finite condition for $Q$. As in the example of $SU(2)$ coupled to $N_f$ flavors, this condition is physically interpreted as the UV completeness requirement that the beta--function of the $SU(2)$ is non--positive: hence the sum of the contribution to the beta function from the $\cn=2$ system(s) represented by the 
$\xymatrix{\bullet\ar@{..}[r]&}$ subquivers should be less or equal to the contribution of $4$ hypermultiplets in the fundamental representation. This observation will allow us to determine the contribution to the gauge beta function of all the possible (complete) $\cn=2$  systems $\xymatrix{\bullet\ar@{..}[r]&}$ (which may have no Lagrangian description, in general).

\vglue 6pt

\textbf{Example.} From figure \ref{ellipticEs} we see that the elliptic $\widehat{\widehat{E}}_r$ quivers correspond to a `weak coupling' regime of the corresponding complete $\cn=2$ theories look as an $SU(2)$ SYM coupled to three decoupled $\cn=2$ systems. For $r=7,8$, one $\cn=2$ system (corresponding to the node in the left side of the figure) is an ordinary hypermultiplet. In section \ref{identificationexceptional} we shall show that the $\widehat{\widehat{E}}_r$ theories have also strongly coupled regimes in which the spectrum consists only of a finite set of BPS hypermultiplets.
\medskip

It is natural to ask how many Kronecker sub-quivers does a quiver have, and how this changes
as the quiver undergoes mutation.
In fact, in a typical mutation--class, most of the quivers have only single--arrows; very few quivers have the maximal number of double--arrows allowed for that class; for instance, for complete $\cn=2$ models which are quiver gauge theories
and for which the matter fields can be massed up (\textit{i.e.}\! all the mass terms are consistent with gauge symmetry),
 there is a \emph{unique} BPS quiver with the maximal number of $2$--arrows equal to the number of $SU(2)$ gauge groups. 
We stress that, in the general case, there is \emph{no} one--to--one correspondence between $SU(2)$ gauge groups and Kronecker subquivers. Even if we take a quiver in the mutation--class with the maximal number of double--arrows, this may be still less than the actual number of $SU(2)$ gauge groups. This happens when we have several $SU(2)$ gauge sectors coupled together by half--hypermultiplets rather than full hypermultiplets, such that the half--hypermultiplets
transform as different representation of the $SU(2)$ gauge groups and cannot receive mass.
In such a case we cannot expect to isolate the pure $SU(2)$ theory, and so we do not expect to have
a corresponding Kronecker subquiver.

As already noted, in principle we can add additional nodes and arrows to a given mutation-finite
quiver and still keep it mutation-finite.  This raises the question
of whether there are maximal mutation-finite quivers for which we cannot add additional
nodes maintaining this property, and their interpretation if they exist.   We will argue  in section \ref{sec:conformal})
that:
\vglue 6pt

\textbf{The graphical conformality criterion:} \textit{A complete $\cn=2$ theory is UV conformal (as contrasted to asymptotically free) if and only if its normalized quiver is either a \underline{maximal} mutation--finite  one, or a vector--less quiver.}
\vglue 6pt

By a \emph{maximal mutation--finite} quiver we mean a quiver which is mutation--finite and not a proper subquiver of any connected mutation--finite quiver. Two basic examples of maximal mutation--finite quivers are the Markov quiver \eqref{markovqui}, corresponding to $SU(2)$ $\cn=2^*$, and the $SU(2)$ $N_f=4$ quiver \eqref{SU(2)Nfquiver}. 
By a \emph{vector--less} quiver we mean a quiver such that no quiver in its mutation class contains multiple arrows; such quivers correspond to $\cn=2$ theories having no phase which looks like SYM (with any gauge group $G$) coupled to some additional matter. In particular, vector--less quiver $\cn=2$ theories have no BPS chambers with {\it  charged} BPS vector multiplets.
\medskip

 To complete the classification of conformally complete $\cn=2$ theories, we need to classify the \emph{vector--less} quivers. Clearly a finite quiver such that all its mutations contain only simple arrows is, in particular, mutation--finite and must be in the
Felikson--Shapiro--Tumarkin list. By inspection, the only classes with this property in the eleven exceptional cases are the three finite Dynkin diagrams $E_6, E_7, E_8$. Likewise, going through the classification of the quivers associated to triangulated surfaces, we see that this property is true only if $\cc$ is the disk with zero or one puncture whose quivers are, respectively, the (finite) Dynkin diagrams of types  $A$ and $D$. Hence, the only $\cn=2$ theories with the properties that all quivers in their mutation classes have only single--lines are the $ADE$ Argyres--Douglas ones (already studied in \cite{cnv}). They are UV conformal.\footnote{By Gabriel theorem \cite{gabriel,MR2197389,auslander,cluster-intro}, these are in one--to--one correspondence with the finite--representation hereditary algebras. This is another confirmation of the deep connection between quiver representation theory and $\cn=2$ theories.}  Note that these are precisely the class
that map to the 2d theories which are minimal, in the sense that they have a UV limit corresponding
to minimal $\cn=2$ conformal theories in 2d (which in particular have $\hat c<1$).

We end this subsection with a remark. The way mutation--finite quivers are classified in the math literature is by studying the maximal ones; once we have identified a maximal mutation--finite quiver, we may rule out all quivers containing it, and, by doing this systematically, we may eliminate all non--mutation--finite ones. Physically, this means that we keep adding `matter' to the $SU(2)^k$ theory until the UV beta functions of all gauge couplings are negative. When we reach a conformal theory we stop, since adding further `matter' will result in a UV incomplete theory. The corresponding quiver is automatically maximal, and we can forget about all quivers containing it. This gives us another way of understanding the correspondence
\begin{equation*}
\text{mutation--finite quivers}\quad\longleftrightarrow\quad\text{complete $\cn=2$ theories.}
\end{equation*}


\section{Identification of a large class of quivers of finite mutation type as generalized Gaiotto theories}\label{gaiottoidesc}

According to the discussion in \S.\,\ref{sec:finitecomplete}, to each mutation--finite class of $2$--acyclic quivers there should correspond a complete $\cn=2$ theory in four dimensions. To make this correspondence more explicit, in the following two sections we identify the supersymmetric theory associated to each mutation--finite class of quivers.
\smallskip
 
The quivers (with at least three nodes) which belong to all, but eleven, mutation--finite classes are adjacency matrices of ideal triangulations of some bordered surface. Therefore we divide the identification process into two steps: First we identify the theories corresponding to the infinite set of quiver classes arising from bordered surfaces $\cc$, and then consider the residual eleven exceptional classes one by one.

The $\cn=2$ models corresponding to the non--exceptional quivers  turn out to be generalizations of the $SU(2)$ theories recently studied by Gaiotto \cite{gaiotto}. The existence of these more general theories already follows from the constructions in sections 3, 8 of \cite{Gaiotto:2009hg}. 

More precisely, as we shall show momentarily, all the non--exceptional complete $\cn=2$ theories may be engineered by compactifing the $A_1$ six dimensional $(2,0)$ theory on a curve $\cc$ of genus $g$ and $n+b$ punctures supplemented with some particular boundary conditions at these punctures. The resulting four dimensional theory will preserve eight supercharges iff the internal $2d$ fields on $\cc$, $(A,\phi)$, satisfy the Hitchin equations \cite{gaiotto,Gaiotto:2009hg}
\begin{gather}\label{Hit1}
F+[\phi,\overline{\phi}]=0\\
\overline{\partial}\phi=0,\label{Hit2}
\end{gather}
with prescribed singularities at the $n+b$ punctures.
The conditions on $(A,\phi)$ are better stated in terms of the spectral cover $\Sigma\rightarrow \cc$ of the Hitchin system \eqref{Hit1}\eqref{Hit2}. $\Sigma$, which is the Seiberg--Witten IR curve of the resulting $4d$  $\cn=2$ theory  \cite{gaiotto,Gaiotto:2009hg}, is the curve in the total space of the cotangent bundle $T^*\cc$ defined by the spectral equation\footnote{\ $y$ is a coordinate along the fiber of $T^*\cc$. The canonical differential $y\,dx$ is identified with the Seiberg--Witten one.}
\begin{equation}\label{spectralcurve}
\det[y-\phi]\equiv y^2-\phi_2=0.
\end{equation}
 The meromorphic quadratic differential $\phi_2$ is required to have (for generic points in the Coulomb branch and values of the parameters) double poles at the ordinary punctures and poles of order $p_i=c_i+2\geq 3$ at the puncture representing the $i$--th boundary component having $c_i$ marked points (section 8  of \cite{Gaiotto:2009hg}). We may think of ordinary punctures as boundary components without marked points.  This is because the quadratic differential $(dz/z)^2$ can be written as $(dw)^2$
where $w={\rm log}z$, and $w$ parameterizes a cylinder.
 When needed, we replace punctures with higher order poles of $\phi_2$ with small circles with $p_i-2$ marked points to reproduce their topological description.\smallskip  

The class of theories studied by Gaiotto in \cite{gaiotto} corresponds to the special case of this construction in which all punctures are just ordinary double poles. This Gaiotto subset consists of models which are superconformal in the limit of zero masses (and Coulomb branch parameters). On the contrary, the general  theory associated to a surface `with boundaries' --- that is, specified by a quadratic differential $\phi_2$ with prescribed higher order poles --- are \emph{not} conformal in the UV but just asymptotically free (AF). The simplest examples \cite{Gaiotto:2009hg} of such AF models are the well--known $SU(2)$ gauge theories with $N_f=0,1,2,3$ fundamental flavors; these theories may also be engineered in the present framework by considering a sphere with two or three punctures having pole orders

\begin{center}
\begin{tabular}{c|c|c|c}
$N_f$ & $\#$ punctures & order of poles & quiver class\\\hline
$0$ & $2$ & $3,3$ & $\widehat{A}_1(1,1)$\\
$1$ & $2$ & $3,4$ & $\widehat{A}_2(2,1)$\\
$2$ & $2$ & $4,4$ & $\widehat{A}_3(2,2)$\\
$2$ & $3$ & $2,2,3$ & 
$\text{``}\widehat{D}_3\text{''}\equiv \widehat{A}_3(1,1)$\\
$3$ & $3$ & $2,2,4$ & $\widehat{D}_4$ 
\end{tabular}
\end{center}
(the $N_f=2$ model has two different, but physically equivalent, realizations in terms of a system of $M$-branes; in terms of the $6d$ $A_1$ $(2,0)$ theory \cite{Gaiotto:2009hg} they correspond to the two surfaces listed in the table; at the quiver level the identity of the two theories expresses the well--known Lie algebra isomorphism $\widehat{SU(4)}\simeq \widehat{SO(6)}$). \medskip 

The identification of the complete $\cn=2$ theories which are UV superconformal is presented in section \ref{sec:conformal}, and agrees with the graphical rule of sect.\,\ref{sec:mutfinvsSU(2)}.
\medskip

It should be stressed, however, that the correspondences
\begin{equation*}
\text{finite--mutation quiver}\ \leftrightarrow\ \text{triangulated surface $\cc$}\ \leftrightarrow\ \text{Gaiotto $\cn=2$ theory} 
 \end{equation*}
require the surface $\cc$ to have at least one puncture to base the triangulation. In ref.\cite{gaiotto} $\cn=2$ models are constructed also for genus $g>1$ surfaces \emph{without} punctures. With the exception of the $g=2$ case (to be discussed in section \ref{identificationexceptional} below), there are no additional mutation--finite quivers to be assigned to these puncture--less theories, given that the theories with at least one puncture already exhaust the full supply of finite--mutation quivers with more than $10$ nodes. Moreover, the  no--puncture $g\geq 3$ theories cannot be equivalent to some other model with punctures already in the classification, since \textit{i)} they are conformal, \textit{ii)} have no flavor charge, \textit{iii)} have $\mathrm{rank}\,\Gamma= 6g-6\geq 12$, and there are no mutation--finite quivers with these three properties. The solution of the puzzle is that these theories, like $\cn=4$, are not quiver theories, in the sense that there are no $D$--tuple of charge vectors $\gamma_a\in \Gamma$ such that all BPS charge vectors may be written as $\pm \sum_a n_a\gamma_a$ with positive $n_a$'s; for these theories the BPS phase are expected to be dense on the unit circle, and thus they do not admit a BPS quiver\footnote{We thank D. Gaiotto for pointing this out.}.
\smallskip

\subsection{$4d/2d$ correspondence and ideal triangulations}\label{2d4dcorresp}

The identification of the non--exceptional $\cn=2$ {complete} theories with the generalized Gaiotto theories, is confirmed by the $4d/2d$ correspondence of ref.\cite{cnv}, reviewed in section \ref{sec:2d4drev}.

Roughly speaking, the $4d/2d$ correspondence says that the quiver of the $4d$ theory is to be identified with (minus) the BPS quiver of the corresponding $2d$ $(2,2)$ theory. At the technical level, things are a bit more involved because of some subtleties with the signs (\textit{i.e.}\! arrow orientations) discussed in \cite{Cecotti:1993rm}. Moreover, as stressed in \cite{cnv}, the classification of $2d$ BPS quivers (modulo $2d$ wall--crossing \cite{Cecotti:1993rm}) is \emph{coarser} than the classification of $4d$ quivers (modulo mutation--equivalence) because there are more $2d$ walls to cross than quiver mutations. 
A more precise dictionary is the following: let
\begin{equation}
S= \prod_\text{half plane}^\curvearrowright \exp(-\mu_\theta)
\end{equation}
be the product of all $SL(D,\Z)$  monodromy group elements
associated to BPS states with phase $\theta$ in the given half--plane\footnote{\ $|(\mu_\theta)_{ij}|$ is equal to the number of BPS solitons connecting the $i$ and $j$ vacua and having BPS phase $\theta$; the sign of $(\mu_\theta)_{ij}=-(\mu_\theta)_{ji}$ follows, up to convention dependent choices, from the rules of ref.\cite{Cecotti:1993rm}.}.  $S$ is related to the monodromy $M$ by the formula $M=(S^{-1})^tS$ \cite{Cecotti:1993rm}. By the $2d$ wall--crossing formula, $S$ is invariant under all wall--crossings except those which make a BPS state exit from the given half--plane (while its PCT conjugate enters from the other side). Then $S$ is defined up to the same mutations as $B$, except that $S$ depends also on the sign conventions of the $2d$ vacua (changing the sign of the $k$--th vacuum makes $\mu_{kj}\rightarrow -\mu_{kj}$). Therefore, the refined statement is that we may choose the $2d$ conventions in such a way that the exchange matrix of the $4d$ quiver is
\begin{equation}
B=S-S^t.
\end{equation}
In the case of complete theories, the corresponding $2d$ models are also complete in the same sense, and we may always reduce ourselves to a convex arrangement of vacua \cite{Cecotti:1993rm}, in which case  we have simply $B=-\mu$, and we may forget about subtleties (at the price of wall--crossing $\mu$ to a suitable $2d$ BPS chamber).

\subsubsection{Lagrangian $A$--branes as ideal triangulations}

We would like to identify the corresponding 2d theory associated with the 4d theory obtained
by 2 5-branes wrapping a Riemann surface with punctures.  We already know that if we have a
type IIB geometry of the form
$$uv-W(y,z)=0$$
The associated 2d theory is a LG theory with superpotential $W(y,z)$ as a function of chiral fields $y,z$.
On the other hand it is also known that this type IIB geometry is dual to a 5-brane as a subspace of
$y,z$ parameterizing $\C^2$ given by wrapping the curve
$$W(y,z)=0$$
and filling the spacetime  \cite{Ooguri:1996wj}.  Now let us consider the Gaiotto theories.  In this
case the 5-brane geometry is captured by the geometry
$$W(y,z)=y^2-\phi_2(z)$$
However, here $y$ has a non-trivial geometry: $y$ is a section of the canonical line bundle
on the Riemann surface.  To make $y$ be ordinary
coordinate we take a reference quadratic differential $\omega_0$, and define
$$\tilde y =y/\omega_0.$$
Under this transformation we get the equation
$$W(\tilde y,z)=\tilde{y}^2 -{\phi_2(z)\over \omega_0}$$
Since the $\tilde{y}^2$ term does not affect the BPS structure and vacua of the theory, this is equivalent
to a $2d$ theory with $(2,2)$ supersymmetry and superpotential
\begin{equation}\label{superpotentialW}
W(z)= \frac{\phi_2(z)}{\omega_0(z)}.
\end{equation}  
Note that near the zeros of $\omega_0$ the $W$ is well defined--it simply corresponds
to regions where the LG potential $|dW|^2$ grows large.
 The meromorphic one--form $dW$ has a number of zeros ($\equiv$ supersymmetric vacua)
\begin{equation}\label{zerocount}
\# \{\text{zeros of }dW\}= 2g-2+ \text{polar degree of }dW= 6g-6+\sum_i(p_i+1),
\end{equation}
where $p_i$ is the order of pole of $\phi_2$ at the $i$--th puncture.
Thus the number of supersymmetric vacua of the two dimensional  theory is equal to $D$, the number of arcs in an ideal triangulation of the corresponding bordered surface. This is no coincidence: let us consider the Lagrangian $A$--branes defined, for this class of $(2,2)$ theories, in \cite{Hori:2000ck}. They are the integral curves $\gamma_{i}$ of the differential equation
\begin{equation}\label{2ddifferential}
\mathrm{Im}(e^{i\theta}\,dW)=0
\end{equation} 
(for some \emph{fixed} but generic value of the angle $\theta$) which start at $t=0$ from the $i$--th zero of $dW$, $X_i$, and approach, as $t\rightarrow \pm \infty$, infinity in the $W$--plane --- that is, a puncture in $\cc$ --- along a direction such that
\begin{equation}
 \mathrm{Re}(e^{i\theta}\, W)\Big|_{t\rightarrow \pm \infty} \rightarrow +\infty.
\end{equation}
We assume $\theta$ to have been chosen so that $\mathrm{Im}(e^{i\theta}\,W(X_i))\neq \mathrm{Im}(e^{i\theta}\,W(X_j))$ for $i\neq j$. Then the branes $\gamma_i$ are distinct.

If
two arcs, $\gamma_i$, $\gamma_j$, cross at some finite value of $t$, they coincide everywhere $\gamma_i\equiv\gamma_j$. Hence the arcs $\gamma_i$ do not cross themselves nor each other, except at the punctures.
This is one of the properties defining the collection of arcs of an ideal triangulation \cite{fomin}. 
To be a compatible collection of arcs on $\cc$, the Lagrangian $A$--branes $\{\gamma_i\}$ should also be non--contractible to a puncture (or boundary arc) and pairwise isotopy inequivalent. If these properties hold, the Lagrangian branes  $\{\gamma_i\}$ form automatically a maximal collection, and hence an ideal triangulation, since their number is the maximal one, given by eqn.\eqref{zerocount}.
In ref.\cite{Hori:2000ck} it was shown that the Lagrangian $A$--branes $\{\gamma_i\}$ span the relative homology group $H_1(\cc,\cb)$ (where $\cb\subset \cc$ is the region near the punctures where $\mathrm{Re}(e^{i\theta}\, W)\gg 1$), so all the  axioms for an {ideal triangulation} are satisfied.
\smallskip

The above construction should be contrasted with the similar, but different, triangulation which arises in the
study of 4d BPS states by considering straight lines on the SW curve, defined by the condition that the phase of the SW differential does not change along the path introduced in  \cite{Klemm:1996bj} and \cite{Shapere:1999xr} and studied extensively in \cite{Gaiotto:2009hg}. There, for the same class of models, one constructs an ideal triangulation using the integral curves of the (real part of) the Seiberg--Witten differential, namely the solutions to  the equation
\begin{equation}\label{4dswdiff}
\mathrm{Im}(e^{i\theta/2}\, y\,dz)=0
\end{equation} 
instead of the one in eqn.\eqref{2ddifferential}. Again, one gets an ideal triangulation, but this time the `vacua', that is the zeros of the Seiberg--Witten differential, are in one--to--one correspondence with the faces of the triangulation, rather than with the arcs. As a check, let us count the number of triangles
\begin{equation}\begin{split}
\#\,\text{triangles}&= 2-2g+\#\,\text{arcs}-\#\,\text{punctures}\\
&= 4g-4+\sum_i p_i \equiv \#\,\text{zeros of }\phi_2.
\end{split}\end{equation}

The adjacency quivers obtained by these two procedures, corresponding to ideal triangulations of the same punctured surface, should be the same up to mutation equivalence. 
\smallskip

Before going to the adjacency quivers, let us illustrate in an example how the $A$--brane ideal triangulation works.

\subsubsection{Example: torus with $n$ ordinary punctures}
\label{sec:torusnpunctures}

We start with the torus with one puncture, which corresponds to $\cn=2^*$ and the Markov quiver \eqref{markovqui}.
There is an essentially unique ideal triangulation
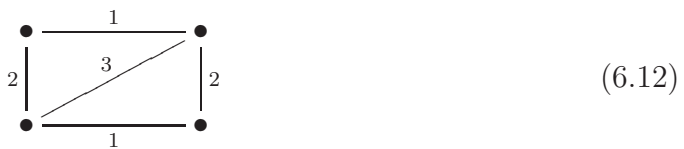
\begin{equation}\label{toruniqtri}
\begin{gathered}\xymatrix{\bullet \ar@{-}[rr]^1\ar@{-}[d]_2 && \bullet\ar@{-}[d]^2\\
\bullet \ar@{-}[rru]^3\ar@{-}[rr]_1 && \bullet}
\end{gathered}\end{equation}
where the opposite sides of the rectangle are identified. The corresponding incidence matrix is
\begin{equation}
B_{1,2}=-2,\quad B_{1,3}=2,\quad B_{2,3}=-2
\end{equation}
giving the Markov quiver \eqref{markovqui}.

To recover this result from the $2d$ perspective,
we go to the universal cover of $\cc$, namely $\C$, and consider the LG model with superpotential $W(X)=i\,\wp(X)$,
identifying the chiral field $X$ with the canonical coordinate of the torus. We take the moduli of the torus
to have $\Z_4$ symmmetry, a square torus of periods $(1,i)$ so that 
\begin{equation}
 (\wp^\prime)^2= 4\,\wp^3-\frac{\Gamma(1/4)^8}{16\pi^2}\, \wp.
\end{equation}
One has $\wp(i\, X)= - \wp(X)$ and $\wp(X)= X^{-2}\,f(X^4)$ with $f(\bar z)=\overline{f(z)}$.
Viewing the torus as a double cover of the plane given by $Z=2\wp(X)$, and the 2-fold cover by $Y=\wp^\prime$
we have
$$Y^2=Z^3-aZ$$
The three classical vacua correspond to the three solutions of $Y=\wp^\prime(X_k)=0$ at finite $Z$, and are at the half--lattice points
\begin{align*}
X_1&= \frac{1}{2}, &X_2&=\frac{1+i}{2},&X_3&= \frac{i}{2}\\
W(X_1)&= i\,\frac{\Gamma(1/4)^4}{8\pi}, & W(X_2)&= 0, & W(X_3)&= -i\,\frac{\Gamma(1/4)^4}{8\pi}.
\end{align*}
The Lagrangian branes map to straight lines on the $W$ plane which in this case correspond to $Z$ plane.
There are a paire of kinks between any pair the three vacua corresponding to the two straight lines which connect them
in the $Z$--plane.
The Lagrangian brane $L_2$ passing through the $\Z_4$ invariant point $X_2$ and going to $\mathrm{Re}(W)=+\infty$ is just the diagonal of the square. Then the two Lagrangian branes $L_{1,3}$ passing through $X_{1,3}$ should correspond to the two $\cs$--shaped curves in the figure (their curvature is exaggerated for drawing purposes)
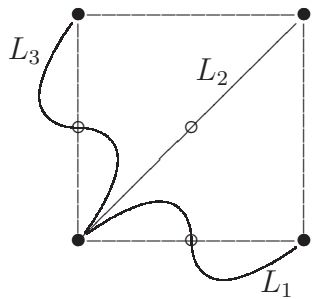
\begin{equation*}
\vbox{\xy {(10,10)*+{\bullet}; (10,40)*+{\bullet} **\crv{}\POS?*^+!L{}};
{(40,10)*+{\bullet}; (40,40)*+{\bullet} **\crv{}\POS?*^+!L{}};
{(10,10)*+{\bullet}; (40,10)*+{\bullet} **\crv{}\POS?*^+!L{}};
{(10,40)*+{\bullet}; (40,40)*+{\bullet} **\crv{}\POS?*^+!L{}};
{(10,10)*+{}; (40,40)*+{} **\crv{}\POS?(0.75)*^+!R{L_2\ }};
{(25,25)*+{\circ}};
{(10,25)*+{\circ}};
{(25,10)*+{\circ}};
{(10,25)*+{}; (10,10)*+{} **\crv{(13,25)&(18,22)&(11.8,12)}\POS?*_+!D{}};
{(10,25)*+{}; (10,40)*+{} **\crv{(7,25)&(2,28)&(8.2,38)}\POS?(0.75)*_+!R{L_3\,}};
{(25,10)*+{}; (10,10)*+{} **\crv{(25,13)&(22,18)&(12,11.8)}\POS?*_+!D{}};
{(25,10)*+{}; (40,10)*+{} **\crv{(25,7)&(28,2)&(38,8.2)}\POS?(0.77)*_+!U{\; L_1}};
\endxy}\end{equation*}
comparing with eqn.\eqref{toruniqtri} we see that the three Lagrangian branes $L_i$ are (up to isotopy) the same as the ideal triangulation arcs.

The Landau--Ginzburg model with $W(X)=i\,\wp(X)$ was solved in ref.\cite{ising} (it corresponds to the three--point functions of the Ising model). It has two BPS states connecting each pair of vacua related by the symmetry $X(t) \leftrightarrow -X(t)$ (modulo periods) which fixes the three classical vacua.
The $S$ matrix is
\begin{equation}
S=\begin{pmatrix}1 & -2 & 2\\ 
0 & 1 & -2\\
0 & 0 & 1\end{pmatrix}
\end{equation}
(the eigenvalues of $M= (S^{-1})^tS$ are $-1,1,-1$) and
\begin{equation}
B=S-S^t=\left(
\begin{array}{ccc}
 0 & -2 & 2 \\
 2 & 0 & -2 \\
 -2 & 2 & 0
\end{array}
\right)
\end{equation}
which is the exchange matrix of the Markov quiver \eqref{markovqui}. 

\medskip

A torus  with $n>1$ punctures has many different ideal triangulations. The one with the more transparent physical interpretation has
the adjacency quiver with maximal number of double--arrows (Kronecker subquivers), namely $n$. This triangulation is the \emph{zig--zag} one (\emph{a.k.a.}\! the \emph{snake} triangulation): See the figure 
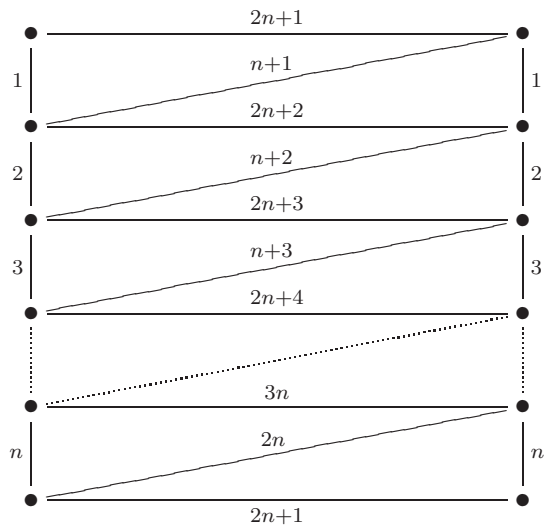
\begin{equation*}
\xymatrix{\bullet \ar@{-}[rrrrrr]^{2n+1} \ar@{-}[d]_1& & & & & & \bullet\ar@{-}[d]^1\\ 
\bullet\ar@{-}[d]_2\ar@{-}[urrrrrr]^{n+1}\ar@{-}[rrrrrr]^{2n+2} & & & & &&\bullet\ar@{-}[d]^2\\
\bullet\ar@{-}[d]_3\ar@{-}[urrrrrr]^{n+2}\ar@{-}[rrrrrr]^{2n+3} & & & & &&\bullet\ar@{-}[d]^3\\
\bullet\ar@{..}[d]\ar@{-}[urrrrrr]^{n+3}\ar@{-}[rrrrrr]^{2n+4} & & &  &&&\bullet\ar@{..}[d]\\
\bullet\ar@{-}[d]_n\ar@{..}[urrrrrr]\ar@{-}[rrrrrr]^{3n} & & & & & &\bullet\ar@{-}[d]^n\\
\bullet\ar@{-}[rrrrrr]_{2n+1}\ar@{-}[urrrrrr]^{2n}& & & &&&\bullet}
\end{equation*}
where corresponding segments of the sides should be identified (in the figure, identified segments carry the same label). 
The arc labelled $k$, with $1\leq k\leq n$ shares \emph{two} triangles with the arc labelled $n+k$. The two triangles have the same orientation, so the corresponding entries of the adjacency matrix are
\begin{equation}
B_{k,n+k}= -2,\quad k=1,2,\dots, n.
\end{equation} 
On the other hand, the $k$--th arc shares a single triangle with the arcs $2n+k$ and $2n+k+1$. One has
\begin{gather}
B_{k,2n+k}= +1\\
B_{k,2n+k+1}=+1.
\end{gather}
Finally, the arc $n+k$ shares a triangle with the arcs $2n+k$ and $2n+k+1$. Then 
\begin{gather}
B_{n+k,2n+k}= -1\\
B_{n+k,2n+k+1}=-1.
\end{gather}
All other entries of the adjacency matrix vanish. In particular, we have $n$ double arrows, as anticipated. All triangles in the quiver are oriented. According to our discussion in section \ref{sec:mutfinvsSU(2)} these quivers correspond to a closed chain of $n$ Kronecker subquivers (\textit{i.e.}\! $SU(2)$ gauge groups) coupled to each other by bi--fundamental hypermultiplets (represented by the nodes $\odot$, $\circledast$ on the figure) 
\begin{gather}\label{toruspunctquive}
\begin{gathered}\xymatrix{ &\circ\ar@<-0.4ex>[dd]\ar@<0.4ex>[dd] && \circ \ar@<-0.4ex>[dd]\ar@<0.4ex>[dd]&& \circ\ar@<-0.4ex>[dd]\ar@<0.4ex>[dd] &&& \circ \ar@<-0.4ex>[dd]\ar@<0.4ex>[dd] &\\
\circledast\ar[ru]&& \odot\ar[lu]\ar[ru] && \odot\ar[lu]\ar[ru] && \odot\ar[lu] \ar@{..}[r] &\odot\ar[ru] &&\circledast\ar[lu]\\
&\circ\ar[ul]\ar[ru] && \circ\ar[lu] \ar[ru] && \circ\ar[lu]\ar[ru] &&& \circ\ar[lu]\ar[ru] &}\end{gathered}
\end{gather} 
where the two bi--fundamentals denoted by the symbol $\circledast$ should be identified. Thus, these $\cn=2$ models correspond to quiver $SU(2)$ gauge theories with underlying graph the affine Dynkin diagram $\widehat{A}_{n-1}$, as expected for the Gaiotto theory engineered by a torus with $n$--punctures. Notice that by the topological theorem \eqref{topologicaltheorem} this $\cn=2$ model has precisely $n$ flavor charges, corresponding to the $n$ bi--fundamentals.
\smallskip

The above snake triangulation may be easily recovered from the two--dimensional point of view. One consider the same Landau--Ginzburg model with Weiertrass superpotential as before, except that we now identity the field $X$ up to multiple periods
\begin{equation}
X\sim X+ a\,n+ b\, i, \ \text{where } a,b\in \Z,
\end{equation}   
so that now we have $3n$ distinct vacua and hence $3n$ distinct $A$--branes which are just the translation by $k=0,1,2,\dots, n-1$ of the basic ones for $n=1$. The case $n=3$ is represented in the figure
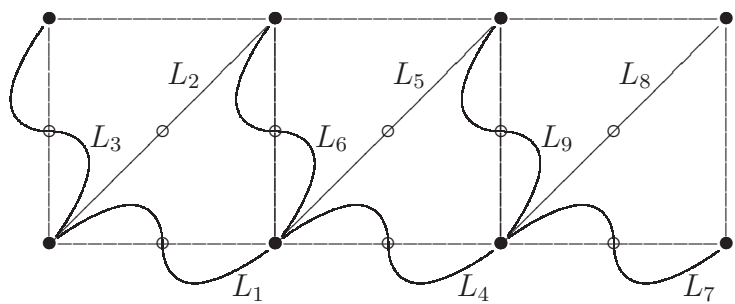
\begin{equation*}
\xy {(10,10)*+{\bullet}; (10,40)*+{\bullet} **\crv{}\POS?*^+!L{}};
{(40,10)*+{\bullet}; (40,40)*+{\bullet} **\crv{}\POS?*^+!L{}};
{(10,10)*+{\bullet}; (40,10)*+{\bullet} **\crv{}\POS?*^+!L{}};
{(10,40)*+{\bullet}; (40,40)*+{\bullet} **\crv{}\POS?*^+!L{}};
{(10,10)*+{}; (40,40)*+{} **\crv{}\POS?(0.75)*^+!R{L_2\ }};
{(25,25)*+{\circ}};
{(10,25)*+{\circ}};
{(25,10)*+{\circ}};
{(10,25)*+{}; (10,10)*+{} **\crv{(13,25)&(18,22)&(11.8,12)}\POS?(0.22)*_+!L{\,L_3}};
{(10,25)*+{}; (10,40)*+{} **\crv{(7,25)&(2,28)&(8.2,38)}\POS?(0.75)*_+!R{}};
{(25,10)*+{}; (10,10)*+{} **\crv{(25,13)&(22,18)&(12,11.8)}\POS?*_+!D{}};
{(25,10)*+{}; (40,10)*+{} **\crv{(25,7)&(28,2)&(38,8.2)}\POS?(0.77)*_+!U{\; L_1}};
{(40,10)*+{\bullet}; (40,40)*+{\bullet} **\crv{}\POS?*^+!L{}};
{(70,10)*+{\bullet}; (70,40)*+{\bullet} **\crv{}\POS?*^+!L{}};
{(40,10)*+{\bullet}; (70,10)*+{\bullet} **\crv{}\POS?*^+!L{}};
{(40,40)*+{\bullet}; (70,40)*+{\bullet} **\crv{}\POS?*^+!L{}};
{(40,10)*+{}; (70,40)*+{} **\crv{}\POS?(0.75)*^+!R{L_5\ }};
{(55,25)*+{\circ}};
{(40,25)*+{\circ}};
{(55,10)*+{\circ}};
{(40,25)*+{}; (40,10)*+{} **\crv{(43,25)&(48,22)&(41.8,12)}\POS?(0.22)*_+!L{\,L_6}};
{(40,25)*+{}; (40,40)*+{} **\crv{(37,25)&(32,28)&(38.2,38)}\POS?(0.75)*_+!R{}};
{(55,10)*+{}; (40,10)*+{} **\crv{(55,13)&(52,18)&(42,11.8)}\POS?*_+!D{}};
{(55,10)*+{}; (70,10)*+{} **\crv{(55,7)&(58,2)&(68,8.2)}\POS?(0.77)*_+!U{\; L_4}};
{(70,10)*+{\bullet}; (70,40)*+{\bullet} **\crv{}\POS?*^+!L{}};
{(100,10)*+{\bullet}; (100,40)*+{\bullet} **\crv{}\POS?*^+!L{}};
{(70,10)*+{\bullet}; (100,10)*+{\bullet} **\crv{}\POS?*^+!L{}};
{(70,40)*+{\bullet}; (100,40)*+{\bullet} **\crv{}\POS?*^+!L{}};
{(70,10)*+{}; (100,40)*+{} **\crv{}\POS?(0.75)*^+!R{L_8\ }};
{(85,25)*+{\circ}};
{(70,25)*+{\circ}};
{(85,10)*+{\circ}};
{(70,25)*+{}; (70,10)*+{} **\crv{(73,25)&(78,22)&(71.8,12)}\POS?(0.22)*_+!L{\,L_9}};
{(70,25)*+{}; (70,40)*+{} **\crv{(67,25)&(62,28)&(68.2,38)}\POS?(0.75)*_+!R{}};
{(85,10)*+{}; (70,10)*+{} **\crv{(85,13)&(82,18)&(72,11.8)}\POS?*_+!D{}};
{(85,10)*+{}; (100,10)*+{} **\crv{(85,7)&(88,2)&(98,8.2)}\POS?(0.77)*_+!U{\; L_7}};
\endxy\end{equation*}
From the figure it is clear that the $A$--branes $L_1,\cdots, L_{3n}$  give precisely the snake triangulation.

Again, the adjacency quiver of the triangulation may be read from the $2d$ BPS spectrum. Between vacua $X=1/2+k$ and $X=1/2(\tau+1)+k$, $k=0,1,2,\dots, n-1$, we have still two solitons, going opposite way along the $B$--cycle, but the vacuum at $\tau/2+k$ is connected to the vacua $1/2+(k-1)$, $1/2+k$, $(\tau+1)/2+(k-1)$ and $(\tau+1)+k$ by just one BPS soliton.
\textit{E.g.}\! for $n=2$ the $S$ matrix is
\begin{equation}
S=\left(
\begin{array}{cccccc}
 1 & -2 & 1 & 0 & 0 & 1 \\
 0 & 1 & -1 & 0 & 0 & -1 \\
 0 & 0 & 1 & -1 & 1 & 0 \\
 0 & 0 & 0 & 1 & -2 & 1 \\
 0 & 0 & 0 & 0 & 1 & -1 \\
 0 & 0 & 0 & 0 & 0 & 1
\end{array}
\right),
\end{equation}
(eigenvalues\footnote{\ In general, the monodromy $M$ for the $n$--punctured torus is equal, up to conjugacy, to the direct sum of $n$ copies of the $n=1$ monodromy.} of $M$: $-1,-1,1,1,-1,-1$) and $B=S-S^t$ is precisely the exchange matrix of the quiver \eqref{toruspunctquive} for $n=2$.

\subsubsection{Adjacency matrix \textit{vs.} $2d$ BPS spectrum}

In the above examples we saw that the adjacency quiver of the triangulation is given by the BPS quiver of the $2d$ $(2,2)$ system whose $A$--branes triangulate the surface $\cc$, in agreement with the basic idea of the $4d/2d$ correspondence.
The examples discussed so far correspond to simple situations where certain sign subtleties play no role. The equality will be verified in many examples below, including some non-trivial cases, as the one discussed in detail in appendix  \ref{app:weierstrass-prime}, where the subtleties of two--dimensional physics do play a significant role.

Let us consider the situation where $\cc=\C$ (\textit{i.e.}\! a sphere with a pole of order $p$ at $z=\infty$), the exchange matrix $B_{ij}$ of the $2d$ BPS quiver is given by the intersection number of  the corresponding arcs (up to mutations) in the $\{\gamma_i\}$ ideal triangulation, see ref.\cite{Hori:2000ck}
\begin{equation}\label{Bij=gigj}
B_{ij}=\pm \gamma_i\cdot \gamma_j.
\end{equation} 
Since the $A$--branes cross only at infinity, to get the correct counting of the intersection number one has to resolve the puncture by replacing it with a small circle with $p-2$ marked points, as required to interpret the family $\{\gamma_i\}$ as an ideal triangulation. Then the intersection $\gamma_i\cdot\gamma_j$ is given by the signed sum of $\pm 1$ over all triangles with sides $\gamma_i$, $\gamma_j$. In the case $\cc=\C$, or, topologically, a disk with $p-2$ marked points on the boundary, the quiver with exchange matrix $\gamma_i\cdot\gamma_j$ is, by the Milnor fiber theorem \cite{milnor}, given by the $A_{p-3}$ Dynkin quiver (up to equivalence), which is the same as the adjacency quiver of the disk with $p$ marked points \cite{fomin}.

In the general case, the intersection $\gamma_i\cdot\gamma_j$ again is concentrated at the poles, which, if irregular, must be resolved into boundary components. Locally, the situation is as in the previous case, and the counting still apply. It remains, however, the problems of specifying the signs \eqref{Bij=gigj} which are not determined at this level of analysis (except for the requirement that they must be compatible with the mutation--finiteness).  There are two sources of signs: the classical sign of the $A$--brane curve, and the quantum sign given by the sign of the determinants in the quantization around that configuration. The methods of ref.\cite{Cecotti:1993rm}, are very convenient to fix the signs (up to conventional choices) and  in all examples we analyzed we get quivers consistent with the 4d/2d correspondence.
 
\medskip

The identification of the $4d$ BPS  quiver of a generalized Gaiotto theory with the topological adjacency quiver of an ideal triangulation of the corresponding bordered surface has a few immediate payoffs.
\smallskip

First of all, it follows from the above correspondence that any mutation invariant of the Dirac pairing matrix, $B_{ij}$ is also a chamber--independent property of the four dimensional $\cn=2$ theory. The simplest such invariant is the \emph{corank} of the matrix $B_{ij}$, that is the number of independent charge vectors $v\in \Gamma$ which have vanishing Dirac pairing with all the charges in the theory. Physically, such vectors should be seen as \emph{flavor} charges, whereas the ones having non--trivial Dirac pairings have electric/magnetic nature. We shall, therefore, refer to the corank of $B$ as the \emph{number of flavor charges}.  For quivers arising from triangulation of surfaces, the number of flavor charges is given by
the number of punctures where $\phi_2$ is allowed to have poles of \emph{even} order \cite{fomin} (in particular, all ordinary double poles will contribute).

This result may also be understood in terms of the geometry of the Seiberg--Witten curve $\Sigma$.  Since $\Sigma$ is a double cover of $\cc$ its genus is given by
\begin{equation}
g(\Sigma)= 2g-1+\frac{1}{2}\, n_B,
\end{equation}   
where $n_B$ is the number of branch points. The branch points are given by \textit{i)} the zeros of $\phi_2$ (there are $4g-4+\sum_i p_i$ of them), \textit{ii)} the poles of $\phi_2$ of odd order. Then
\begin{equation}
g(\Sigma)= 4g-3+\frac{1}{2}\sum_{p_i\ \text{even}}p_i+\frac{1}{2}\sum_{p_i\ \text{odd}}(p_i+1).
\end{equation}
$g(\Sigma)$ is the number of linearly independent holomorphic one forms on $\Sigma$; however, $g$ of them are just pull--backs of holomorphic forms on the Gaiotto curve $\cc$. These are even under the cover group $\Z_2$, while the remaining $g(\Sigma)-g$ are odd. Dually, the number of odd $1$--cyles is $2g(\Sigma)-2g$.
Given that the canonical one--form, $y\, dx$, is $\Z_2$ odd, we get that the total number of electric \emph{and} magnetic charges is
\begin{equation}
2g(\Sigma)-2g=6g-6+\sum_{p_i\ \text{even}}p_i+\sum_{p_i\ \text{odd}}(p_i+1)= \mathrm{rank}\, B,
\end{equation}
as predicted by the Dirac quiver/triangulation quiver identification.
\medskip

The second obvious pay--off is a very convenient way of constructing (and understanding) complicated theories in terms of simpler ones. 
Indeed, having related a large class of $\cn=2$ theories to surfaces with punctures and boundaries, one can easily take two such theories, view them as two decoupled sectors of a more complicated  theory, and couple them by some suitable $\cn=2$ supersymmetric interactions. At the geometric level, this process of couplings various sub--sectors to construct a new model corresponds to surgery of triangulated surfaces. This viewpoint leads directly to simple rules for gluing together the sub--quivers associated to each sector into the quiver of the fully coupled theory. Thus one may get the quivers of  complicated models without going through the triangulation process or the $4d/2d$ correspondence. There exist different kinds of surgery, corresponding to physically different ways of coupling together the various sub--sectors. The geometrical rules of triangulation guarantee that only couplings which are fully consistent at the quantum level may be realized by a sequence of these surgical operations on quivers. For complicated models, which have no regime in which all couplings are simultaneously weak, this would be hard to check directly. Surgery processes are described in detail in section \ref{surgeries} below. 

\subsection{Ideal triangulations vs. Gaiotto $SU(2)$ theories}

We start by considering the original Gaiotto theories, namely closed surfaces with only ordinary punctures. 
Let $\cc$  be a surface of genus $g$ with $n$ ordinary punctures. The corresponding $\cn=2$ theory has a gauge group $SU(2)^{n+3g-3}$ \cite{gaiotto}, and hence a charge lattice $\Gamma$ generated by $3g-3+n$ electric charges, $3g-3+n$ magnetic ones, and $n$ flavor charges associated to the residues of $\sqrt{\phi_2}$ at the $n$ punctures. Thus,
\begin{equation}
\mathrm{rank}\,\Gamma= 6g+3n-6, 
\end{equation}
which is equal to the number of arcs in an ideal triangulation of the surface, and the number of nodes in its adjacency quiver.

From the description in section \ref{sec:adjacencyquivers} it follows that we may simplify, for this class of surfaces, the rules to construct the adjacency quivers by gluing blocks. We may start with a collection of of quiver blocks of just three kinds 
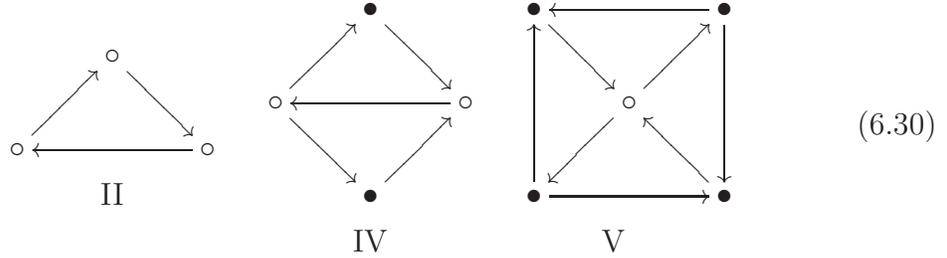
\begin{equation}
\begin{gathered}\xymatrix{& \circ\ar[dr] & \\
\circ\ar[ur] & & \circ\ar[ll]}\\
\mathrm{II}\end{gathered}\quad
\begin{gathered}\xymatrix{& \bullet\ar[dr] & \\
\circ\ar[ur]\ar[dr] && \circ\ar[ll]\\
& \bullet\ar[ur] &}\\
\mathrm{IV}\end{gathered}
\begin{gathered}\quad
\xymatrix{\bullet\ar[dr] & & \bullet\ar[ll]\ar[dd]\\
& \circ \ar[ur]\ar[dl] &\\
\bullet\ar[uu]\ar[rr] & & \bullet\ar[lu]}\\
\mathrm{V}
\end{gathered}
\end{equation} 
and then glue them together by identifying \emph{all} white nodes $\circ$ in pairs, this last condition being equivalent to $\partial\cc=\emptyset$ (\textit{i.e.}\! only ordinary punctures). 

The topological invariants $g$ and $n$ may be  read directly from the exchange matrix $B$ of the quiver: $n$ is just the corank $f$ of $B$ ($=$ the number of flavor charges) and
\begin{equation}\label{formgenus}
g= \frac{D-3f+6}{6},
\end{equation}
where $D$ is the size of the matrix $B$, equal to the number of nodes in the quiver.

Now we discuss a few examples. The case of $n$--punctured torus was considered in section \ref{sec:torusnpunctures}.

\subsubsection{Example: the sphere with $4$ punctures} 

The quiver for the sphere with four punctures, corresponding to $SU(2)$ gauge theory coupled to four flavors in the fundamental representation, is easy to construct. Just take two copies of the type IV block, and glue them together by identifying the white nodes $\circ$ in such a way that the orientations of the arrows connecting them match. We get the quiver\footnote{\ Here and below, we use vertical bars $|$ to denote the decomposition of a quiver into its basic blocks.}
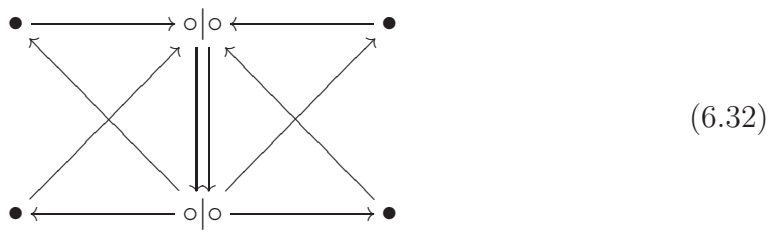
\begin{equation}\label{nffour}\begin{gathered}
\xymatrix{\bullet \ar[rr] && \circ|\circ \ar@<-0.45ex>[dd]\ar@<0.45ex>[dd] &&\bullet \ar[ll]\\
&&&&\\
\bullet \ar[uurr] &&\circ|\circ \ar[uurr]\ar[ll]\ar[uull]\ar[rr] &&\bullet\ar[uull]}
\end{gathered}
\end{equation}
equal to \eqref{SU(2)Nfquiver} for $N_f=4$. The underlying graph corresponds to Saito's $\widehat{\widehat{D}}_4$ elliptic root system \cite{saito}.

The exchange matrix $B$ has four zero eigenvalues: the corresponding eigenvectors are obtained by attaching a weight $1/2$ to the two white nodes, a weight $1$ to any one of the blacks ones, and zero to the other three nodes. Then the corank of $B$ is  $4$, and the quiver represents a triangulation of a surface with $(g,n)=(0,4)$ (see \! eqn.\eqref{formgenus}).

The mutation--class of the quiver \eqref{nffour} contains four essentially distinct quivers, as it is easy to check with the help of Keller's quiver mutation Java applet \cite{kellerapp}. The one shown in \eqref{nffour} is the one relevant in a weakly coupled chamber; it may be interpreted as the result of the coupling of four heavy electric hypermultiplets, represented by the black nodes, each carrying its own flavor charge, to the pure $SU(2)$ gauge theory, represented by the Kronecker subquiver, $\xymatrix{\circ \ar@<0.3ex>[r]\ar@<-0.3ex>[r]&\circ}$. In this limit, the two white nodes correspond to the dyon of charge $(e_1,m_1)=(2,-1)$ and the monopole of charge $(e_2,m_2)=(0,1)$ with Dirac pairing\footnote{Although the results of ref.\cite{Gaiotto:2009hg} are not stated in the language of quivers, many of their findings may be rephrased in the present formalism, with full agreement, except that their discussion in section 10.7 corresponds to a quiver differing from \eqref{nffour} by the orientation of two arrows. That quiver is not of finite--mutation type, and hence does not correspond to a complete theory in our sense. It does appear, however, in another kind of finite--type classification for $\cn=2$ quivers, namely those which admit a chamber with a finite BPS spectrum consisting only of hypermultiplets. The quiver \eqref{nffour} does not seem to have this last property.}
\begin{equation}
 \langle(e_1,m_1), (e_2,m_2)\rangle\equiv e_1m_2-m_1e_2 =2.
\end{equation}

\smallskip

According to the $4d/2d$ correspondence, the quiver \eqref{nffour} may be obtained as the BPS quiver of the $2d$ theory on the sphere with (say) the usual Fubini--Study K\"ahler potential, $K=-\log(1+|Y|^2)$, and superpotential
\begin{equation}
W(Y)= \frac{1}{Y^2+Y^{-2}},
\end{equation}
which is symmetric under the interchange of the two poles $Y\leftrightarrow Y^{-1}$ of the sphere, as well as under $Y\leftrightarrow -Y$. One has
\begin{equation}
W^\prime(Y)= -\frac{2Y-2Y^{-3}}{(Y^2+Y^{-2})^2}\equiv 2\frac{Y^5-Y}{(Y^4+1)^2}
\end{equation}
From which we see that the classical vacua are the four roots of unity $Y=i^k$, the south pole $Y=0$, and --- by the $Y\leftrightarrow Y^{-1}$ symmetry --- the north pole $Y=\infty$.
In total, we have \emph{six} vacua, as expected.

The critical values of the superpotential are $W=0$ for the two polar vacua,  and $W=y^2/2\equiv \pm 1/2$ for the vacua at the roots of unity. In the $W$--plane all soliton are just segments along the real axis \cite{fendleyetal}. Thus the BPS equation, $W(Y)=t$, reduces to the quadratic equation in $Y^2$
\begin{equation}
(Y^2)^2 -\frac{1}{t}\, Y^2+1.
\end{equation}
In the relevant interval of the real axis, $-1/2<t<1/2$, the discriminant is positive, and we have two real roots $Y^2$. As $t\rightarrow -1/2$ both roots go to $Y^2=-1$; analogously for $t\rightarrow 1/2$ both roots approach  $Y^2=1$. As $t\rightarrow 0$ one solution goes to zero and one to $\infty$. In conclusion, in each interval $-1/2\leq W\leq 0$, and $0\leq W\leq 1/2$, \emph{both} roots of the quadratic equation in $Y^2$ do correspond to solitons: one going to the vacuum at the north pole, $Y^2=\infty$, and the other to the vacuum at the south pole $Y^2=0$. Recalling that each solitonic solution in terms of $Y^2$ corresponds to \emph{two} solutions in terms of $Y$ related by the $\Z_2$ symmetry $Y\leftrightarrow -Y$: each starts at one of the two root--of--unity vacua sharing the given critical value (these two vacua are interchanged by $\Z_2$) and ends at one of the two polar vacua (which are $\Z_2$ invariant).
Then the BPS quiver has the form
(we label the vertices by the value of $Y$) 
\begin{equation}\label{quivOM}
\begin{gathered}\begin{xy} 0;<1pt,0pt>:<0pt,-1pt>:: 
(144,0) *+{1} ="0",
(0,52) *+{-i} ="1",
(0,0) *+{i} ="2",
(144,52) *+{-1} ="3",
(72,0) *+{0} ="4",
(72,52) *+{\infty} ="5",
"4", {\ar"0"},
"0", {\ar"5"},
"1", {\ar"4"},
"5", {\ar"1"},
"2", {\ar"4"},
"5", {\ar"2"},
"4", {\ar"3"},
"3", {\ar"5"},
\end{xy}\end{gathered}
 \end{equation}
 corresponding to the $S$ matrix
 \begin{equation}
 S=\left(
\begin{array}{cccccc}
 1 & 1 & 0 & 0 & -1 & 0 \\
 0 & 1 & 1 & -1 & 0 & 1 \\
 0 & 0 & -1 & 0 & 1 & 0 \\
 0 & 0 & 0 & 1 & -1 & 0 \\
 0 & 0 & 0 & 0 & 1 & -1 \\
 0 & 0 & 0 & 0 & 0 & 1
\end{array}
\right)
 \end{equation}
 ($2d$ monodromy spectrum given by $-1,1,1,1,1,-1$, the four $1$'s being associated to the four flavor charges).
The quiver \eqref{quivOM} is mutation equivalent to \eqref{nffour}.

\subsubsection{Example: the sphere with $n\geq 5$ punctures}

The extension to an arbitrary number $n\geq 4$ of ordinary punctures is straightforward. One takes two type IV blocks and $2(n-4)$ type II blocks and glue them together as in the figure
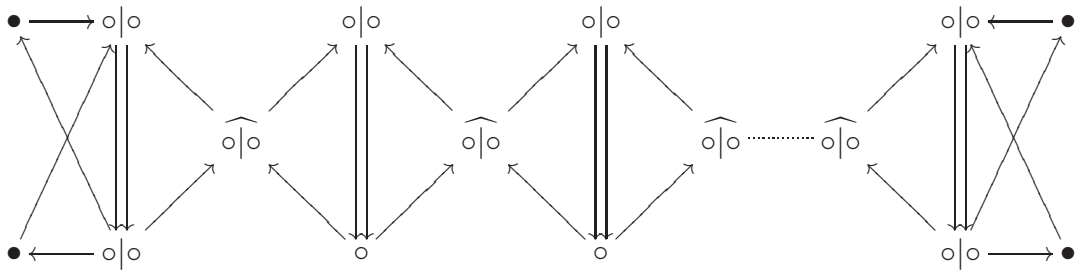
\begin{gather*}
\xymatrix{\bullet \ar[r] &\circ|\circ\ar@<-0.4ex>[dd]\ar@<0.4ex>[dd] && \circ |\circ\ar@<-0.4ex>[dd]\ar@<0.4ex>[dd]&& \circ|\circ\ar@<-0.4ex>[dd]\ar@<0.4ex>[dd] &&& \circ|\circ \ar@<-0.4ex>[dd]\ar@<0.4ex>[dd] &\bullet\ar[l]\\
&& \widehat{\circ|\circ}\ar[lu]\ar[ru] && \widehat{\circ|\circ}\ar[lu]\ar[ru] && \widehat{\circ|\circ}\ar[lu] \ar@{..}[r] &\widehat{\circ|\circ}\ar[ru] &\\
\bullet\ar[ruu] &\circ|\circ\ar[l]\ar[luu]\ar[ru] && \circ\ar[lu] \ar[ru] && \circ\ar[lu]\ar[ru] &&& \circ|\circ\ar[lu]\ar[r]\ar[ruu] &\bullet\ar[luu]}
\end{gather*} 
The incidence matrix of this quiver has $n$ zero eigenvectors, corresponding to attaching a weight $1$ to any one of the nodes $\bullet$ or $\widehat{\circ|\circ}$, weight $1/2$ to the nodes $\circ|\circ$ connected to it by an arrow, and zero everywhere else. Since the total number of nodes is $D=3n-6$, from eqn.\eqref{formgenus}, we see that the above quiver corresponds to a surface with numerical invariants $(g,n)=(0,n)$. 

The nodes $\bullet$ and $\widehat{\circ|\circ}$ are in one--to--one correspondence with the flavor charges (\textit{i.e.}\! zero eigenvectors of the incidence matrix $B$). Then they are interpreted as hypermultiplets carrying their own flavor charge and having electric charge $-1$ (that is, in the fundamental representation) with respect to each of the $SU(2)$ gauge groups (represented by the double--arrow Kronecker sub--quivers) connected to it by the arrows. Indeed, the node from which a double arrow starts/ends have charges $(e,m)= (2,-1)/(0,1)$ with respect to the corresponding gauge group and the arrows in the figure are consistent with the Dirac pairings 
\begin{equation}
 \langle (2,-1),(-1,0)\rangle= -1,\qquad \langle (0,1),(-1,0)\rangle= 1.
\end{equation}
The charge vectors in the kernel of $B$,
\begin{equation}
 \gamma_{\bullet_a} +\frac{1}{2} \sum_{\circ|\circ \rightleftarrows  \bullet_a}\gamma_{\circ|\circ}, \qquad\quad 
\gamma_{\widehat{\circ|\circ}_b} +\frac{1}{2} \sum_{\circ|\circ \rightleftarrows  \widehat{\circ|\circ}_b}\gamma_{\circ|\circ}\in \Gamma,
\end{equation}
then correspond to purely flavor ones.

 Thus the nodes $\bullet$ represent fundamental hypermultiplets, while the $\widehat{\circ|\circ}$ nodes stand for bi--fundamental ones. The above figure is the BPS quiver parallels the linear quiver representation of this gauge theory
\begin{equation}
\xymatrix{\text{\fbox{2}}\ar@{-}[r]&*++[o][F]{2}\ar@{-}[r]&*++[o][F]{2}\ar@{-}[r]&*++[o][F]{2}\ar@{..}[r]&*++[o][F]{2}\ar@{-}[r]&\text{\fbox{2}} }
\end{equation} 

The number of distinct quivers in the mutation class of the one represented in the figure grows quite rapidly with $n$. The first few numbers are\footnote{\ These numbers refer to the distinct quivers modulo sink/source equivalence as built in Keller's mutation applet \cite{kellerapp}.}
\begin{center}
\begin{tabular}{|c|c|c|c|c|}\hline
number of punctures & 4 & 5 &6 &7\\\hline
$\#$ of distinct mutation--equivalent quivers &  4 & 26 & 191 & 1904\\\hline
\end{tabular}
\end{center}

Since the theories are complete, each different quiver in the equivalence class corresponds to a physical regime of the $\cn=2$ theory. For genus zero surfaces with only ordinary double poles, one finds only one quiver in the mutation--class with the maximal number of double arrows (\textit{i.e.}\! Kronecker subquivers), namely $n-3$, which is the one we have drawn above, and which corresponds to the standard regime admitting a Lagrangian description.

\subsubsection{Example: genus $g>1$ with $n\geq 1$ punctures}\label{examplegL1npunct}

\begin{figure}
\begin{equation*}
\begin{gathered}\xymatrix{& \bullet \ar@{-}[r]^{1}\ar@{-}[rrrrdddd]_{\!\!12}\ar@{-}[ldddd]^{\!15}\ar@{-}[ddddd]^{\!14}\ar@{-}[dddddrrr]_{\!13} &\bullet\ar@{-}[r]^{2} \ar@{-}[rrrddd]^{\!\!\!10}\ar@{-}[rrrdddd]^{\!\!\!11} & \bullet\ar@{-}[r]^{3}\ar@{-}[rrdd]^{\!\!\!8}\ar@{-}[rrddd]^{\!\!9}&\bullet\ar@{-}[ddr]^{\!7} & \\ 
\bullet \ar@{-}[ur]^6  &&&& &\bullet\ar@{-}[ul]_4\\
& & & & & \bullet\ar@{-}[u]_3\\
& & & & & \bullet\ar@{-}[u]_2\\
\bullet\ar@{-}[uuu]^5& & & & & \bullet\ar@{-}[u]_1\\
& \bullet \ar@{-}[ul]^6\ar@{-}[rrr]_5 & & & \bullet \ar@{-}[ur]_4 &}
\end{gathered}\end{equation*}
\caption{\label{pipio} A $g=2$ $n=3$ `snake' ideal triangulation.}
\end{figure}
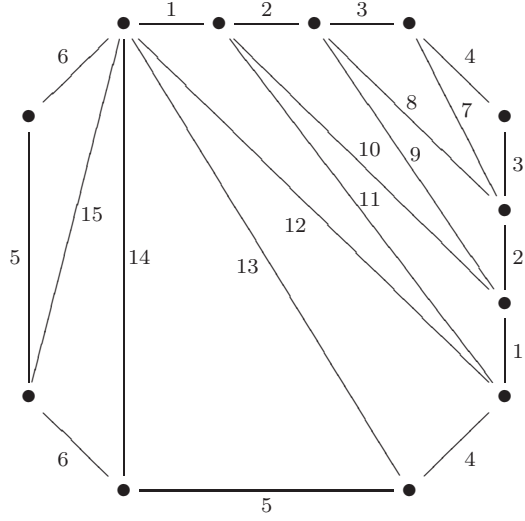

\begin{figure}
\begin{equation*}
\begin{xy} 0;<1pt,0pt>:<0pt,-1pt>:: 
(11,179) *+{1} ="0",
(0,92) *+{2} ="1",
(11,0) *+{3} ="2",
(122,17) *+{4} ="3",
(199,17) *+{5} ="4",
(321,88) *+{6} ="5",
(80,49) *+{7} ="6",
(31,60) *+{8} ="7",
(69,92) *+{9} ="8",
(31,125) *+{10} ="9",
(80,138) *+{11} ="10",
(122,160) *+{12} ="11",
(161,88) *+{13} ="12",
(199,160) *+{14} ="13",
(241,88) *+{15} ="14",
"9", {\ar"0"},
"0", {\ar@<-0.25ex>"10"},
"0", {\ar@<0.25ex>"10"},
"11", {\ar"0"},
"7", {\ar"1"},
"1", {\ar@<-0.25ex>"8"},
"1", {\ar@<0.25ex>"8"},
"9", {\ar"1"},
"3", {\ar"2"},
"2", {\ar@<-0.25ex>"6"},
"2", {\ar@<0.25ex>"6"},
"7", {\ar"2"},
"6", {\ar"3"},
"11", {\ar"3"},
"3", {\ar"12"},
"4", {\ar"5"},
"12", {\ar"4"},
"4", {\ar"13"},
"14", {\ar"4"},
"13", {\ar"5"},
"5", {\ar@<-0.25ex>"14"},
"5", {\ar@<0.25ex>"14"},
"6", {\ar"7"},
"8", {\ar"7"},
"8", {\ar"9"},
"10", {\ar"9"},
"10", {\ar"11"},
"12", {\ar"11"},
"13", {\ar"12"},
"14", {\ar"13"},
\end{xy}
\end{equation*}
\caption{\label{figg=2n=3} The adjacency quiver corresponding to the ideal triangulation \ref{pipio} of a $g=2$ surface with three punctures. The numeration of the nodes corresponds to the numeration of arcs in \ref{pipio}. In the left side of the quiver we see the `segment of a quiver $SU(2)$ theory' associated to the three punctures.}
\end{figure}
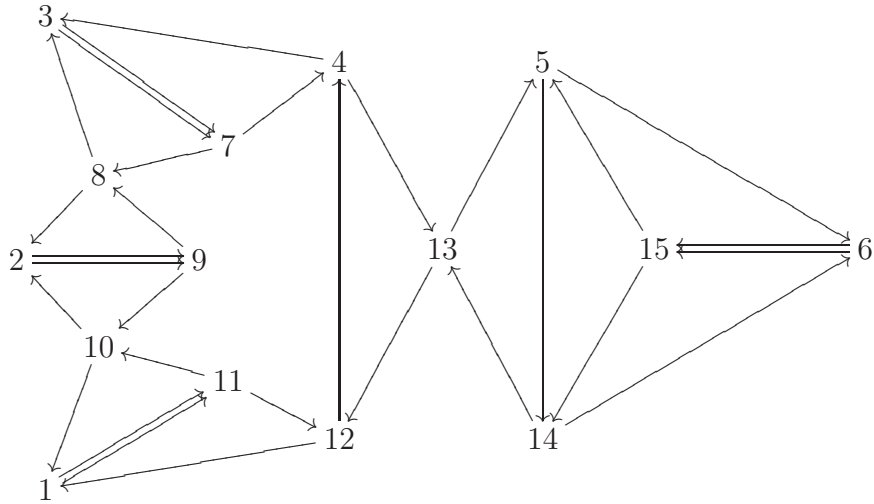

The analogue of the snake triangulation (see sect.\ref{sec:torusnpunctures}) for higher genus surface would be to cut open the surface to get a hyperbolic $4g$--gon with sides pairwise identified, taking care to choose one of the cuts in such a way that it goes through all the $n$ punctures. See  figure \ref{pipio} for the example with $g=2$, $n=3$.
Then one starts doing the snake triangulation from the side on which the punctures lie (see the upper right corner of the figure). From that part of the triangulation we get $n$ double arrows; for the example in the figure, they correspond to the following entries of the adjacency matrix 
\begin{equation}
B_{3,7}=+2,\quad B_{2,9}=+2,\quad B_{1,11}=+2.
\end{equation}
Then it remains to perform the triangulation of a $[4(g-1)+2]$--gon with the first $4(g-1)$ sides identified pairwise in the form $s_1,s_2, s_1, s_2, s_3, s_4, s_3,....$ while the last two sides are not identified. In the figure this corresponds to the part of the surface below arc $12$. Let $c(g)$ be the maximal number of double arrows that we may get from such a triangulation. Then we have a triangulation with at most
\begin{equation}
n+c(g)
\end{equation}  
double arrows. It is easy to convince oneself that $c(g)=g-1$.
See figure \ref{figg=2n=3} for the quiver corresponding to the ideal triangulation \ref{pipio}.

\begin{figure}
\begin{equation*}
\xymatrix{& \bullet \ar@{-}[r]^{1}\ar@{-}[rrrrdddd]_{\!\!12}\ar@{-}[dddddrrr]_{\!\!13} &\bullet\ar@{-}[r]^{2} \ar@{-}[rrrddd]^{\!\!\!10}\ar@{-}[rrrdddd]^{\!\!\!11} & \bullet\ar@{-}[r]^{3}\ar@{-}[rrdd]^{\!\!\!8}\ar@{-}[rrddd]^{\!\!9}&\bullet\ar@{-}[ddr]^{\!7} & \\ 
\bullet \ar@{-}[ur]^6\ar@{-}[ddddrrrr]_{\!\!14}  &&&& &\bullet\ar@{-}[ul]_4\\
& & & & & \bullet\ar@{-}[u]_3\\
& & & & & \bullet\ar@{-}[u]_2\\
\bullet\ar@{-}[uuu]^5\ar@{-}[drrrr]^{\!15}& & & & & \bullet\ar@{-}[u]_1\\
& \bullet \ar@{-}[ul]^6\ar@{-}[rrr]_5 & & & \bullet \ar@{-}[ur]_4 &}\qquad\quad 
\xymatrix{& \bullet \ar@{-}[r]^{1}\ar@{-}[rrrrdddd]_{\!\!12}\ar@{-}[ldddd]^{\!15}\ar@{-}[ddddd]^{\!14} &\bullet\ar@{-}[r]^{2} \ar@{-}[rrrddd]^{\!\!\!10}\ar@{-}[rrrdddd]^{\!\!\!11} & \bullet\ar@{-}[r]^{3}\ar@{-}[rrdd]^{\!\!\!8}\ar@{-}[rrddd]^{\!\!9}&\bullet\ar@{-}[ddr]^{\!7} & \\ 
\bullet \ar@{-}[ur]^6  &&&& &\bullet\ar@{-}[ul]_4\\
& & & & & \bullet\ar@{-}[u]_3\\
& & & & & \bullet\ar@{-}[u]_2\\
\bullet\ar@{-}[uuu]^5& & & & & \bullet\ar@{-}[u]_1\\
& \bullet \ar@{-}[ul]^6\ar@{-}[rrr]_5\ar@{-}[urrrr]_{\!13} & & & \bullet \ar@{-}[ur]_4 &}
\end{equation*}
\caption{\label{otherg2} Inequivalent `snake' triangulation of the same surface.}
\end{figure}
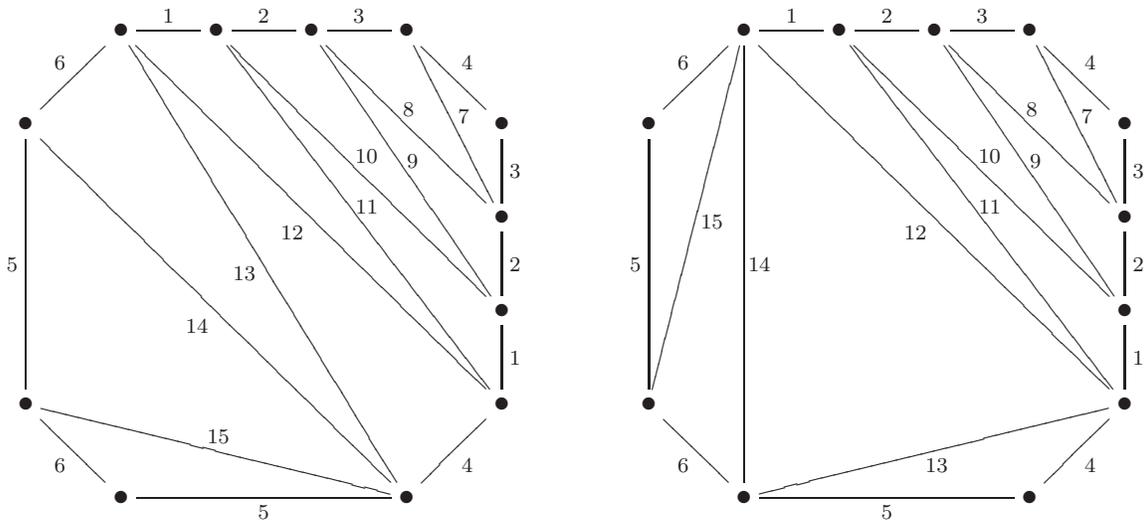

Therefore, for $g>1$, the maximal number of double--arrows, $n+g-1$, is less than the number of $SU(2)$ gauge groups, namely $n+3g-3$.  As discussed before,  this means that these theories have no chamber in which all the matter
multiplets can be massed up.  Indeed we will later show this is the case, by showing that there are no
gauge invariant mass terms that can mass up all the matter fields.
On the other hand, for $g>1$ the quiver with a maximal number of double arrows is not unique. For instance, for $g=2$, $n=3$ the triangulations in figure \ref{otherg2} also lead to four double arrows.

\subsection{Generalized Gaiotto theories}

The quivers of generalized Gaiotto theories are constructed by gluing together all five kinds of blocks, and there is no need to pair up every white node.

Generically, each quiver may be decomposed into blocks in a unique way. In this case, there is a unique bordered surface associated to the mutation-class of the quiver. There are a few exceptions to the uniqueness of the correspondence, and all these exceptions have a simple physical explanation: basically, these theories have more than one string/$M$--theory engineering, and each of these realizations corresponds to a bordered surface. The quiver--mutation class, however, should be (and it is) independent of the geometrical realization. The typical example is $SU(2)$ with two flavors which has two such realizations \cite{gaiotto,Gaiotto:2009hg,Gaiotto:2009fs}.  

\subsubsection{Example: $A_n$ and $D_n$ Argyres--Douglas models}

From the ideal triangulation point of view \cite{fomin}, the $A_n$, $D_n$ models correspond, respectively, to the disk with $n+3$ marked point on the boundary and to the punctured--disk with $n$ marked points, that is to a sphere equipped with a quadratic differential $\phi_2$ having one pole of degree $n+5$, and, respectively, two poles of degrees $2$ and $n+2$.
\smallskip

These models are easily understood from the point of view of the $4d/2d$ correspondence. For the $A_n$ series we choose a reference quadratic differential $\omega_0$ having a  pole of order $4$ at infinity, while for the $D_n$ series we pick an $\omega_0$ having a third order pole at infinity and a simple pole at the ordinary puncture.
We get the superpotentials:
\begin{align}
&A_n\colon\quad W(X)= X^{n+1}+\text{lower terms}\\
&D_n\colon\quad W(X)= \frac{1}{X}+X^{n-1}+\text{lower terms}.
\label{supDn}\end{align}
The first superpotential is just the usual one for the $A_n$ minimal models \cite{min1,min2}, and we know that, in some chamber, the BPS quiver is just the $A_n$ Dynkin diagram with some orientation of the edges (which orientation being immaterial, since all orientations are mutation--equivalent for a tree quiver). This is, of course, the correct quiver for the $A_n$ Argyres--Douglas model obtained by compactifing the Abelian six dimensional $(2,0)$ theory on a complex curve of equation
\begin{equation}
y^2= X^{n+1}+\text{lower terms},
\end{equation}
unfolding the minimal $A_n$ singularity.
\smallskip

On the other hand, eqn.\eqref{supDn} does not look like the usual superpotential for the $D_n$ minimal models,
\begin{equation*}
W(X,Z)= X^{n-1}+X\,Z^2+\text{lower terms}.
\end{equation*}
However, to identify the BPS quiver we are free to deform the theory by adding
`lower terms' in $W(X,Z)$ in any convenient way. We take them to have the form,
\begin{equation}
W(X,Z)= X^{n-1}+X\, Z^2-2\lambda Z.
\end{equation} 
Now the chiral superfield $Z$ is massive, and since it appears quadratically can be integrated out, giving
\begin{equation}\label{Dnmod}
W(X)= X^{n-1}- \lambda^2/X,
\end{equation}
in agreement with eqn.\eqref{supDn}. Hence the BPS quiver is the same as the $D_n$ minimal model one, that is (up to mutation equivalence) the $D_n$ Dynkin diagram with some orientation of the arrows (again, all orientations are equivalent).
\smallskip

The four--dimensional $\cn=2$ models of these series are studied in detail in ref.\cite{cnv}.

\subsubsection{Example: $SU(2)$ with $N_f=0,1,2,3$}

\noindent$\bullet$ \textit{Pure $SU(2)$}\vglue 12pt

The quadratic differential for the $N_f=0$ theory has the general form
\begin{equation}
\phi_2 = \left(\frac{A}{z^3}+\frac{B}{z^2}+\frac{C}{z}\right)dz^2,
\end{equation}
which has poles of order $3$ at the north and south pole of $\mathbb{P}^1$. Then its quiver should correspond to the triangulation of the annulus with a marked point on each boundary component, which is the Kronecker quiver, that is the affine
$\widehat{A}_1(1,1)$ quiver.

Let us check this result from the $4d/2d$ correspondence. We choose $\omega_0$ equal to $dz^2/z^2$, and write $z=e^X$ with $X$ taking value in the cylinder, \textit{i.e.}\! $X\sim X+2\pi i$. The resulting Landau--Ginzburg model is 
\begin{equation}W(X)= A\, e^{-X}+B +C\, e^X,
\end{equation}
which is equivalent to the $\mathbb{CP}^1$ sigma--model, whose BPS spectrum was computed in refs.\cite{Cecotti:1991me,exact,Cecotti:1993rm}: the model has two vacua connected by two BPS particles, and hence its BPS quiver is $\widehat{A}_1(1,1)$.
\bigskip

\noindent$\bullet$ \textit{$N_f=1$}\vglue 12pt

The $N_f=1$ quadratic differential is
\begin{equation}
\phi_2 = \left(\frac{A}{z^4}+\frac{B}{z^3}+\frac{C}{z^2}+\frac{D}{z}\right)dz^2.
\end{equation}
It has a pole of order $4$ at the south pole $z=0$ and one of order $3$ at the north pole $z^{-1}=0$; hence it corresponds to the triangulation of an annulus with one marked point on one boundary and two on the other, whose adjacency quiver is (up to equivalence) equal to the affine quiver $\widehat{A}_2(2,1)$.

The same conclusion is obtained from the $4d/2d$ correspondence.
Choosing $\omega_0$ as in the $N_f=0$ case, we get the Landau--Ginzburg model on the cylinder
\begin{equation}\label{ddbb}
W(X)= A e^{-2X}+ B e^{-X}+ C +D e^{X},
\end{equation}
which was solved in refs.\cite{Cecotti:1991me,Cecotti:1993rm}. From the solution, one sees that BPS quiver of the model \eqref{ddbb} is $\widehat{A}_2(2,1)$.
\bigskip

\noindent$\bullet$ \textit{$N_f=2$. First realization}\vglue 12pt

$N_f=2$ has two brane engineerings \cite{gaiotto,Gaiotto:2009hg,Gaiotto:2009fs} which correspond to ideal triangulations of \emph{different} bordered surfaces. The two triangulations, corresponding to the same physical theory, have the same adjacency quiver (up to mutation); indeed, this is one of the few cases in which the same mutation--class of quivers corresponds to a pair of topologically distinct surfaces, namely an annulus with two marked points on each boundary, and a disk with one ordinary puncture and three marked points on the boundary. The equality becomes less mysterious if we recall that the first surface has the $\widehat{A}_3(2,2)$ affine Dynkin quiver, whereas the second should have the  $\widehat{D}_3$ affine Dynkin quiver, and the two quivers are identified by the Lie algebra isomorphism $\mathfrak{su}(4)\simeq \mathfrak{so}(6)$.
\smallskip

The $\phi_2$ for the first realization is
\begin{equation}
\phi_2 = \left(\frac{A}{z^4}+\frac{B}{z^3}+\frac{C}{z^2}+\frac{D}{z}+E\right)dz^2, 
\end{equation}
which indeed corresponds to an annulus with two marks on each boundary.
The corresponding LG model, defined on the cylinder, has superpotential
\begin{equation}
W(X)= A e^{-2X}+ B e^{-X}+ C +D e^{X}+E e^{2X},
\end{equation}
Again, to compute the equivalence class of the BPS quiver we may adjust the constants to convenient values. Setting $B=D=0$, we recover the $\sinh(2X)$ model solved in ref.\cite{Cecotti:1991me}. From the explicit solution we see that the BPS quiver is $\widehat{A}_3(2,2)$, as predicted by the $4d/2d$ correspondence.
\bigskip

\noindent$\bullet$ \textit{$N_f=2$. Second realization}\vglue 12pt

The $\phi_2$ of the second realization is
\begin{equation}\label{nf2sec}
\phi_2 = \left(\frac{A}{z^2}+\frac{B}{(z-1)^2}+\frac{C}{z(z-1)}+\frac{D}{z}\right)dz^2,
\end{equation}
which manifestly corresponds to a disk with two punctures and one mark on the boundary.
The corresponding LG model has superpotential
\begin{equation}W(X)= A+ \frac{B\, e^{2X}}{(e^{X}-1)^2}+ \frac{C\,e^X}{(e^X-1)} +D e^{X}.
\end{equation}

The check that the BPS quiver of the Landau--Ginzburg model 
\eqref{nf2sec} is mutation equivalent to $\widehat{A}_3(2,2)$ is confined in appendix \ref{APPdetailsLS}.
\bigskip

\noindent$\bullet$ \textit{$N_f=3$}\vglue 12pt

This model has the quadratic differential
\begin{equation}
\phi_2= \left(\frac{A}{z^2}+ \frac{B}{(z-1)^2}+ \frac{C}{z}+ \frac{D}{z-1}+E\right)dz^2
\end{equation}
corresponding to the twice--punctured disk with 2 marked points on the boundary, whose adjacency quiver is the
affine $\widehat{D}_4$.

The
 LG model is
\begin{equation}
W(X)= e^{2X}+ \frac{1}{(1-e^{-X})^2}\equiv e^{2X}\: \frac{(e^X-1)^2+1}{(e^X-1)^2}.
\end{equation}
In appendix \ref{APPdetailsLS} it is checked that the BPS quiver of the $2d$ theory is in the mutation class of $\widehat{D}_4$.

\subsubsection{Example: other affine $\widehat{A}, \widehat{D}$ models}

$SU(2)$ gauge theory with $N_f=0,1,2,3$ give the first examples of four--dimensional $\cn=2$ models whose BPS quiver is of the affine $\widehat{A}$ or $\widehat{D}$ type.

The general affine $\widehat{A}$ model corresponds to a quadratic differential on the sphere having two poles of order $n+2$ and $m+2$, with $n,m\geq 1$, that is, to an annulus $\mathscr{A}_{n,m}$ with $n$ (resp.\! $m$) marked points on the first (resp.\! second) boundary. The adjacency quiver of $\mathscr{A}_{n,m}$ is $\widehat{A}_{n+m-1}(n,m)$, \textit{i.e.}\! the $\widehat{A}_{n+m-1}$ Dynkin graph with $n$ arrows pointing in the positive direction and $m$ in the negative one.

\begin{figure}
\begin{equation*}
\begin{gathered}
\xymatrix{ & & & & \bullet \ar@<0.4ex>[dd]\ar@<-0.4ex>[dd] & & & & &\\
\bullet_{n-1} & \cdots\ar[l] & \bullet_2 \ar[l] & \bullet_1\ar[l]\ar[ur] & & \circ_1\ar[ul]\ar[r] & \circ_2 \ar[r] & \circ_3 \ar[r] &\cdots \ar[r] & \circ_{m-1}\\
& & & & \bullet \ar[ur]\ar[ul] & & & & & &} 
\end{gathered}
\end{equation*}
\caption{\label{PPPiii} The quiver mutation--equivalent to the affine Dynkin quiver $\widehat{A}(n,m)$ (with $n,m\geq 1$) having a Kronecker subquiver.}
\end{figure}
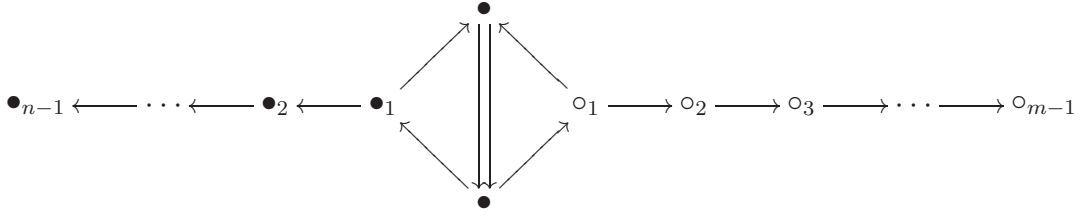

The quiver with the maximal number ($=1$) of Kronecker subquivers in the mutation--class of the Dynkin quiver $\widehat{A}_{n+m-1}(n,m)$
is represented in figure \ref{PPPiii};
this quiver may be interpreted as an $SU(2)$ gauge sector coupled to two disconnected $\cn=2$ systems in the sense of section \ref{sec:mutfinvsSU(2)}. Taking $n,m=1,2$ we recover $SU(2)$
 with $N_f=0,1,2$.
 \medskip
 
The corresponding $2d$ theory is 
\begin{equation}
W(X)= e^{nX}+e^{-mX}.
\end{equation} 
Its BPS spectrum is given by the second case of example 4 in section 8.1 of ref.\cite{Cecotti:1993rm} ($n$ of that reference corresponds to the present $n+m$, while $k_0$ is to be identified with $m$) corresponding to an affine $\widehat{A}_{n+m-1}$ Dynkin graph.  As a further check, we note that the conjugacy class of the $2d$ quantum monodromy computed in  \cite{Cecotti:1993rm} precisely agrees with that of minus the Coxeter element of the $\widehat{A}(n,m)$ quiver computed in ref.\cite{kellrecrel}.
\medskip

The affine quivers $\widehat{D}_{n-1}$ correspond to a triangulation of a disk with two punctures and
$(n-3)$ marked points on the boundary. 
The mutation--equivalent quiver with the maximal number (one) of Kronecker subquivers is obtained by gluing one block of type IV, one of type II, and $n-5$ blocks of type I,
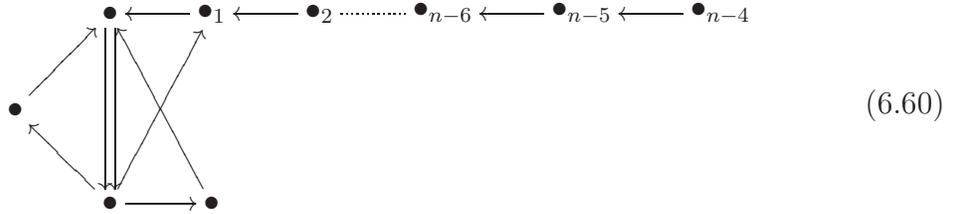
\begin{equation}\label{Dnaff}
\begin{gathered}
\xymatrix{ & \bullet\ar@<-0.35ex>[dd]\ar@<0.35ex>[dd] &\bullet_1\ar[l]& \bullet_2\ar[l]&\bullet_{n-6}\ar@{..}[l]& \bullet_{n-5}\ar[l] &\bullet_{n-4}\ar[l]\\
\bullet\ar[ur] & & & & &\\
& \bullet \ar[lu]\ar[ruu]\ar[r] & \bullet\ar[uul] &&&&}
\end{gathered}
\end{equation}
which corresponds to the coupling of $SU(2)$ to three $\cn=2$ systems, two of which being ordinary hypermultiplets.
$n=5$ reproduces $SU(2)$ with $N_f=3$. 
\smallskip

The $2d$ model is a generalization of the one for $SU(2)$ with three flavors. 
\smallskip

There are some exceptional cases. From $SO(6)\simeq SU(4)$, we see that $\widehat{D}_3\simeq \widehat{A}_3(2,2)$, and the same quiver represent both the triangulation of  a twice--punctured  $1$--gon and of an annulus with two marks on each boundary. As we have remarked these two surfaces correspond to two different $M$--theory realizations of $SU(2)$ coupled to two fundamental flavors.

\subsubsection{Example: a remarkable \emph{unique--quiver} AF model}\label{sec:uniquequiver}

$\cn=2$ and $\cn=4$ $SU(2)$ super--Yang--Mills share a rare property, namely their quivers --- respectively the Kronecker and the Markov ones --- are the \emph{only} element of their mutation class.
In this section, we illustrate a third $\cn=2$ theory with this uniqueness property: the generalized Gaiotto model on the torus with a pole of order three (\textit{i.e.}\! a boundary with a marked point).
Cutting open the torus, we have the ideal triangulation in the figure
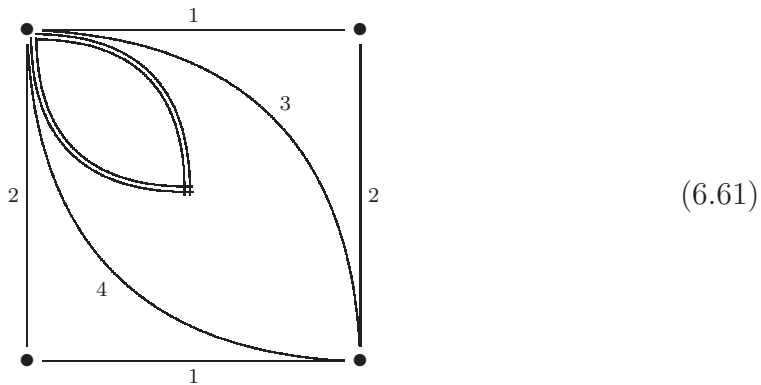
\begin{gather}
\begin{gathered} \xymatrix{\bullet \ar@{=}@<-0.7ex>@/^2pc/[ddrr]
 \ar@{=}@<0.7ex>@/_2pc/[ddrr]\ar@{-}[rrrr]^1 \ar@{-}@/^3.5pc/[ddddrrrr]^3\ar@{-}@/_3.5pc/[ddddrrrr]_4 & & & &\bullet\ar@{-}[dddd]^2\\
 & & & &\\
 && &&\\
 &&&&\\
 \bullet\ar@{-}[uuuu]^2 &&&&\bullet\ar@{-}[llll]^1 }
\end{gathered}\end{gather}
where the double line stands for the boundary of the surface. With the numbering of arcs in figure, the adjacency matrix reads
\begin{align}
&B_{1,2} =+2 && B_{1,3} =-1 && B_{1,4}=-1\\
& B_{2,3}=+1 && B_{2,4} =+1 && B_{3,4} =+1.
\end{align}
corresponding to the quiver
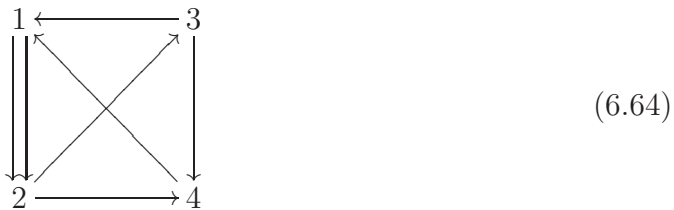
\begin{equation}\label{remquiver}
\begin{gathered}
\xymatrix{1\ar@<-0.5ex>[dd]\ar@<0.5ex>[dd] && 3\ar[ll]\ar[dd]\\
& &\\
2 \ar[rr]\ar[rruu] && 4\ar[uull]}
\end{gathered}
\end{equation}
Using Keller's mutation applet \cite{kellerapp}, one checks that this quiver is the only one in its mutation class.  This theory has no flavor charge, and it is not UV conformal according to our discussion in section \ref{sec:conformal}, as well as the graphical rule of section \ref{sec:mutfinvsSU(2)}; indeed, \eqref{remquiver}
is a proper subgroup of the finite--mutation quiver obtained
by gluing three type II blocks
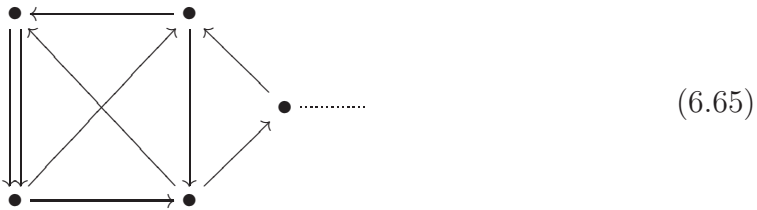
\begin{equation}
\begin{gathered}
\xymatrix{\bullet\ar@<-0.4ex>[dd]\ar@<0.4ex>[dd] && \bullet\ar[ll]\ar[dd] &&\\
& & &\bullet\ar[ul]\ar@{..}[r]&\\
\bullet \ar[rr]\ar[rruu] && \bullet\ar[uull]\ar[ru] &}
\end{gathered}
\end{equation}

In section \ref{sec:newdualities} we give an alternative definition of this theory as $SU(2)$ SYM gauging the diagonal $SU(2)$ subgroup of the $SU(2)\times SU(2)$ global symmetry of a composite $\cn=2$ system.
\medskip

From the $4d/2d$ correspondence perspective, the simplest Landau--Ginzburg superpotential corresponding to this geometry is
\begin{equation}
W(X)=\wp^\prime(X).
\end{equation}
One has
\begin{equation}
W^\prime(X)= \wp^{\prime\prime}(X)= 6\, \wp(X)^2- \frac{1}{2}\, g_2,
\end{equation}
which gives four supersymmetric vacua at $\pm X_\pm$, where $\wp(X_\pm)=\pm \sqrt{g_2/12}$. This $2d$ model has all the subtleties we alluded to before; luckily, they were understood in  \cite{Cecotti:1993rm}. The detailed analysis is presented in  appendix
\ref{app:weierstrass-prime}. The $2d$ computation 
confirms the quiver \eqref{remquiver}.

\subsection{Surface/quiver surgeries}\label{surgeries}

From the general discussion in \S.\,\ref{sec:mutfinvsSU(2)} as well as the examples in the previous two subsections, we see that the process of coupling several basic $\cn=2$ systems to construct more complicated ones is reflected at the quiver level in a kind of graphical gluing process. In the case of generalized Gaiotto theories, this gluing process should be related to a topological surgery of the corresponding bordered surface, triangulated in a such a way that the triangulation of the resulting surface may be easily related to those of the several pieces we glue.

The surface surgery process is important from Gaiotto's duality point of view \cite{gaiotto}, where $SU(2)$ gauge sectors are described, in their weak coupling limit, as long plumbing tubes connecting punctures in standard degeneration limit of Riemann surfaces. The plumbing parameter is given by $q=e^{2\pi i \tau}$, where $\tau$ is the complexified $SU(2)$ coupling. Thus the surgery processes allow us to fill a gap in the discussion of  \S.\,\ref{sec:mutfinvsSU(2)} by showing that a Kronecker subquiver $\mathbf{Kr}$ may be identified with a plumbing tube, which may be taken to be tiny, thus setting the corresponding $SU(2)$ coupling to small values where a Lagrangian description is meaningful.

There are many possible surgery processes, corresponding to the variety of `fundamental' $\cn=2$ systems and of possible supersymmetric couplings between them. Here we limit to the basic ones, without any claim to the completeness of the list. They are the ones with the more transparent physical interpretation.

\subsubsection{Massive flavor surgery}

Suppose we are in the following situation. In some regime, the Gaiotto theory associated to the closed surface $\cc_{g,n}$ looks like two distinct sectors weakly coupled through some bi--fundamental hypermultiplet, carrying its own flavor charge, whose $SU(2)\times SU(2)$ symmetry is weakly gauged by vectors belonging to both of the above sectors. Giving mass to the coupling hypermultiplet, and taking the limit $m\rightarrow\infty$, the theory completely decouples into two distinct $\cn=2$ systems, each  corresponding to a piece of the original surface $\cc_{g,n}$ which gets broken in two parts in the infinite mass limit. We are interested in understanding the
$\cn=2$ physical systems encoded in each surface piece, and their relation to the coupled $\cn=2$ model engineered  
by the original surface $\cc_{g,n}$.
Then we wish to learn
 how to revert the process and couple together the sub--systems by gluing various elementary `pieces' to produce the higher genus surface $\cc_{g,n}$.
\smallskip

The connected surface pieces arising from the $m\rightarrow\infty$ limit are necessarily surfaces {with boundaries} (\textit{i.e.}. whose Gaiotto construction has irregular poles). Indeed, the original theory was conformal, and hence the $\beta$--functions of all $SU(2)$ groups vanished, including the $SU(2)$'s gauging the symmetries of the hypermultiplet whose mass we take to infinity. When this last field is decoupled, the corresponding $\beta$--functions will not be  zero any longer, but equal to minus the original contribution from the massive hypermultiplet. Therefore, neither of the two remaining decoupled sectors may be superconformal, and hence they cannot correspond to a closed surface. However, since the surgery is local, and only a puncture is involved, the two pieces will have just one boundary component each, and the original puncture associated to the massive flavor will remain as a marked point on each boundary.
\smallskip

\begin{figure}
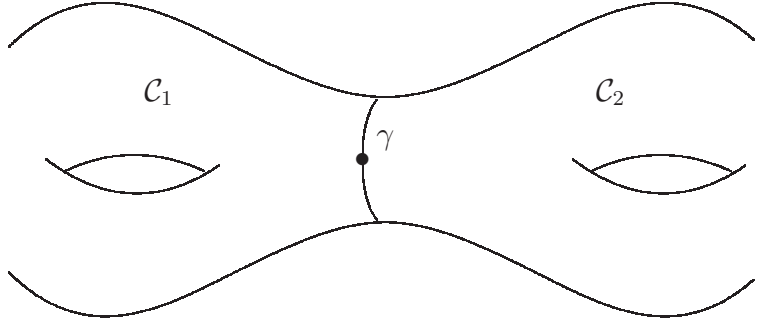

\begin{equation*}
\xy {(20,20)*+{}; (120,20)*+{} **\crv{(30,10)&(50,20)&(70,30)&(90,20)&(110,10)}\POS?*^+!L{}};
{(20,50)*+{}; (120,50)*+{} **\crv{(30,60)&(50,50)&(70,40)&(90,50)&(110,60)}\POS?*^+!L{}};
{(25,35)*+{}; (49,35)*+{} **\crv{(37,26)}\POS?*^+!L{}};
{(95,35)*+{}; (119,35)*+{} **\crv{(107,26)}\POS?*^+!L{}};
{(27.5,33.5)*+{}; (47,33)*+{} **\crv{(37,38)}\POS?*^+!L{}};
{(97.5,33.5)*+{}; (117,33)*+{} **\crv{(107,38)}\POS?*^+!L{}};
{(67,35)*+{}; (70,26)*+{} **\crv{(67,28)}\POS?*^+!L{}};
{(67,35)*+{\bullet}};
{(67,35)*+{}; (70,44)*+{} **\crv{(67,42)}\POS?(0.2)*^+!L{\,\gamma}};
{(40,44)*+{\cc_1}}; {(100,44)*+{\cc_2}};
\endxy\end{equation*}
\caption{\label{flavorarc} A separating arc $\gamma$ passing through a `massive' ordinary puncture $\bullet$}
\end{figure}

From the point of view of the ideal triangulation, this is described as follows. The triangulation has an arc $\gamma$, starting and ending at the `massive' (ordinary) puncture, which separates the surface into parts (see figure \ref{flavorarc}). We cut along the arc $\gamma$ and separate the surface into two components $\cc_1$ and $\cc_2$. The arc $\gamma$ then becomes --- on both pieces $\cc_1,\cc_2$ --- a boundary with a marked point at the position of the original puncture. Notice that this process is essentially local, so our discussion applies also to the case in which cutting the separating arc $\gamma$ will not disconnect the surface, but rather produce two boundaries each with a marked point.\smallskip

The two pieces are of the form $\cc_{g_1,n_1,1,1}$ and $\cc_{g_2,n_2,1,1}$ with
\begin{gather}
g=g_1+g_2\\
n=n_1+n_2+1.
\end{gather}
The original quiver associated to the closed surface $\cc_{g,n}$ had rank $6g-6+3n$, whereas the rank of each of the resulting subquivers is $6 g_i+3n_i-2$ so
\begin{equation}
\mathrm{rank}(\cc_{g,n})= \mathrm{rank}(\cc_{g_1,n_1,1,1})+\mathrm{rank}(\cc_{g_2,n_2,1,1})+1
\end{equation}
which is the correct number since we lose one flavor charge in the infinite mass limit. Instead, if the surface remains connected after cutting $\gamma$, it has the form $\cc_{g-1,n-1,2,2}$ whose rank $6g-6+3n-1$ is again one less that the original one.
\smallskip

From the quiver point of view, the process of breaking the surface into two parts is straightforward. One simply eliminates the separating node $\bullet$ and all the arrows connecting it to the rest of the quiver, thus obtaining two disconnected components corresponding to the ideal triangulations of the two pieces $\cc_1$, $\cc_2$  of the surface $\cc_{g,n}$ or a connected quiver corresponding to a surface $\cc_{g-1,n-1,2,2}$ having {two} boundaries each with a marking.\smallskip  

The inverse process, the massive flavor surgery, is also easy to describe. Suppose we are given the quivers, $Q_1$ and $Q_2$, associated to the two pieces each corresponding to a surface $\cc_i$ with a boundary $\gamma_i$ having a single marked point (or the connected adjacency quiver of a surface with two boundary components with one marking each).  In the triangulation of $\cc_1$, the boundary segment $\gamma_1$ is either a side of an ordinary triangle, or of a punctured $2$--gon, or of a twice--punctured $1$--gon (this last possibility occurring only if $\cc_1$ itself is a twice--punctured $1$--gon). In the block decomposition of $Q_1$, the first two possibilities correspond to a `boundary block' of type, respectively, I or III. In the third case $Q_1\equiv \widehat{A}_3(2,2)$. The same applies to $Q_2$.

\begin{figure}
\begin{gather*}
 \begin{gathered}\xymatrix{\circ\\ \\ \circ\ar[uu]}\\
\mathrm{I}\end{gathered}\ \longrightarrow\ \begin{gathered}\xymatrix{\circ\ar[dr] &\\
& \circ\ar[dl]\\
\circ\ar[uu] &}\\ \mathrm{II}\end{gathered}\\
\begin{gathered}\xymatrix{& \bullet\\ \circ \ar[ur]\ar[dr] &\\ &\bullet}\\
\mathrm{III}\end{gathered}\ \longrightarrow\ \begin{gathered}\xymatrix{& \bullet \ar[dr] &\\
\circ\ar[ur]\ar[dr] & & \circ\ar[ll]\\
& \bullet\ar[ur] &}\\ \mathrm{IV}\end{gathered}\\
\begin{gathered}\widehat{A}_3(2,2)\end{gathered}\ \longrightarrow\ \begin{gathered}\xymatrix{\bullet\ar[dd]\ar[rr]& & \bullet\ar[ld]  \\
& \circ\ar[ul]\ar[dr]& \\
\bullet\ar[ur]&&  \bullet\ar[uu]\ar[ll]}\\ \mathrm{V}\end{gathered}
\end{gather*}
\caption{\label{ddd1} Quiver block replacements in massive flavor surgery.}
\end{figure}
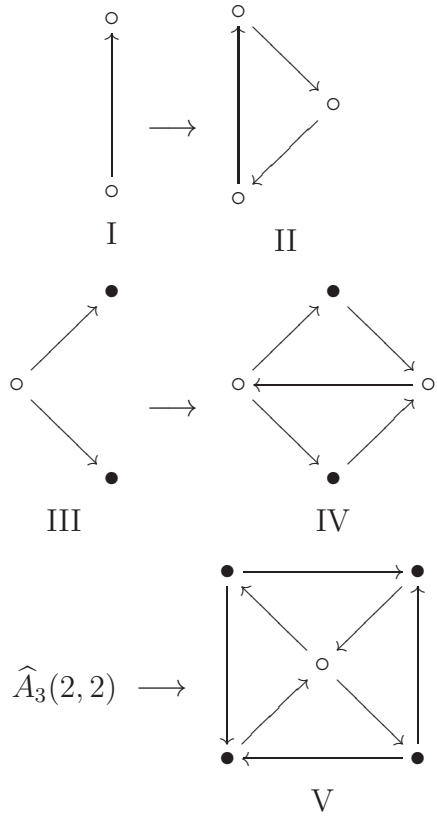

The rule to glue together $Q_1$, $Q_2$ `in the massive flavor way'  is just to replace, in the block decomposition of each $Q_i$,  the block associated to the boundary $\gamma_i$ with a block having one more white node $\circ$ according to figure \ref{ddd1}.

Finally, we identify the white nodes $\circ$ added to the two quivers $Q_i$ getting a connected quiver $Q$ with rank
$D(Q)=D(Q_1)+D(Q_2)+1$. The extra node produced by the process is the massive flavor charge of the coupling hypermultiplet.

\subsubsection{Examples} 

To simplify the figures, we represent double arrows as single arrows with a 2 in a box.

\textbf{1}. The $g=2$ $n=1$ quiver
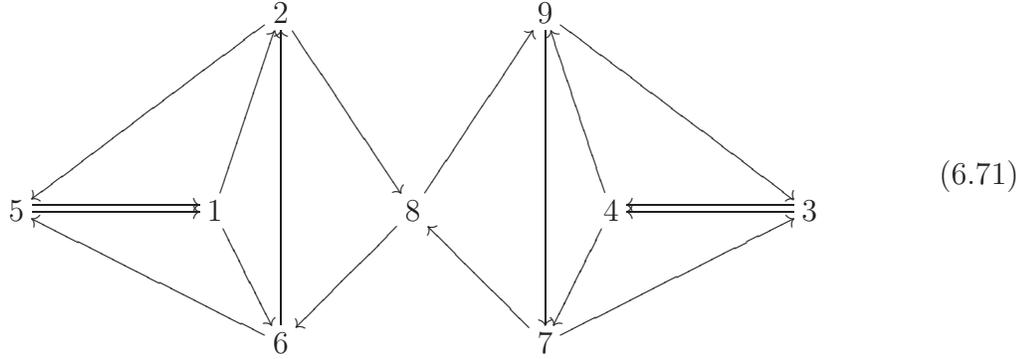
\begin{equation}\begin{gathered}
\begin{xy} 0;<1pt,0pt>:<0pt,-1pt>:: 
(75,75) *+{1} ="0",
(100,0) *+{2} ="1",
(300,75) *+{3} ="2",
(225,75) *+{4} ="3",
(0,75) *+{5} ="4",
(100,125) *+{6} ="5",
(200,125) *+{7} ="6",
(150,75) *+{8} ="7",
(200,0) *+{9} ="8",
"0", {\ar"1"},
"4", {\ar@<-0.05ex>"0"},
"4", {\ar@<0.45ex>"0"},
"0", {\ar"5"},
"1", {\ar"4"},
"5", {\ar"1"},
"1", {\ar"7"},
"2", {\ar@<-0.45ex>"3"},
"2", {\ar@<0.05ex> "3"},
"6", {\ar"2"},
"8", {\ar"2"},
"3", {\ar"6"},
"3", {\ar"8"},
"5", {\ar"4"},
"7", {\ar"5"},
"6", {\ar"7"},
"8", {\ar"6"},
"7", {\ar"8"},
\end{xy}\end{gathered}\label{firrrr}
\end{equation} 
has a separating node, namely $8$. Erasing this and the associated arrows, we get two disconnected copies of the quiver associated with a torus with a boundary having a marked point, eqn.\eqref{remquiver}.
\smallskip

\textbf{2.} The $g=2$, $n=2$ quiver
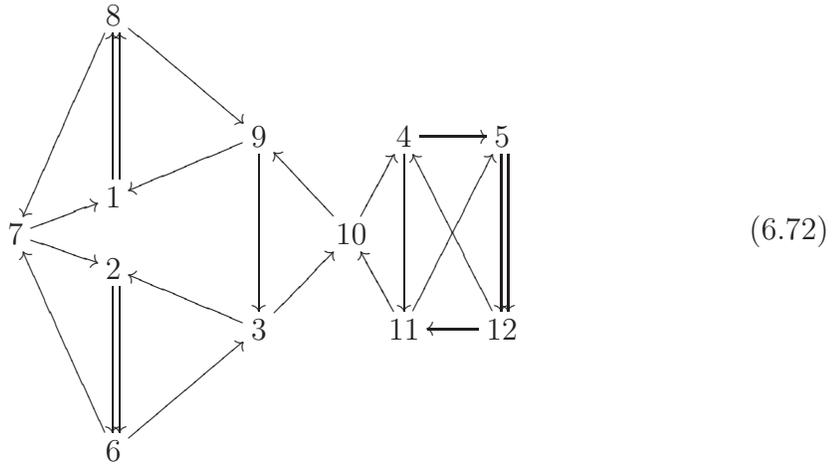
\begin{equation}\begin{gathered}
\begin{xy} 0;<1pt,0pt>:<0pt,-1pt>:: 
(37,69) *+{1} ="0",
(37,96) *+{2} ="1",
(92,119) *+{3} ="2",
(147,46) *+{4} ="3",
(184,46) *+{5} ="4",
(37,165) *+{6} ="5",
(0,83) *+{7} ="6",
(37,0) *+{8} ="7",
(92,46) *+{9} ="8",
(127,83) *+{10} ="9",
(147,119) *+{11} ="10",
(184,119) *+{12} ="11",
"6", {\ar"0"},
"0", {\ar@<-0.45ex>"7"},
"0", {\ar@<0.05ex>"7"},
"8", {\ar"0"},
"2", {\ar"1"},
"1", {\ar@<0.45ex>"5"},
"1", {\ar@<-0.05ex>"5"},
"6", {\ar"1"},
"5", {\ar"2"},
"8", {\ar"2"},
"2", {\ar"9"},
"3", {\ar"4"},
"9", {\ar"3"},
"3", {\ar"10"},
"11", {\ar"3"},
"10", {\ar"4"},
"4", {\ar@<-0.05ex>"11"},
"4", {\ar@<0.45ex>"11"},
"5", {\ar"6"},
"7", {\ar"6"},
"7", {\ar"8"},
"9", {\ar"8"},
"10", {\ar"9"},
"11", {\ar"10"},
\end{xy}\end{gathered}\label{g2n2}
\end{equation} 
has a separating node, $10$. Deleting it and its arrows we get on the right the quiver of a un--punctured torus having one boundary with a marked point, and on the left the quiver of a once--punctured torus with a boundary with a marked point.

\subsubsection{Gauge surgery: the tube case}

Assume we have a quiver with a Kronecher sub--quiver attached to two oriented triangles as in the figure
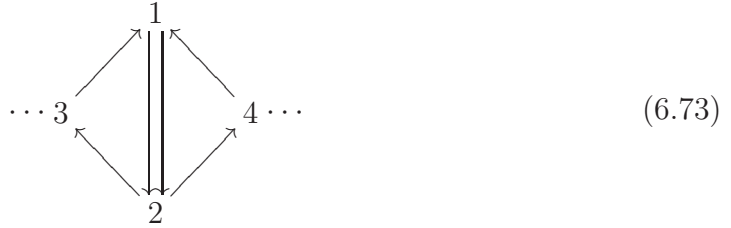
\begin{equation}\label{figuresym2}
\begin{gathered}\xymatrix{&& 1 \ar@<-0.5ex>[dd]\ar@<0.5ex>[dd] & &\\
\cdots\!\!\!\!\!\!\!\!\!\!\!\!\!\!\!\!\!\!\!\! & 3 \ar[ur] & & 4 \ar[ul] &\!\!\!\!\!\!\!\!\!\! \!\!\!\!\!\!\!\!\!\!\cdots \\
& & 2\ar[ul]\ar[ur] &&}\end{gathered}
\end{equation}
where the ellipsis $\cdots$ means that the nodes $3$, $4$ are attached to the rest of the quiver by any number of arrows consistent with the quiver being of the triangulation type. In practice, this means that the nodes $3$, $4$ should be identified with a white node of some block of the rest of the quiver.

Figure \eqref{figuresym2} corresponds to one of the three ways a Kronecker subquiver may appear in a finite--mutation quiver (see \S.\,\ref{sec:mutfinvsSU(2)}), and is physically interpreted as
 an $SU(2)$ SYM  gauging the $SU(2)$ symmetries of the $\cn=2$ systems represented by the subquivers $\cdots\, 3$  and $4\,\cdots$.
 
As we shall see momentarily,   from the triangulation viewpoint the subquiver \eqref{figuresym2} represents a tube region of the surface $\cc_{g,n,b,c}$. Of course, this is nothing else than Gaiotto's descriptions of $SU(2)$ gauge groups as plumbing tubes \cite{gaiotto}.
 Then we can borrow his analysis of the relation between the (complexified) $SU(2)$ coupling $\tau$ and the plumbing parameter $q=e^{2\pi i\tau}$.
 The weak coupling  limit then corresponds to a tube in the Riemann surface $\cc_{g,n,b,c}$ which becomes infinitely long. In the limit $q= e^{2\pi i\tau}\rightarrow 0$, the tube pinches, and we remain either with two disconnected surfaces, $\cc_{g_1,n_1,b_1}$ and $\cc_{g_2,n_2, b_2}$, where
\begin{equation}
g_1+g_2=g,\qquad n_1+n_2=n+2,\qquad b_1+b_2=b,
\end{equation}  
or with a connected surface $\cc_{g^\prime,n^\prime, b^\prime}$ with
\begin{equation}
 g^\prime=g-1,\qquad n^\prime=n+2,\qquad b^\prime=b.
\end{equation}
In either cases, the total number of nodes in the (possibly disconnected) quiver is conserved.

By the very concept of complete $\cn=2$, the decoupled $q\rightarrow 0$ theories should be also complete, and their quivers of finite--mutation type. Thus the coupling/decoupling process may be expressed in the quiver--theoretical language. 
\smallskip

\begin{figure}
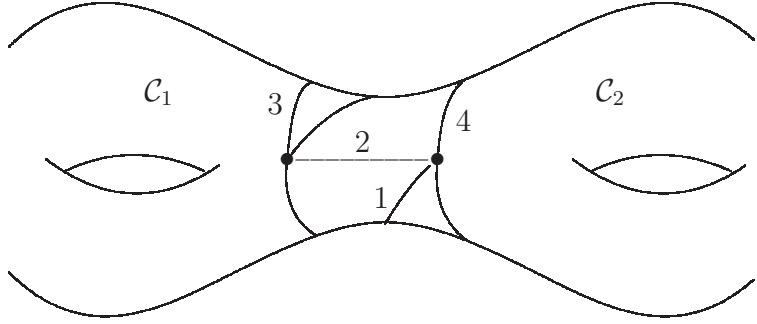

\begin{equation*}
\xy {(20,20)*+{}; (120,20)*+{} **\crv{(30,10)&(50,20)&(70,30)&(90,20)&(110,10)}\POS?*^+!L{}};
{(20,50)*+{}; (120,50)*+{} **\crv{(30,60)&(50,50)&(70,40)&(90,50)&(110,60)}\POS?*^+!L{}};
{(25,35)*+{}; (49,35)*+{} **\crv{(37,26)}\POS?*^+!L{}};
{(95,35)*+{}; (119,35)*+{} **\crv{(107,26)}\POS?*^+!L{}};
{(27.5,33.5)*+{}; (47,33)*+{} **\crv{(37,38)}\POS?*^+!L{}};
{(97.5,33.5)*+{}; (117,33)*+{} **\crv{(107,38)}\POS?*^+!L{}};
{(57,35)*+{}; (62,24.3)*+{} **\crv{(56,27)}\POS?*^+!L{}};
{(57,35)*+{\bullet}; (77,35)*+{\bullet} **\crv{}\POS?(0.5)*^+!D{2}};
{(57,35)*+{}; (62,44.5)*+{} **\crv{(58,48)}\POS?(0.38)*^+!R{3\,}};
{(77,35)*+{}; (82,46.3)*+{} **\crv{(77.5,45)}\POS?(0.3)*^+!L{\,4}};
{(40,44)*+{\cc_1}}; {(100,44)*+{\cc_2}};
{(77,35)*+{}; (82,23.5)*+{} **\crv{(76,27)}\POS?*^+!L{}};
{(57,35)*+{}; (70,43.6)*+{} **\crv{(63,43)}\POS?*^+!R{}};
{(70,26.5)*+{}; (77,35)*+{} **\crv{(74,33)}\POS?(0.3)*^+!R{1\,}};
\endxy\end{equation*}
\caption{\label{tubesurgery} The punctures and arcs corresponding to the subquiver \eqref{figuresym2}.}
\end{figure}

In the triangulation of $\cc_{g,n,b,c}$, the sub--quiver \eqref{figuresym2} corresponds to a tube or, more precisely, to a cylinder $C_{1,1}$ with a marked point on each boundary which is glued through its two boundary arcs --- corresponding to nodes $3$ and $4$ in \eqref{figuresym2} --- to the rest of the surface $\cc_{g,n,b,c}$ in such a way that the two markings on the boundaries of $C_{1,1}$ correspond to two (ordinary) punctures of the surface $\cc_{g,n,b,c}$.

The cylinder with a marking on each boundary, $C_{1,1}$, is precisely the surface corresponding to pure $SU(2)$ $\cn=2$ super--Yang--Mills.
We represent the cylinder $C_{1,1}$ as a rectangle with the two vertical sides identified. Then an ideal triangulation looks like
\begin{equation}\label{tube}
\begin{gathered}\xymatrix{\bullet \ar@{-}[rr]^3 && \bullet\\
\bullet\ar@{-}[u]^2\ar@{-}[rru]^1\ar@{-}[rr]_4 && \bullet\ar@{-}[u]_2 }
\end{gathered}\end{equation}
or, equivalently figure \ref{tubesurgery}, where the arcs are numbered as the nodes in the subquiver \eqref{figuresym2}.

To do the surgery, we cut away the cylinder $C_{1,1}$ along the two separating arcs $3$ and $4$. This operation produces two boundaries each with a marked point $\bullet$. Next we glue to each of these two boundaries a self--folded triangle along its external arc $ext$
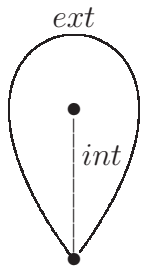
\begin{equation*}
\xy {(20,0)*+{\bullet}; (20,20)*+{\bullet} **\crv{}\POS?(0.7)*^+!L{int}};
{(20,0)*+{}; (20,0)*+{} **\crv{(5,20)&(20,40)&(35,20)}\POS?*_+!D{ext}}
\endxy
\end{equation*}
which introduces the extra puncture replacing the pinched tube.

The net result of gluing the self--folded triangle, is replacing the block attaching the node $3$ (resp.\! $4$) to the rest of the quiver in the following way\vglue 9pt

\begin{equation}\label{gaugesurgeryta}\text{
 \begin{tabular}{c|c}\hline
  type original block & type replacing block (*)\\\hline
  I & III\\\hline
II & IV\\\hline
III & $\widehat{A}_3(2,2)$\\\hline
IV & V\\\hline
V & excep. trian. $4$--punct. sphere\\\hline
 \end{tabular}}
\end{equation}
 \begin{quote} \textsc{Table:} gauge tube surgery.
 
 (*) Attaching blocks of type III and V are possible only for $\cc_1$ equal to the twice--punctured $1$--gon and, respectively, the $4$--punctured sphere (with its exceptional triangulation).\end{quote}\vglue 9pt
 
In the last step each of the two adjacency quivers of $\cc_1$, $\cc_2$ gets an extra node (associated to the internal arc $int$ of the glued self--folded triangle); since in the process we have lost the two nodes
associated to arcs $1$ and $2$ in figure \eqref{tube}, the total number of nodes is conserved, as expected.
\smallskip

The inverse process \textit{(gluing)} is also easy. One takes two surfaces, $\cc_{g_1,n_1,b_1,c_1}$ and $\cc_{g_2,n_2,b_2,c_2}$, triangulated in such a way that the corresponding quivers have one of the blocks in the second column of table \eqref{gaugesurgeryta}. These blocks correspond to a `puzzle piece' of the triangulation containing a self--folded triangle. Then one cuts away the self--folded triangles from the corresponding `puzzle pieces' of the two triangulated surfaces, producing a boundary  with one marked point on each surface $\cc_{g_1,n_1,b_1,c_1}$, $\cc_{g_2,n_2,b_2,c_2}$. Finally one glues these boundaries to the boundaries of the cylinder \eqref{tube} identifying the marked points.
\medskip

The above `tube' surgery is only a special instance of the coupling of two $\cn=2$ theories by replacing a pair of punctures by a thin tube. It works under the special assumption that both surfaces to be glued are triangulated in such a way that a self--folded triangle exists (in particular, each surface must have at least either \emph{two} punctures or a puncture and a boundary). There are more general ways of gluing quivers, which make sense under weaker assumptions on the two surfaces to be glued.
We may glue, for instance, the quivers of a higher genus surface with one puncture to that of a surface with two punctures. However, it is not possible to relax this milder condition. The point is that, otherwise, we could
get the quiver of a puncture--less surface by gluing two once--punctured lower genus ones. But this is clearly impossible.

\subsubsection{Example: generalized hypermultiplet gaugings}

Assume that the $SU(2)$ SYM associated to the tube to be pinched is coupled to the other sectors by two generalized `hypermultiplets'. At the quiver level, this means that we have a full subquiver of the form
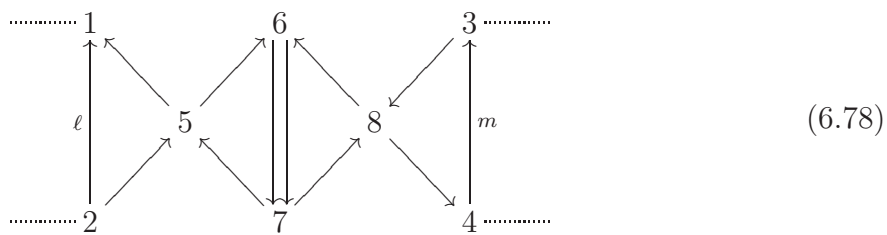
\begin{equation}\label{figuresym}
\begin{gathered}\xymatrix{ & 1\ar@{..}[l] && 6 \ar@<-0.5ex>[dd]\ar@<0.5ex>[dd] & &3\ar[dl]\ar@{..}[r]&\\
& & 5 \ar[ur]\ar[ul] & & 8 \ar[ul]\ar[dr] &&&\\
& 2\ar@{..}[l] \ar[uu]^\ell \ar[ur] & & 7\ar[ul]\ar[ur] &&4\ar[uu]_m \ar@{..}[r]&}\end{gathered}
\end{equation}
where the $\xymatrix{\ar@{..}[r]&}$ stands for any number of arrows connecting the four nodes 1,2,3,4 of the subquiver to the nodes of the rest of the quiver, while the nodes $2$ and $1$ (resp.\! $4$ and $3$) are connected by $\ell$ arrows (resp.\! $m$ arrows). 

The triangles $1,2,5$ and $3,4,8$ correspond to blocks of type II. Decoupling the $SU(2)$, they get replaced by  type IV blocks
(cfr.\! table \eqref{gaugesurgeryta}). Then, as $\tau\rightarrow 0$ we get
\begin{equation}\label{finalquiv}
\xymatrix{\cdots\!\!\!\!\!\!\!\!\!\!\!\!\!\!\!\!\!\!\!\!& 1 && 5a\ar[ll] && & 8a \ar[ddrr] & &3\ar[ll]\ar[lldd]&\!\!\!\!\!\!\!\!\!\!\!\!\!\!\!\!\!\cdots\\
& &  &  & &  &&&&\\
\cdots\!\!\!\!\!\!\!\!\!\!\!\!\!\!\!\!\!& 2\ar[uu]^\ell \ar[rr]\ar[rruu]  && 5b\ar[uull] && & 8b\ar[rr]&&4\ar[ll]\ar[uu]_{m} &\!\!\!\!\!\!\!\!\!\!\!\!\!\!\!\!\!\cdots}
\end{equation}
(the full quiver may or may not be disconnected).

If $\ell=m=2$, corresponding to an ordinary bi--fundamental hypermultiplet, we break the tube by replacing a gauge group and two bi--fundamentals by two pairs of fundamental hypermultiplets coupled to the two $SU(2)$'s  associated to the pairs of nodes $1,2$ and $3,4$, respectively.

\subsubsection{Examples: gauging $\cn=2$ subsystems}\label{exoticgaugings}

From the above we see that we can couple the  $SU(2)$ gauge system to any Gaiotto $\cn=2$ system whose surface $\cc$ has at least one ordinary puncture (subject to the condition that the glued surface has at least one puncture --- if we wish to
have a theory with a well--defined quiver).  Such a system admits an $SU(2)$ global symmetry which can be gauged\footnote{We thank D. Gaiotto for a discussion on this point.}.

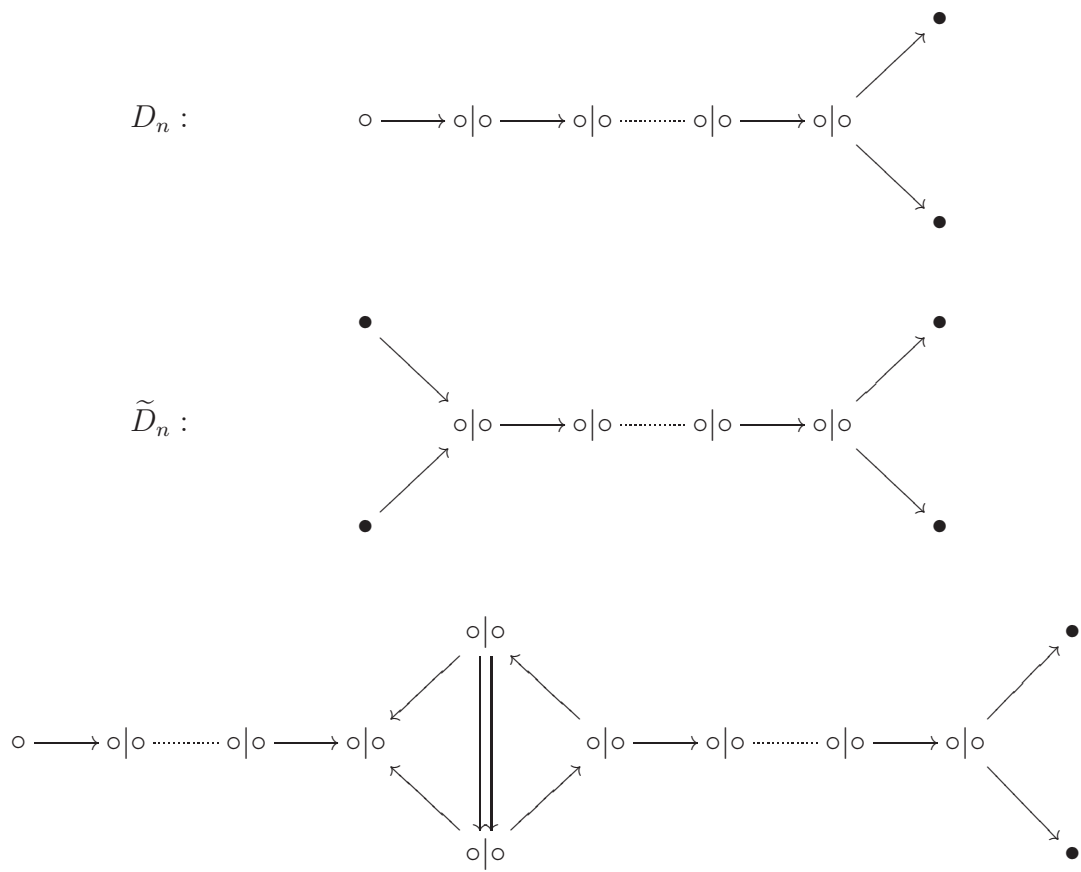
\begin{figure}
\begin{align*}
&D_n: &&\begin{gathered}
\xymatrix{&&&&& \bullet\\
\circ\ar[r] & \circ|\circ \ar[r]& \circ|\circ\ar@{..}[r]&\circ|\circ\ar[r]&\circ|\circ\ar[ru]\ar[rd]&\\
&&&&& \bullet}\end{gathered}\end{align*}
\begin{align*}&\widetilde{D}_n: &&\begin{gathered}
\xymatrix{\bullet\ar[rd]&&&&& \bullet\\
&\circ|\circ\ar[r] & \circ|\circ\ar@{..}[r]&\circ|\circ\ar[r]&\circ|\circ\ar[ru]\ar[rd]&\\
\bullet\ar[ur]&&&&& \bullet}\end{gathered}\end{align*}
\begin{align*}
\begin{gathered}
\xymatrix{&&&&\circ|\circ\ar[dl]\ar@<0.4ex>[dd]\ar@<-0.4ex>[dd]&&&&& \bullet\\
\circ\ar[r]&\circ|\circ\ar@{..}[r]&\circ|\circ\ar[r]&\circ|\circ&& \circ|\circ\ar[lu] \ar[r]& \circ|\circ\ar@{..}[r]&\circ|\circ\ar[r]&\circ|\circ\ar[ru]\ar[rd]&\\
&&&&\circ|\circ\ar[ru]\ar[lu]&&&&& \bullet}\end{gathered}\end{align*}
\caption{\label{quivIIIB}The $D_n$, $\widehat{D}_n$ and $\Gamma_{n,m}$ quivers decomposed into blocks: the last block on the right is of type III.  The blocks are divided by the vertical line $|$; the two $\circ$'s separated by a vertical line should be identified to get back the original quiver.}
\end{figure}

The more elementary such surfaces $\cc$ are:
\begin{itemize}
\item the punctured disk with $m$ marked points on the boundary whose adjacency quiver is (up to mutation equivalence) the Dynkin quiver $D_m$;
\item the twice--punctured  disk with $m$ marking on the boundary
corresponding to the affine $\widehat{D}_{m+2}$ quivers, mutation equivalent to \eqref{Dnaff};
\item the punctured annulus with $(n,m)$ marking on the boundaries, last quiver in figure \ref{quivIIIB}.  
\end{itemize}

In their standard (Dynkin) form, the corresponding quivers contain one type III block (two for $\widehat{D}_{m+2}$, associated to the two ordinary punctures), as can be seen from the block decompositions in figure \ref{quivIIIB}
.

By the rules of the gauge tube surgery, we may replace that type III block by a type I, and couple the `new' white node to a Kronecker subquiver via an oriented triangle. We describe this process as a `gauging' of the system described by the original surface $\cc$.

Graphically, the gauging procedure looks as follows
\begin{equation}\label{Dgaugingf}
\begin{gathered}
\xymatrix{&& \bullet\\
&\circ \ar@{..}[l]\ar[ur]\ar[dr]&\\
&& \bullet }
\end{gathered}
\  \boldsymbol{\Rightarrow}\ 
\begin{gathered}
\xymatrix{&&\\
&\circ \ar@{..}[l]\ar[r]&\circ\\
&& }
\end{gathered}\ \boldsymbol{\Rightarrow}\ 
\begin{gathered}
\xymatrix{&& & \circ\ar@<-0.4ex>[dd]\ar@<0.4ex>[dd] &\ar@{..}[l]\\
&\circ \ar@{..}[l]\ar[r]&\circ\ar[ur]& &\\
&& &\circ \ar[ul] &\ar@{..}[l]}
\end{gathered}
\end{equation} 

There is a field theory explanation of the above surgery. The idea is that each block of type III in an adjacency quiver $Q$ carries a global $SU(2)$ symmetry, and the surgery is just gauging it. Indeed, in the presence of a type III block we have a special flavor charge\footnote{\ Recall that a flavor charge is a vector in $\Gamma$ which is a zero eigenvector of the exchange matrix $B$.} $J$ with weights $+1$ and $-1$ for the two black nodes of the type III block and zero elsewhere.
The quiver (and hence the physics) is symmetric under the simultaneous interchange of the two black nodes and the corresponding mass parameters. This $\Z_2$  symmetry acts on the above charge as $J\rightarrow -J$, so the natural interpretation is that $J$ is the Cartan generator of $\mathfrak{su}(2)$ and $\Z_2$ its Weyl group.

We can check this interpretation in a special case. From figure \eqref{Dgaugingf} we see that the gauging of an ordinary fundamental hypermultiplet corresponds to the gauging of the $D_2\sim A_1\times A_1$ Argyres--Douglas system: a fundamental hypermultiplet is \emph{two} free hypermultiplets each with its own $SU(2)$ flavor charge.  In other words
we can consider the subquiver consisting of the two end nodes of the quiver, which corresponds to two decoupled
hypermultiplets, which can be gauged by the $SU(2)$.  In this way the BPS quiver keeps only one
of the two fundamentals (as discussed in the context of BPS quivers of $SU(2)$ coupled to one fundamental),
as the other one can be obtained by the combination of elements of $SU(2)$'s Kronecker quiver, and one of the two
fundamental states.  This explains why effectively we get rid of one of the two end nodes of the quiver diagram
and connect the remaining node to the Kronecker quiver in the standard way.

\medskip

A preliminary discussion of the physical properties of these gauged $\cn=2$ systems are presented in section \ref{exoticpro}.

\subsection{Vector--less quivers}\label{sec:vectorless}

In this section we show why the only `vector--less' quivers are the $ADE$ Dynkin ones.  By this we mean
that this is the only class which does not have any double arrows in any mutation of the corresponding quiver.
For the eleven exceptional classes the fact that there are double lines in the quiver follows from direct inspection. It remains to consider the adjacency quiver of bordered surfaces.

The example in \S.\ref{examplegL1npunct} shows that all surfaces with $g\geq 1$ and at least one puncture have a triangulation with at least one double--arrow. On  the other hand, suppose  we have a surface with $g\geq 1$ and $b\geq 1$. We may cut open the surface to get a hyperbolic $4g$--gon and start triangulating as in the figure
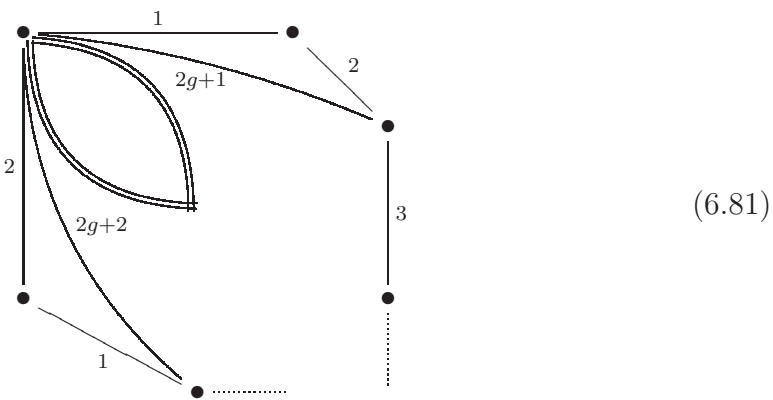
\begin{gather}
\begin{gathered} \xymatrix{\bullet \ar@{=}@<-0.7ex>@/^2pc/[ddrr]
 \ar@{=}@<0.7ex>@/_2pc/[ddrr]\ar@{-}[rrr]^1 \ar@{-}@/^0.5pc/[drrrr]_{2g+1}\ar@{-}@/_1.5pc/[ddddrr]^{2g+2} & &  &\bullet\ar@{-}[dr]^2 &\\
 & & & &\bullet\ar@{-}[dd]^3\\
 && & &\\
\bullet\ar@{-}[uuu]^2 &&& &\bullet\ar@{..}[d]\\
 &&\bullet\ar@{-}[llu]^1\ar@{..}[r] &&  }
\end{gathered}\end{gather}
which gives $B_{12}=+2$. Hence all $g\geq 1$ triangulation quivers are mutation--equivalent to ones having at least one double--arrow.

For $g=0$, all surfaces with $n\geq 4$ or $b\geq 2$ have quivers in the mutation--class with double arrows.
Taking into account the restrictions on $n$, $b$, $c$ for $g=0$ \cite{fomin}, we remain with the possibility $b=1$. 
If $b=1$ and $n= 2$ we have affine--$\widehat{D}$ quivers which are mutation--equivalent to those in figure  
\eqref{Dnaff} having a Kronecker subquiver.

We remain with surfaces with $b=1$, $n=0$, corresponding to the mutation class of the $A_r$ Dynkin quivers, and
$b=1$, $n=1$, associated to the mutation class of the $D_r$ Dynkin ones. These finite Dynkin quivers are known to be vector--free.

\section{Identification of the exceptional theories}\label{identificationexceptional}

It remains to identify the complete $\cn=2$ theories associated to the eleven exceptional mutation classes which are mutation--finite but not associated to the ideal triangulation of any surface. They may be divided in four families (we write a standard representative for each mutation--class):
\begin{enumerate}
  \item finite--type Dynkin quivers of type $E_6, E_7, E_8$;
\item affine--type Dynkin quivers of type $\widehat{E}_6, \widehat{E}_7, \widehat{E}_8$;
\item Saito's \cite{saito} elliptic--type Dynkin quiver (with oriented triangles) of type $\widehat{\widehat{E}_7},
\widehat{\widehat{E}_7}, \widehat{\widehat{E}_8}$;
\item the Derksen--Owen quivers $X_6$ and $X_7$ \cite{derksen}. \end{enumerate}

The models associated to the first family, $E_6, E_7, E_8$, were already discussed in \cite{cnv}.
They are a generalization of the Argyres--Douglas model corresponding to the world--sheet theory of a M5--brane compactified to four dimension on a complex curve with equation the corresponding  $E$--type minimal singularity
\begin{equation}\begin{array}{c|c}\
E_6 & y^3+x^4=0\\\hline
E_7 & y^3+ y\, x^3=0\\\hline
E_8 & y^3+x^5=0
\end{array}\end{equation}

They are UV conformal, and vector--less.

\subsection{Elliptic and affine $E$--models}
\begin{figure}
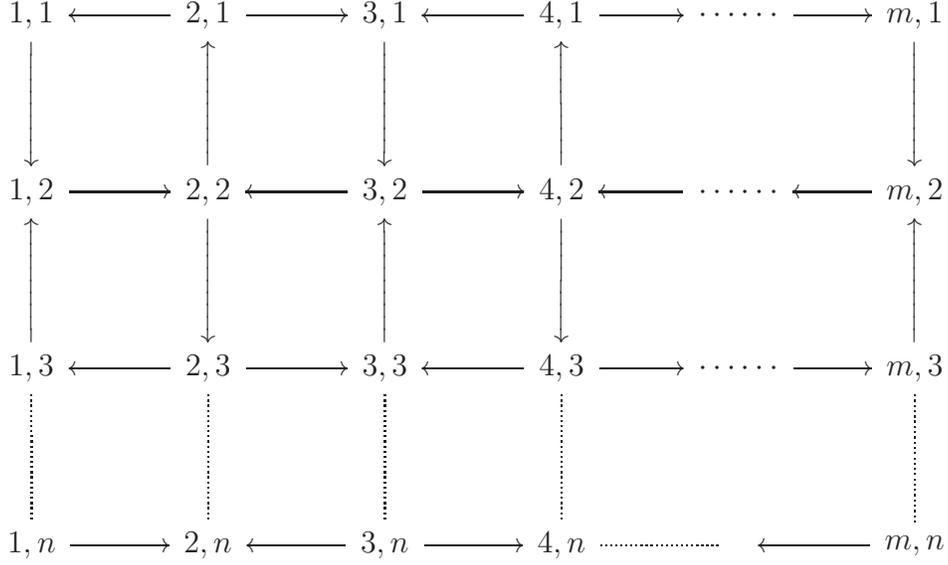

 \begin{equation*}
 \begin{diagram}
  \node{1,1}\arrow{s}\node{2,1}\arrow{w}\arrow{e}\node{3,1}\arrow{s} \node{4,1}\arrow{w} \arrow{e} \node{\cdots\cdots}\arrow{e}\node{m,1}\arrow{s}\\
 \node{1,2} \arrow{e} \node{2,2}\arrow{n}\arrow{s} \node{3,2}\arrow{w}\arrow{e} \node{4,2}\arrow{n}\arrow{s}
  \node{\cdots\cdots }\arrow{w}\node{m,2}\arrow{w}\\ 
 \node{1,3}\arrow{n} \node{2,3}\arrow{w}\arrow{e}\node{3,3}\arrow{n} \node{4,3}\arrow{w}\arrow{e} \node{\cdots\cdots}\arrow{e}\node{m,3}\arrow{n}\\
 \node{1,n}\arrow{e}\arrow{n,..,-} \node{2,n}\arrow{n,..,-} \node{3,n}\arrow{n,..,-}\arrow{w}\arrow{e} \node{4,n}\arrow{n,..,-} \arrow{e,..,-} \node{\ }\node{m,n}\arrow{w}\arrow{n,..,-}
 \end{diagram} 
 \end{equation*} 
\caption{\label{squaremnquiver} The $A_m\,\square\,A_n$ quiver.}
\end{figure}

The elliptic $E$--models turn out to be special instances of the class of models studied in \cite{cnv} which are labelled by a pair $(G,G^\prime)$ of simply--laced Dynkin graphs ($G,G^\prime=ADE$). They correspond to the $4d$ $\cn=2$ theory obtained by
compactifying Type IIB superstring on the local Calabi--Yau hypersurface $\mathscr{H}\subset \C^4$ of equation
\begin{equation}
\mathscr{H}\colon\qquad W_G(x_1,x_2)+W_{G^\prime}(x_3,x_4)=0,
\end{equation}
where $W_G(x_1, x_2)+x_0^2$ is the canonical surface singularity associated to the given Dynkin diagram $G$.
The quiver of the $(G, G^\prime)$ model is given by the \emph{square tensor product} of the Dynkin graphs of $G$ and $G^\prime$, $G\,\square\, G^\prime$ (for the product orientation rule see refs.\cite{keller-periodicity, cnv}). The quiver $A_m\,\square\,A_n$ is represented in figure \ref{squaremnquiver}.

\smallskip

Up to mutation--equivalence one has the following identifications \cite{fomin}:
\begin{gather}
\label{simEhh1}\widehat{\widehat{E}_6} \sim A_2\,\square\, D_4\\ 
\widehat{\widehat{E}_7} \sim A_3\,\square\, A_3\\
\label{simEhh2}\widehat{\widehat{E}_8} \sim A_2\,\square\, A_5.
\end{gather}
The first one may be further simplified using $D_4\sim A_2\,\square\, A_2$ \cite{cnv}. Hence the corresponding $4d$ $\cn=2$ models may be engineered by Type IIB on the hypersurface $\mathscr{H}$:

\begin{equation}\label{doublehatsing}\begin{array}{c|c|c}\
\text{quiver} & \text{CY hypersurface } \mathscr{H} & (n_1,n_2,n_3)\\\hline 
\widehat{\widehat{E}_6}\phantom{\Bigg|} & x_0^2+x_1^3+x_2^3+x_3^3+ax_1x_2x_3=0 & (2,2,2)\\\hline
\widehat{\widehat{E}_7}\phantom{\Bigg|} & x_0^2 + x_1^4+x_2^4+x_3^2+ax_1x_2x_3=0 & (3,3,1)\\\hline
\widehat{\widehat{E}_8}\phantom{\Bigg|} & x_0^2+x_1^3+x_2^6+x_3^2+ax_1x_2x_3=0 & (2,4,1)
\end{array}\end{equation}
Notice that the section $x_0=0$ of each hypersurface is a quasi--homogeneous cone over an elliptic curve embedded in some weighted projective space. Indeed the Saito's elliptic root systems are related to elliptic singularities.
The only other elliptic Dynkin diagram which is a finite--mutation quiver is $\widehat{\widehat{D}_4}$ which corresponds to $SU(4)$ with $N_f=4$ (\textit{i.e.}\! the sphere with four punctures).
\smallskip

In a $\widehat{\widehat{E}}_r$ mutation class there are many quivers having a transparent physical interpretation. First of all, we have the tensor product quivers
$G\,\square\,G^\prime$, $G^\prime\,\square\,G$,  $G^\prime\,\boxtimes\,G$, and
$G\,\boxtimes\,G^\prime$, which using the results of ref.\cite{cnv} imply that the model is UV conformal with a quantum monodromy $\mathbb{M}(q)$ of order\footnote{\ Eqn.\eqref{orderMM} holds for the groups in eqns.\eqref{simEhh1}--\eqref{simEhh2} but not in general. For the general case see \cite{ceclectures}.}
\begin{equation}\label{orderMM}
 r=\frac{h(G)+h(G^\prime)}{\gcd\{h(G),h(G^\prime)\}} =\begin{cases}
            2 & \text{for }\widehat{\widehat{E}}_7\\
3 & \text{for }\widehat{\widehat{E}}_6,\, \widehat{\widehat{E}}_8, 
                                                      \end{cases}
\end{equation}
which means, in particular, that the the UV $U(1)_R$ charges $r_i$ of the primary operators are of the form $\frac{1}{r}\,\mathbb{N}$. Moreover, the $(G,G^\prime)$ $\cn=2$ model has two special chambers with a \emph{finite} BPS spectrum consisting only of hypermultiplets. In the first such chamber they have charges \cite{cnv}
\begin{equation}
\alpha_i \otimes \sum_a n_a^{(s)}\, \beta_a \in \Gamma_G\otimes \Gamma_{G^\prime}\simeq \Gamma_{G\,\square\, G^\prime},
\end{equation}
where $\alpha_i\in \Gamma_G$ are the \emph{simple} roots of $G$ and $\sum_a n_a^{(s)}\,\beta_a\in \Gamma_{G^\prime}$ are all the positive roots. In the second chamber the two Dynkin diagrams interchange roles $G\leftrightarrow G^\prime$. On the other hand, the $\widehat{\widehat{E}}_r$ quivers are not vector--less and hence have regimes described by mutation--equivalent quivers containing Kronecker subquivers; indeed the usual elliptic Dynkin forms have one Kronecker sub--quiver, see figure  \ref{ellipticEs}, and they correspond to pure $SU(2)$ coupled to three $\cn=2$ $D$--systems of the kind discussed in sections \ref{sec:mutfinvsSU(2)} and \ref{exoticgaugings}.  

The family of coupled three  $\cn=2$ $D$--systems has quivers of
the suggestive form
\begin{multline}
Q(n_1,n_2,n_3)=\\ =\begin{gathered}\xymatrix{ & & & \bullet\ar@<0.4ex>[dd]\ar@<-0.4ex>[dd] & b_1\ar[l]\ar@{..}[r] &b_{n_2-1} & b_{n_2}\ar[l]  \\
a_{n_1}\ar[r] & a_{n_1-1}\ar@{..}[r] & a_{1} \ar[ru] && & & &\\
& & &  \bullet \ar[ul]\ar[r]\ar[ruu] & c_1\ar[uul]\ar@{..}[r] &c_{n_3-1} & c_{n_3}\ar[l]}
\end{gathered}\label{Qn1n2n3quiv}
\end{multline}
(notice that the quiver is symmetric under the interchanging of the nodes with $a$, $b$ and $c$ labels.)
$Q(1,1,1)\simeq \widehat{D}_4$ is just the quiver of $SU(2)$ with three flavors.
\smallskip

The $\widehat{\widehat{E}}_r$ $\cn=2$ models are engineered by Type IIB on the CY hypersurface
$x_0^2+W_{n_1,n_2,n_3}(x_1,x_2,x_3)=0$, where $W_{n_1,n_2,n_3}(x_1,x_2,x_3)$ is the equation of the elliptic curve in weighted projective space
\begin{equation}\label{Wn1n2n3}
W_{n_1,n_2,n_3}(x_1,x_2,x_3)\equiv  x_1^{n_1+1}+x_2^{n_2+1}+x_3^{n_3+1}+\lambda\, x_1\,x_2\,x_3
\end{equation}
and the integers $(n_1,n_2,n_3)$ are as in the table \eqref{doublehatsing}.
The corresponding quiver is simply $Q(n_1,n_2,n_3)$ for the same triplet of integers. Following our discussion in section \ref{sec:mutfinvsSU(2)}, we expect that these models have BPS chambers, different from the two finite--spectrum ones analyzed in ref.\cite{cnv}, with BPS vector multiplets in the spectrum weakly coupled to the supersymmetric $D$--systems.

This completes the identification for the elliptic--$E$ $\cn=2$ models as the models obtained by compactifying Type IIB on the corresponding CY hypersurface, see table \eqref{doublehatsing}. 
\medskip

More generally, we may ask for which triplet of integers $(n_1,n_2,n_3)$ --- besides the ones in table \eqref{doublehatsing} --- the quiver $Q(n_1,n_2,n_3)$ is of the finite--mutation type.
Not surprisingly, the condition turns out to be
\begin{equation}\label{coxrefl}
 \frac{1}{n_1+1}+\frac{1}{n_2+1}+\frac{1}{n_3+1}\geq 1,
\end{equation}
in one--to--one correspondence with Coxeter reflection groups for the sphere and the plane. The $\cn=2$ theories for which the inequality $\geq$ in eqn.\eqref{coxrefl} is replaced by equality  $=$ are actually UV superconformal (see next section). 

The solutions to condition \eqref{coxrefl} are listed in table \ref{tablen1n2n3}.
\begin{table}
\begin{equation}
\text{\begin{tabular}{c|c|c}\hline
$n_1,n_2,n_3$ &  equivalent Dynkin quiver & \\\hline
$1,1,s$ &  $\widehat{D}_{s+3}$ & disk with $n=2$, $c=s+1$\\
$1,2,2$ & $\widehat{E}_6$ & asymptotically free \\
$1,2,3$ & $\widehat{E}_7$ & asymptotically free\\
$1,2,4$ & $\widehat{E}_8$&  asymptotically free\\
 $2,2,2$ & $\widehat{\widehat{E}}_6$& superconformal\\
 $1,3,3$ & $\widehat{\widehat{E}}_7$ &superconformal\\
 $1, 2,5$ & $\widehat{\widehat{E}}_8$ &superconformal \\\hline
\end{tabular}}
\end{equation}
\caption{\label{tablen1n2n3} The solutions $(n_1,n_2,n_3)$ to condition \eqref{coxrefl} and the Dynkin quiver mutation--equivalent to the quiver $Q(n_1,n_2,n_3)$. }
\end{table}
\smallskip

From the table we infer an interpretation of the affine--$\widehat{E}$ quivers. They are precisely the asymptotically free, complete $\cn=2$ models associated to Type IIB on the (UV fixed point of the) hypersurface
\begin{equation}\label{hyperCCYY}
 x_0^2+x_1^{n_1+1}+x_2^{n_2+1}+x_3^{n_3+1}+\lambda\, x_1\,x_2\,x_3=0
\end{equation}
where $n_1,n_2,n_3$ are as specified in the table \ref{tablen1n2n3}.

Table \ref{tablen1n2n3} gives us also an alternative construction of affine-$\widehat{D}$ models in terms of Type IIB enineering. \smallskip

As a further check of the identifications for the affine $\widehat{D}_r$, $\widehat{E}_r$ models in  table
\ref{tablen1n2n3}, let us consider it from the point of view of the $4d/2d$ correspondence. 
The above identifications gives the $2d$ Landau--Ginzburg model with superpotential $W_{n_1,n_2,n_3}(x_1,x_2,x_3)$
in eqn.
\eqref{Wn1n2n3}. The $\widehat{D}_r$, $\widehat{E}_r$ affine Dynkin diagrams correspond to the triplets of integers
$(n_1,n_2,n_3)$ with
\begin{equation}\label{affinetriplet}
 \frac{1}{n_1+1}+\frac{1}{n_2+1}+\frac{1}{n_3+1} >1.
\end{equation}
 The identification requires the Witten index of the two dimensional model to be equal to the rank of the corresponding affine Lie algebra, \textit{i.e.}\! $r+1$. A direct computation shows that, under the condition  
\eqref{affinetriplet}, one has
\begin{equation}
 \text{$2d$ Witten index }=\ n_1+n_2+n_3+2 =\begin{Bmatrix} s+4 & \text{for } \widehat{D}_{s+3}\\
7 & \text{for } \widehat{E}_6\\
8 & \text{for } \widehat{E}_7\\
9 & \text{for } \widehat{E}_8
\end{Bmatrix}\equiv r+1.
\end{equation}
This result supplements the classification of $2d$  $\cn=2$ affine models \cite{Cecotti:1993rm}.

\subsection{The Derksen--Owen quivers $X_7$, $X_6$}

There remain only two mutation--finite classes: $X_7$ and $X_6$.

\subsubsection{$X_7$}

The mutation class of $X_7$ consists of just two distinct quivers \cite{derksen}. The one with double--arrows is

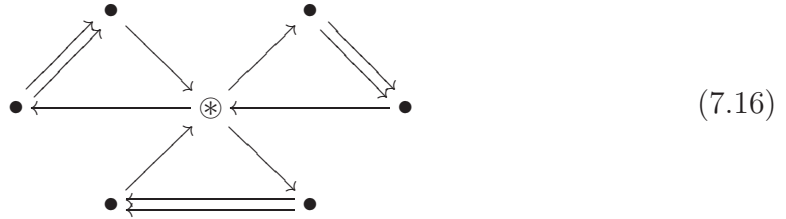
\begin{equation}\label{fig:X7}
\begin{gathered} \xymatrix{& \bullet\ar[rd] & & \bullet\ar@<0.4ex>[dr]\ar@<-0.4ex>[dr] &\\
\bullet\ar@<0.4ex>[ur]\ar@<-0.4ex>[ur]&&\circledast\ar[ll]\ar[ur]\ar[rd] && \bullet\ar[ll]\\
& \bullet\ar[ru] && \bullet \ar@<0.4ex>[ll]\ar@<-0.4ex>[ll] &}\end{gathered}
\end{equation}\vglue 6pt

The quiver \eqref{fig:X7} is maximal finite--mutation (\textbf{Theorem 13} of \cite{derksen}), and hence it is expected to correspond to an UV conformal $\cn=2$ theories (this prediction will be confirmed momentarily). \smallskip

$X_7$ has one flavor charge, associated to the node in \eqref{fig:X7} represented by the symbol $\circledast$. The corresponding vector in the charge lattice is
\begin{equation}\label{flavchX7}
 \text{flavor charge vector}= \gamma_\circledast+\frac{1}{2}\sum \gamma_\bullet.
\end{equation}

The physical interpretation of this quiver is straightforward. Associated to the above flavor charge we have a mass parameter $m$. Taking $m\rightarrow\infty$, we approach a limit where a weakly coupled Lagrangian description is adequate: We have a \emph{full} hypermultiplet
in the quaternionic (pseudoreal) representation
\begin{equation}
 (\mathbf{2},\mathbf{2},\mathbf{2})_{+1}\oplus (\mathbf{2},\mathbf{2},\mathbf{2})_{-1}
\end{equation}
of its symmetry group $SU(2)\times SU(2)\times SU(2)\times SO(2)$ and the three $SU(2)$'s are weakly gauged by three copies of $SU(2)$ SYM represented by the three Kronecker subquivers, $\xymatrix{\bullet \ar@<0.4ex>[r]\ar@<-0.4ex>[r]&\bullet}$, in figure \eqref{fig:X7}. Its unique flavor charge \eqref{flavchX7} corresponds to the $SO(2)$ symmetry of the hypermultiplet with mass parameter $m$.

Taking $m\rightarrow 0$, this model reduces to the conformal Gaiotto model with $g=2$ and \emph{no} puncture.
Indeed, in some corner of its moduli space, the genus two curve with no punctures may be physically interpreted  as in 
the figure
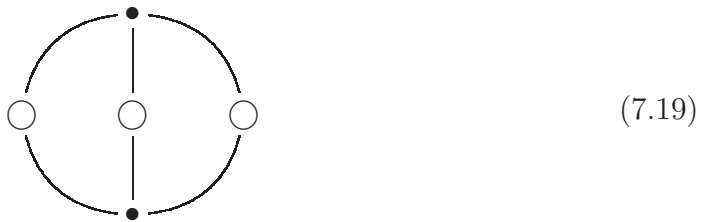
\begin{equation}\label{X7g2}
\begin{gathered}\xymatrix{ &\bullet \ar@{-}[d]& \\
\bigcirc\ar@{-}@/^1pc/[ur] &\bigcirc\ar@{-}[d] &\bigcirc \ar@{-}@/_1pc/[ul]\\
& \bullet\ar@{-}@/_1pc/[ur] \ar@{-}@/^1pc/[ul]&}\end{gathered}
\end{equation}
where the $\bigcirc$'s stand for $SU(2)$ gauge groups and the $\bullet$'s for tri--fundamental \emph{half}--hypermultiplets. The two half--hypermultiplets have the same quantum numbers with respect to all gauge groups, and so we may combine them into a \emph{complete} hypermultiplet in the $(\mathbf{2},\mathbf{2},\mathbf{2})$ of the $SU(2)^3$ gauge group. This process introduces --- in the above Lagrangian corner of the moduli space --- an emergent $SO(2)$ symmetry --- not present in the original Gaiotto construction --- which is the one associated to the node $\circledast$ of the $X_7$ quiver. In particular, the relation with Gaiotto's $g=2$ theory shows that the $X_7$ $\cn=2$ is UV conformal, as expected from the graphical rule.
\medskip

\begin{figure}
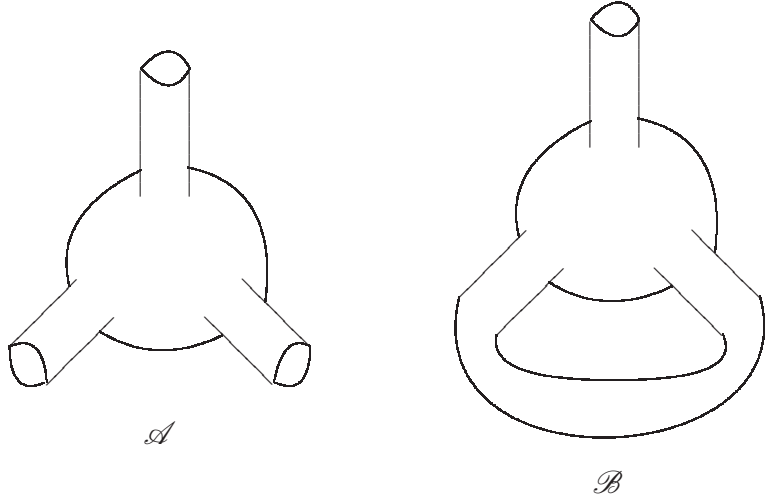

\begin{equation*}\begin{gathered}
\xy {(5,10)*+{};(15,20)*+{} **\crv{}\POS?*^+L!{}};
{(10,5)*+{};(20,15)*+{} **\crv{}\POS?*^+L!{}};
{(17,12)*+{};(34.3,12)*+{} **\crv{(25,7)}\POS?*^+L!{}};
{(5,10)*+{};(10,4.25)*+{} **\crv{(9.78,12.1)}\POS?*^+L!{}};
{(5,10)*+{};(10.67,5.76)*+{} **\crv{(5.2,2.4)}\POS?*^+L!{}};
{(45,10)*+{};(35,20)*+{} **\crv{}\POS?*^+L!{}};
{(13,18)*+{};(23.6,34)*+{} **\crv{(10.4,28)(12,33.5)(17,33.8)}\POS?*^+L!{}};
{(40,5)*+{};(30,15)*+{} **\crv{}\POS?*^+L!{}};
{(39,16)*+{};(27.67,34)*+{} **\crv{(41,32)}\POS?*^+L!{}};
{(45,10)*+{};(40,4.25)*+{} **\crv{(41.22,12.1)}\POS?*^+L!{}};
{(45,10)*+{};(39.33,5.76)*+{} **\crv{(44.8,2.4)}\POS?*^+L!{}};
{(22.5,31)*+{};(22.5,47)*+{} **\crv{}\POS?*^+L!{}};
{(29,31)*+{};(29,47)*+{} **\crv{}\POS?*^+L!{}};
{(22.5,47)*+{}; (29.5,45.9)*+{} **\crv{(27,52)}\POS?*^+L!{}};
{(22.5,47)*+{}; (29.5,48.1)*+{} **\crv{(27,42)}\POS?*^+L!{}};
\endxy\\
\mathscr{A}
\end{gathered}
\qquad\qquad
\begin{gathered}
{\xy  {(5,40)*+{};(15,50)*+{} **\crv{}\POS?*^+L!{}};
{(10,35)*+{};(20,45)*+{} **\crv{}\POS?*^+L!{}};
{(17,42)*+{};(34.3,42)*+{} **\crv{(25,37)}\POS?*^+L!{}};
%
{(45,40)*+{};(35,50)*+{} **\crv{}\POS?*^+L!{}};
{(13,48)*+{};(23.6,64)*+{} **\crv{(10.4,58)(12,33.5)(17,63.8)}\POS?*^+L!{}};
{(40,35)*+{};(30,45)*+{} **\crv{}\POS?*^+L!{}};
{(39,46)*+{};(27.67,64)*+{} **\crv{(41,62)}\POS?*^+L!{}};
%
{(22.5,61)*+{};(22.5,77)*+{} **\crv{}\POS?*^+L!{}};
{(29,61)*+{};(29,77)*+{} **\crv{}\POS?*^+L!{}};
{(22.5,77)*+{}; (29.5,75.9)*+{} **\crv{(27,82)}\POS?*^+L!{}};
{(22.5,77)*+{}; (29.5,78.1)*+{} **\crv{(27,72)}\POS?*^+L!{}};
{(5,40)*+{};(44.65,41.3)*+{} **\crv{(2,27)&(25,15)&(49,28.5)}\POS?*^+L!{}};
{(10,35)*+{};(39.45,36.2)*+{} **\crv{(9,30)& (25,28) & (43.6,30)}\POS?*^+L!{}};
\endxy}
\\
\mathscr{B}
\end{gathered}
\end{equation*}
\caption{\label{g==2}$\mathscr{A}$: A trifundamental half--multiplet corresponds to a thrice--punctured sphere attached to three long plumbing tubes. $\mathscr{B}$: Two of the three punctures may be connected by a long tube making a handle.}
\end{figure}

This emergence of a flavor symmetry is special to $g=2$, and does not generalizes to $g>2$. This explains why $X_7$ is an isolated exception without higher rank analogues. Indeed, in the Gaiotto framework \cite{gaiotto}, the
degeneration of a genus $g>1$ surface without punctures into three--punctured spheres connected by long cyclinders corresponds to a Lagrangian description in which each punctured sphere corresponds to a trifundamental \emph{half}--hypermultiplet  in the representation $(\mathbf{2}, \mathbf{2}, \mathbf{2})$ of $SU(3)^3$, which  has \emph{no} flavor symmetry (see figure \ref{g==2}.\,$\mathscr{A}$), while each long cylinder corresponds to a weakly coupled $SU(2)$ SYM. In order to have a flavor symmetry, we need at least two such half--hypermultiplets in the same representation of the gauge group. This may happen only if the three punctures of the sphere representing the second half--hypermultiplet are connected to the same three tubes as the sphere representing the first one. Then the two punctured spheres and the three tubes connecting them form a $g=2$ surface disconnected from the rest.
The only other possibility is that two punctures of the same sphere are connected together to form a handle (as in figure \ref{g==2}$.\,\mathscr{B}$). This also leads to $g=2$ (see next section).      

\smallskip

From the figure \eqref{X7g2} it is obvious that the model is UV conformal: Indeed, each $SU(2)$ `sees' four fundamental hypermutliplets, and hence has a vanishing $\beta$--function.

\subsubsection{$X_6$}

The $X_6$ exceptional mutation class consists of five distinct quivers \cite{derksen}. Two of them have double arrows (they are source/sink equivalent, and hence represent essentially the same physics), 
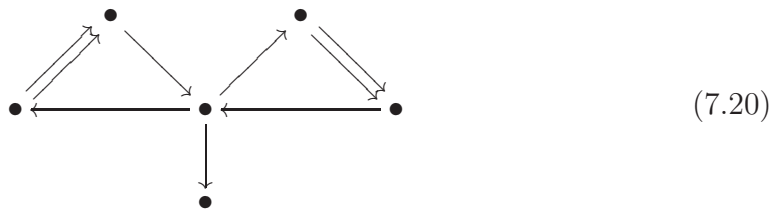
\begin{equation}\label{fig:x6}
\begin{gathered} \xymatrix{& \bullet\ar[rd] && \bullet\ar@<0.4ex>[dr]\ar@<-0.4ex>[dr]  &\\
\bullet \ar@<0.4ex>[ur]\ar@<-0.4ex>[ur] && \bullet \ar[ll]\ar[ur]\ar[d] && \bullet\ar[ll]\\
&& \bullet &&}\end{gathered}
\end{equation}
 
The $X_6$ quiver has no flavor charge. $X_6$ is not maximal mutation--finite, but the only mutation--finite quiver containing it is $X_7$ itself (\textbf{Theorem 12} of \cite{derksen}). Hence the corresponding $\cn=2$ theory must be UV asymptotically free, and must arises as a particular decoupling limit of the $X_7$ $\cn=2$ model.   In fact, as already discussed
any subsystem of a quiver can be viewed as arising in a particular limit of moduli space of that theory.  Thus $X_6$ which
is a subquiver of $X_7$ obtained by deleting one of the nodes of a double line can be obtained from a limit of $X_7$
theory.
\smallskip

That the $X_7$ theory has such a limit may be understood more explicitly.
By the very concept of complete $\cn=2$ theory, the $X_7$ model has enough quantum consistent deformations that we may actually realize as sensible QFT all of its formal geometric limits. In particular, in the deformation space of $X_7$ there should be contained all relevant/marginal deformations of any theory related by Gaiotto dualities to the $g=2$ conformal theory which is the $m\rightarrow 0$ limit of $X_7$. 

Between the Gaiotto dual theories, we have the one corresponding to the degeneration of the $g=2$ surface shown in the right hand side of the figure
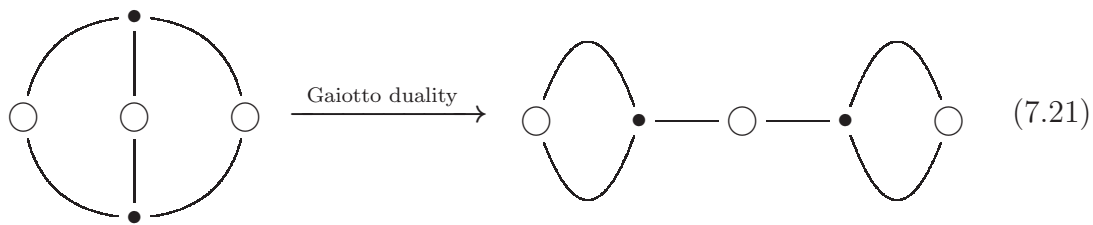
\begin{equation}
\begin{gathered}\xymatrix{ &\bullet \ar@{-}[d]& \\
\bigcirc\ar@{-}@/^1pc/[ur] &\bigcirc\ar@{-}[d] &\bigcirc \ar@{-}@/_1pc/[ul]\\
& \bullet\ar@{-}@/_1pc/[ur] \ar@{-}@/^1pc/[ul]&}\end{gathered}\ \boldsymbol{\xrightarrow{\ \text{Gaiotto duality }\ }}
\ \begin{gathered}\xymatrix{& &&&\\
\bigcirc\ar@{-}@/^2.5pc/[r]\ar@{-}@/_2.5pc/[r] & \bullet\ar@{-}[r]&\bigcirc \ar@{-}[r]&\bullet \ar@{-}@/^2.5pc/[r]\ar@{-}@/_2.5pc/[r]&\bigcirc\\
& &&& }\end{gathered}
\end{equation} 
where, again, $\bigcirc$'s stand for $SU(2)$ gauge groups and $\bullet$'s for \emph{half}--hypermultiplets $h^{(a)}_{\alpha\dot\alpha \ddot\alpha}$ in the
$(\mathbf{2},\mathbf{2}, \mathbf{2})$ of $SU(2)^3$.

In the second limit, the same $SU(2)$ SYM gauges the first pair $\alpha, \dot\alpha$ of $SU(2)$ indices, so that the matter representation content in terms of the three gauge groups is
\begin{equation}
\Big((\mathbf{3},\mathbf{1},\mathbf{2})\oplus (\mathbf{1},\mathbf{1},\mathbf{2})\Big)\oplus \Big((\mathbf{1},\mathbf{3},\mathbf{2})\oplus (\mathbf{1},\mathbf{1},\mathbf{2})\Big).
\end{equation}
In this duality limit, we have two half--hypermultiplets with the same quantum numbers under all gauged symmetries, namely $(\mathbf{1},\mathbf{1},\mathbf{2})$, and hence an $SO(2)$ flavor symmetry rotating them. To this $SO(2)$ symmetry we may associate a mass defomation, $\mu$. Since the $X_7$ theory is complete, this deformation should correspond to a region in its coupling space. 

At this point we take the decoupling limit $\mu\rightarrow \infty$.
We get an $\cn=2$ theory with a charge lattice of rank $6$, no flavor charge,
which is asymptotically free.  Assuming there is a BPS quiver for this theory, it should be mutation--finite and contained in $X_7$. 
There is only one such quiver, namely $X_6$.
\smallskip

\section{Conformal, complete theories}\label{sec:conformal}

\subsection{$U(1)_R$ symmetry}

The  $\cn=2$ theories corresponding to mutation--finite quivers, being UV complete QFT, in the ultra--violet are either conformal or asymptotically free. In the first case there is a point in their parameter space (belonging to some specific chamber and hence corresponding to a particular quiver in the given mutation--class) in which the full superconformal invariance is restored.

In this section we address the question of classifying the subset of complete $\cn=2$ theories which have such a superconformal point. In $4d$,  a necessary condition for $\cn=2$ superconformal invariance is the existence of a conserved $U(1)_R$ current. More precisely, the $U(1)$ associated to the overall phase of the Seiberg--Witten differential $\lambda$ should become a symmetry at the conformal point.

For a generalized Gaiotto model, this $U(1)$ acts on the quadratic differential as 
\begin{equation}
\phi_2\rightarrow e^{2i\theta}\, \phi_2.   
\end{equation}
Hence, for this class of models, we have a conserved $U(1)_R$ symmetry iff there exists a complex automorphism of the surface $\cc$, $f_\theta\colon\cc\rightarrow\cc$, such that
\begin{equation}\label{confcondition}
f_\theta^*\phi_2\Big|_{\text{conformal}\atop \text{point}}=
e^{2i\theta}\,\phi_2\Big|_{\text{conformal}\atop \text{point}}.\end{equation}

For the kind of punctured bordered surfaces of interest here, we have a continuous group of automorphisms only if $\cc$ is a sphere with one or two punctures, where we may have either ordinary double poles or higher ones.   Except for these special cases, \eqref{confcondition} may be satisfied only by setting
\begin{equation}
\phi_2\Big|_{\text{conformal}\atop \text{point}}=0.
\end{equation}
Moreover, this should be achieved by finite deformation of the theory (otherwise, we would simply
have an asymptotically free theory, which is conformal at infinite distance in Coulomb branch).  For poles
higher than order 2, there will always be some Coulomb branch vevs which correspond to residues of the poles,
and using the metric\footnote{This follows from special geometry and the fact
that the Kahler function is given, in the standard notation, by $\sum_i (a^i {a^i_D}^*-c.c.) $.} $\int| \delta \lambda_{SW}|^2$ we find that this leads to infinite distance, where $\lambda_{SW}$
denotes the Seiberg-Witten differential $ydx$.  The regular poles can be set to zero by setting the corresponding
mass to zero.   Thus, the only superconformal $\cn=2$ theories associated to surfaces with $g>0$ or $g=0$ with at least three punctures (ordinary or otherwise) are the regular Gaiotto ones without higher order poles.

The sphere with a single puncture is a well--defined $\cn=2$ quiver theory only if we have a pole of order  $p\geq 6$ --- corresponding to a disk with $(p-2)$ marked points \textit{i.e.}\! a $(p-2)$--gon. This corresponds to the $A_{p-5}$ Argyres--Douglas models which are known to have a superconformal point. 

Likewise, the sphere with an ordinary puncture and one pole of order $p$ is associated to $D_p$ Argyres--Douglas theory which also has a superconformal point.

Instead, the sphere with two higher order poles is associated to an annulus with marked points on both boundaries. This theory is just asymptotically free: special instances are $SU(2)$ gauge theory coupled to $N_f=0,1,2$ fundamental flavors. The fact that they are not superconformal is particularly evident from the $4d/2d$ perspective: they correspond to the $2d$ models
\begin{equation}
W(X)= e^{nX}+ e^{-mX},
\end{equation} 
which has no continuous symmetry since the approximate $U(1)_R$ symmetries around the north and south poles do not agree in the intermediate region. In the language of ref.\cite{Cecotti:1993rm}, this corresponds to a unipotent non--semisimple $2d$ monodromy.

It remains to discuss the 11 exceptional models. The models associated to the ordinary $E_6,E_7,E_8$ Dynkin quivers are a kind of exceptional Argyres--Douglas theories, already studied in \cite{cnv}. They are known to have a conformal point. 

The $\cn=2$ theories associated to affine and elliptic $E$--type Dynkin quivers are best studied by the Type IIB geometrical engineering described in section \ref{identificationexceptional}. Then the conformal $U(1)_R$ should arise from a $U(1)$ symmetry of the local Calabi--Yau hypersurface which acts on the holomorphic $3$--form $\Omega$ as
$\Omega\rightarrow e^{i\theta}\,\Omega$. In this way we see that the affine $\widehat{E}$--models have no conformal point, and thus are UV asymptotically free. This was to be expected, given that the affine $\widehat{A}$-- and affine $\widehat{D}$--models are UV asymptotically free, and affine $\widehat{A}\widehat{D}\widehat{E}$ models form a family with uniform properties.

The elliptic $\widehat{\widehat{E}}$--models, instead, have a conformal regime which was  studied in detail in ref.\cite{cnv} and reviewed in \S.\,\ref{identificationexceptional}. Notice that the only other elliptic Dynkin diagram which gives a mutation--finite quiver, namely $\widehat{\widehat{D}}_4$, corresponds to $SU(2)$ with $N_f=4$, and it is also UV superconformal.

Finally $X_7$ has a conformal limit, corresponding to $m\rightarrow 0$, as we may check from its Lagrangian formulation. In this limit the theory coincides with the $g=2$ Gaiotto model, so --- as a conformal theory --- it is already in the surface list, and we don't get a new model.
$X_6$ is not UV conformal.\smallskip

In conclusion, the full list of complete $\cn=2$ theories which have a UV superconformal limit are
\begin{itemize}
\item Gaiotto theories;
\item $ADE$ Argyres--Douglas theories;
\item elliptic $\widehat{\widehat{E}}_6, \widehat{\widehat{E}}_7,\widehat{\widehat{E}}_8$ theories;
\item $X_7$.
\end{itemize}

\subsection{Proof of the graphical rule}

Finally, we wish to show that the above list coincides with the set of all normalized mutation--finite quivers which are either vector--less or maximal.

The rule holds for the 11 exceptional classes by inspection: affine $\widehat{E}_r$ and $X_6$ are neither maximal nor vector--free, and are non--conformal; the others are either vector--free, $E_r$, or maximal, $\widehat{\widehat{E}}_r$, $X_7$, and are conformal.

Then to prove the graphical rule it is enough to show that a \emph{normalized} (non--exceptional) mutation--finite quiver which is maximal is the triangulation of a surface without boundaries (that is with only ordinary punctures).

Indeed, if a surface $\cc$ has a boundary component $S^1$, we may glue to it another surface $\cc^\prime$ with an $S^1$ boundary, and hence $\cc$ is not maximal. More precisely, at the level of block decomposition of the adjacency quiver, the $S^1$ boundary component  corresponds to one of the following three possibilities\footnote{\ In case of a boundary with many marked points, we have typically many of the following quiver blocks, and hence many possible extensions of the quiver which keep it mutation--finite.}: \textit{i)} a free unpaired white
node $\circ$; \textit{ii)} a block of type II; \textit{iii)} a block of type III. To normalize the quiver, we replace the blocks of type III with 
 a type II and a type I with arrows pointing in opposite directions, so case \textit{iii)} is eliminated by the normalization assumption.

In case \textit{i)} we may glue another block at the unpaired $\circ$ node and the quiver is not maximal. In case \textit{ii)} we replace the block II by a block III oriented in the same way, and the quiver is not maximal. 

On the other hand, a surface without boundaries (corresponding to a Hitchin system with only regular singularities) has an adjacency quiver composed by blocks of type II, IV and V with all the white nodes $\circ$ paired up. There is no possibility to attach extra nodes while getting a graph which is still an adjacency quiver. Finally, we have to check that no adjacency quiver of a surface with no--boundary is a subquiver of an exceptional one. This is true by inspection.

\section{Physical properties of gauging $\cn=2$ $D$--sub-systems}\label{exoticpro}

In this paper we have found compelling evidence that many complete $\cn=2$ systems are best understood as a number of $SU(2)$ gauge sectors coupled to some $\cn=2$ systems with $SU(2)$ symmetry. In this section we discuss some physical properties of the \emph{very simplest} examples of such  $\cn=2$ systems, consisting of gauging $\cn=2$ $D$--subsystems.\smallskip

\subsection{$\beta$--functions of $D$--systems}

We first focus our attention on the $\cn=2$ theories associated to the affine quivers $\widehat{A}(m,n)$ with\footnote{\ The affine quiver $\widehat{A}(m,0)$ is mutation equivalent to the finite Dynkin quiver $D_m$.} $m,n\geq 1$,
$\widehat{D}_{n-1}$ and $\widehat{E}_r$. We have seen that they are mutation equivalent, respectively, to
figure \ref{PPPiii}, eqn.\eqref{Dnaff}, and eqn.\eqref{Qn1n2n3quiv} with $(n_1,n_2,n_3)$ as in table \ref{tablen1n2n3}.
They are naturally interpreted as $SU(2)$ coupled to
\begin{itemize}
 \item one  $D_{m+1}$--system for $\widehat{A}(m+1,1)$;
\item one $D_{m+1}$--system and one $D_{m^\prime+1}$--system for $\widehat{A}(m+1,m^\prime+1)$;
\item two fundamental hypermultiplets and one $D_{m+1}$--system for $\widehat{D}_{m+3}$;
\item one fundamental hypermultiplet and two $2$ $D_3$--systems for $\widehat{E}_6$;
\item one fundamental hypermultiplet, a $D_3$--system, and a $D_4$--system for $\widehat{E}_7$; 
\item one fundamental hypermultiplet, a $D_3$--system, and a $D_5$--system for $\widehat{E}_8$;
\end{itemize}
Note that as discussed in \S.\,\ref{exoticgaugings}, a $D_{m+1}$ system couples to an $SU(2)$ Kronecker
quiver by the attachment of the subquiver
\begin{equation}
\begin{gathered}\overbrace{\xymatrix{\circledcirc\ar@{-}[r]&\bullet\ar@{-}[r]&\bullet\ar@{-}[r] &\bullet\ar@{..}[r]&\bullet\ar@{-}[r]& \bullet}}^{m\ \text{nodes}}\end{gathered}
\end{equation}
(the orientation being irrelevant) having a special node, $\circledcirc$, where we attach the oriented triangle coupling the subquiver to the Kronecker one. For $m=1$, we get back the usual hypermultiplet; to get more elegant formulae, it is convenient to extend the definition to $m=0$, corresponding to the empty $\cn=2$ system.

As we saw in the previous section, all affine complete $\cn=2$ theories are asymptotically free. Hence the $\beta$--function of the $SU(2)$ has to be negative. Comparing with the above list, we get that
\emph{the contribution to the $SU(2)$ $\beta$--function from the coupling to an $D_{m+1}$ $\cn=2$ theory is less than twice the contribution from a fundamental hypermultiplet.}

To get a precise formula for the $\beta$--function contribution of a $D_{m+1}$ system we have to look at the 
\emph{elliptic} complete $\cn=2$ models: $\widehat{\widehat{D}}_4$,
$\widehat{\widehat{E}}_6$, $\widehat{\widehat{E}}_7$, $\widehat{\widehat{E}}_8$, which may also be described as $SU(2)$ coupled to $D_{m+1}$--system (see figure \ref{ellipticEs} on page \pageref{ellipticEs}).
These theories are UV superconformal,
and hence have a vanishing $\beta$--function. These results are reproduced by taking the
$\beta$--function of the $D_{m+1}$ system to be
\begin{equation}\label{exoticbeta}
  2\left(1-\frac{1}{m+1}\right)
\end{equation}
times that of a fundamental hypermultiplet. Note that this formula gives the correct result for $m=0$ and $m=1$,
and it is always less than $2$, as required.

Eqn.\eqref{exoticbeta} has a simple heuristic interpretation in terms of the string world--sheet theory.
$SU(2)$ coupled to three $D_{m+1}$--system, is engineered by Type IIB on the hypersurface 
\eqref{hyperCCYY}, and the world--sheet theory is the Landau--Ginzburg model with the \textsc{rhs} of
\eqref{hyperCCYY} as superpotential \emph{with Liouville superfield dependent couplings} (in order to get
$2d$ superconformal invariance) \cite{Ooguri:1996wj}. The world--sheet Liouville couplings reflect the $4d$ $\beta$--function.
These couplings, and hence the $\beta$--function, are proportional to $(\hat c-1)$. In particular
\begin{equation}
\lambda\, X_1 X_2 X_3 \rightarrow \lambda_0\, e^{(1-\hat c)\phi}\, X_1X_2X_3
\end{equation}
$\lambda$ being the coupling which, in the conformal case, encodes the modulus of the torus $\tau$.
 Let $b$ the coefficient of the $SU(2)$ $\beta$--function (normalized so that the contribution of a fundamental hypermultiplet is $+1$); then
\begin{equation}
b=-4+2\sum_{i=1}^3 \left(1-\frac{1}{m_i+1}\right) \equiv \sum_{i=1}^3\left(1-\frac{2}{m_i+1}\right)-1 =\hat c-1, 
\end{equation}
and so eqn.\eqref{exoticbeta} is suggestive of another manifestation of the general $4d/2d$ correspondence.
\medskip

\subsection{$\widehat{D}$--systems and new $\cn=2$ dualities}\label{sec:newdualities}

Similar arguments may be applied to other basic $\cn=2$ systems which are conveniently used as building blocks of more complex theories. \textit{E.g.}\! the $\widehat{D_{m+1}}$ theory has
an $SU(2)\times SU(2)$ symmetry, that can be gauged, corresponding the two double ends.
As discussed in  \S.\,\ref{exoticgaugings} this leads to attaching the subquiver
\begin{equation}\label{biexotic}
\begin{gathered}\overbrace{\xymatrix{\circledcirc\ar@{-}[r]&\bullet\ar@{-}[r]&\bullet\ar@{-}[r] &\bullet\ar@{..}[r]&\bullet\ar@{-}[r]& \bullet \ar@{-}[r]& \circledcirc}}^{m\ \text{nodes}}\end{gathered}
\end{equation}
to two Kronecker systems one on each end.
Since we may replace anyone of the type I blocks in figure \eqref{exoticgaugings} with a type II block, any quiver containing the subquiver \eqref{exoticgaugings} is not maximal, and hence corresponds to an UV asymptotically free theory. A naive analogy with the previous case would lead to the wrong conclusion that  the contribution from such an $m\geq 2$ system to the $SU(2)$ $\beta$--functions of both SYM's coupled at the nodes $\circledcirc$  is less than the one from a bi--fundamental hypermultiplet. This is \emph{not} correct: The contribution to the $\beta$--function of the gauging $SU(2)$'s is \emph{equal} to that of a bi--fundamental hypermultiplet. Nevertheless the resulting model cannot be superconformal simply because the $\widehat{D}_{m+1}$ sector is by itself asymptotically free, and the couplings which have negative $\beta$--functions are the ones inside the system described by the subquiver \eqref{biexotic}.
Indeed, we have a dual picture of this $\cn=2$ theory: Up to mutation, the quiver $\widehat{D}_{m+1}$ may be taken in the form \eqref{Dnaff} which is naturally interpreted as an $SU(2)$ SYM coupled to two fundamental hypermultiplets and one $D$-system. The $SU(2)\times SU(2)$ flavor symmetry of the $\widehat{D}_{m+1}$ system may be interpreted simply as the usual  flavor symmetry of the two fundamental hypermultiplets. So, we may think of a model where the $SU(2)\times SU(2)$ symmetry of a $\widehat{D}_{m+1}$ theory is gauged as a theory with one more gauge group, where the extra group gauges a pair of bi--fundamental half--hypermultiplets and a $D$--system.

\smallskip

A new kind of $\cn=2$ duality is obtained from the mutation--equivalence $\Gamma(n,m)\sim \Gamma(m,n)$ for the triangulation of a punctured annulus with $(n,m)$ marking on the boundaries. In term of quivers, this may be seen as an $SU(2)$ which gauges the $SU(2)$ symmetry of a $D$-system and one of the two $SU(2)$ factor subgroups of the $SU(2)\times SU(2)$ symmetry of a $\widehat{D}$--system, while the other subgroup remains as a global symmetry (corresponding to the type III block in the   
$\Gamma(n,m)$ quiver). Again we have a duality interchanging the ranks of the two  systems. This theory may be understood as an $SU(2)^2$ gauge theory  where both $SU(2)$'s gauge the same half--bifundamental, then each of them gauge a $D$--system, and one of the two $SU(2)$'s also gauges a fundamental hypermultiplet.

\medskip

The two $\circledcirc$ nodes of the subquiver \eqref{biexotic} may be gauged by two distinct $SU(2)$ SYM's, or the same SYM may gauge the diagonal $SU(2)$ subgroup of the $SU(2)\times SU(2)$ of the $\widehat{D}$--systems. In the last case we get the quiver
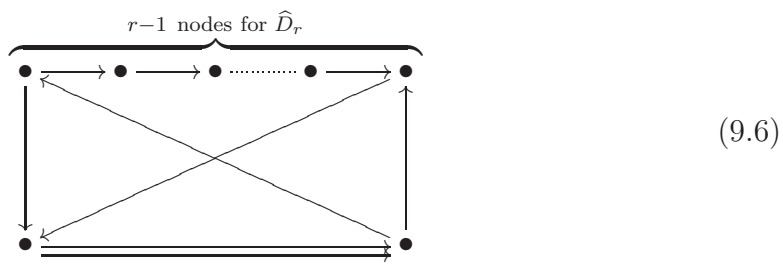
\begin{equation}
\begin{gathered}\overbrace{\xymatrix{\bullet\ar[dd] \ar[r] & \bullet \ar[r] &\bullet \ar@{..}[r] & \bullet\ar[r] & \bullet\ar[ddllll]\\
&&&&\\
\bullet\ar@<-0.8ex>[rrrr]\ar@<-0.2ex>[rrrr] &&&& \bullet\ar[uullll]\ar[uu]}}^{r-1\ \text{nodes for }\widehat{D}_r}\end{gathered}
\end{equation}
\vglue 6pt
$r=2$ gives $\cn=2^*$, $r=3$ the unique--quiver model of section
\ref{sec:uniquequiver}, and more generally, the generalized Gaiotto theory associated to a torus with a boundary having $r-2$ marks.

Note that, since $\widehat{D}_3\sim\widehat{A}_3(2,2)$ the `remarkable' theory of \S.\,\ref{sec:uniquequiver} may be interpreted as $SU(2)$ SYM gauging an $SU(2)$ subgroup of the $SU(2)\times SU(2)$ flavor symmetry of $SU(2)$ SQCD with $N_f=2$. This gives a Lagrangian formulation of the unique--quiver model of section 
\ref{sec:uniquequiver}, confirming that it is an asymptotic free theory without flavor charges.  

\subsection{BPS spectrum of $SU(2)$ SYM coupled to $D$--systems}\label{specexotics}

In this section we determine the BPS spectra of $SU(2)$ SYM coupled to one, two, or three $D_r$--systems. With the exception of the elliptic models\footnote{\ The strong coupling BPS spectrum of the elliptic models is described in \S. It is likely that they have also `weak coupling' chambers with BPS vector multiplets.} $\widehat{\widehat{E}}_r$, these theories are asymptotically free and have an affine quiver of the form $\widehat{A}(m,n)$ ($m,n\geq 1$), $\widehat{D}_r$ or $\widehat{E}_r$.
The first $\cn=2$ models in these series are just $SU(2)$ SQCD with $N_f\leq 3$.

As in section \ref{sec:reptheory} the BPS spectrum is determined by the Kac--Moody representation theory. 

We have a strong coupling BPS chamber with only hypermultiplet dyons, one for each simple root of corresponding Kac--Moody algebra with the charge vector
\begin{equation}
\alpha_i =(0,\cdots, 0,1,0,\cdots 0)
\end{equation}
in the basis of the charge lattice $\Gamma$ in which the quiver has the standard affine Dynkin graph form.

Then we have a weak coupling chamber with an infinite BPS dyon spectrum consisting of hypermultiplet of charge vector
\begin{equation}
\sum_i n_i \,\alpha_i \in \Delta^\text{re}_+
\end{equation}
and a BPS vector multiplet of charge vector equal the indivisible imaginary root
\begin{equation}
\delta= \sum_i a_i\, \alpha_i,
\end{equation}
where $a_i$ are the Dynkin weights, equal, by the McKay correspondence, to the dimensions of the irreducible representations of the corresponding finite subgroup of $SU(2)$.

\section{Conclusions}
In this paper we have seen the unexpected power of BPS quivers in classification of $\cn =2$ supersymmetric
theories in 4 dimensions.  In particular we have given evidence that all the (punctured) $SU(2)$ Gaiotto theories
do admit BPS quivers, and we have identified them.  Furthermore we have seen that complete $\cn=2$ theories,
which are defined to allow arbitrary variations of all the central charges, admit only special BPS quivers:
If they admit BPS quivers, the corresponding BPS quivers have a finite orbit under mutations.  This
allows one to classify all complete $\cn=2$ theories which admit BPS quivers.  Moreover we have
identified them as generalized $SU(2)$ Gaiotto theories, and an additional 11 classes which we have
identified.

These results are rather promising:  Not only they illustrate that the general program of classication of
$\cn=2$ theories are tractable, but that we can identify them with known theories which arise either
in gauge theory or string theory constructions.  Moreover our approach has shown the power of 
4d/2d correspondence in identifying these theories.

 There are a few directions that the results of this
paper can be extended:  The first question is to generalize this classification to other $\cn=2$ theories which are not complete.  Given that these map to 2d $\cn=2$ supersymmetric theories with $1< \hat c\leq 2$,
this may be more tractable than it may appear.  Another question would be to understand the role
that the BPS quivers play in the $\cn=2$ theories in 4d.  Which theories do admit BPS quivers?
We have in particular identified a number of complete $\cn=2$ theories that do not admit BPS quivers.
These include the rank 2 Gaiotto theories at higher genus without punctures.  Are there other complete
$\cn=2$ theories which do not admit BPS quivers?  Or the reverse question:  Why is it that
$\cn=2$ theories generically seem to admit a BPS quiver?  

Clearly we are just at the beginning of a deeper understanding of the $\cn=2$ supersymmetric
theories in 4 dimensions.  It is remarkable how much we have already learned.  At the same time, one feels that the most exciting results in this area lie ahead and are yet to be discovered.

\section*{Acknowledgements}

We thank D. Gaiotto and  B. Keller for many valuable correspondences.  We would also
like to thank C. Cordova for helpful comments on the manuscript.
SC thanks the High Energy Group of Harvard University, where part of this work was completed, for hospitality.

The research of CV was supported in part by NSF grant PHY-0244821.

\appendix

\section{Strong coupling spectra of affine quiver models}\label{app:strongcoupling}

In this appendix we show that the strongly coupled spectrum of any $\cn=2$ theory having an affine quiver without oriented cycles is given by one hypermultiplet per simple root.\smallskip

The basic point about affine quivers without oriented loops is the existence of frieze sequence \cite{kellrecrel}. In particular,  we may number the vertices from $1$ to $D$ in such a way that each vertex $i$ is a source in the full subquiver of vertices $1,\cdots, i$.
Let $\tilde\mu_k$ be the combination of the elementary quiver mutation, $\mu_k$, with the corresponding  change of basis in $\Gamma$ as defined in equations (6.2)(6.3) of \cite{cnv}  (we adopt the same conventions). Then if the product
\begin{equation}\label{seqmutation}
\tilde\mu_1\circ\tilde\mu_{2}\circ\cdots\circ \tilde\mu_D,
\end{equation}
acts on the quantum torus algebra $\mathbb{T}_\Gamma$ as the inversion $I$, then the  corresponding product of elementary quantum cluster mutations
\begin{equation}\label{Kqaffine}
\mathbb{K}(q)=\cq_1\,\cq_{2}\,\cdots\,\cq_D,
\end{equation}
is the quantum half--monodromy (the \emph{omnipop} in the language of \cite{Gaiotto:2009hg}) from which we may read the BPS spectrum\footnote{\ As well as the BPS phase order.} in the corresponding chamber (which is the strongly coupled one) \cite{cnv,ceclectures}.
\smallskip

The above identity follows from the simple observation that the vertex $i$ is a source in the mutated quiver
\begin{equation}
Q_i=\mu_{i+1}\circ\mu_{i+2}\circ\cdots\circ \mu_D(Q),
\end{equation}
so the $i$--th transformation $\tilde\mu_i$ in the sequence
\eqref{seqmutation} just inverts $X_i\rightarrow X_i^{-1}$ while keeping invariant $X_j$ for $j\neq i$. Thus the effect of the product 
\eqref{seqmutation} is just to invert all quantum cluster variables, that is the product in eqn.\eqref{seqmutation} is $I$.

The formula \eqref{Kqaffine} also determines the BPS phase cyclic order in terms of the affine quiver orientation.

\section{Details on some Landau--Ginzburg models}\label{APPdetailsLS}

In this appendix we present some details on the two--dimensional computations for some of the Landau--Ginzburg models mentioned in the main body of the paper.

\subsection{The second form of $N_f=2$}

This realization of $N_f=2$ may be set in relation with the LG model
\begin{equation}
W(X)= e^X+ \frac{1}{(1-e^{-X})^2}.
\end{equation}
This $2d$ theory has four classical vacua. One at $e^{-X}=\infty$, and the other three at the at $e^{-X}$ equal to the three roots of
\begin{equation}
y^3-y^2+3y-1=0,
\end{equation}  
which has one \emph{positive} real root $r=e^{-X_r}$, $X_r>1$, and a pair of complex conjugate ones $\rho, \bar \rho$.
The critical values are
\begin{align}
&W_\infty =0,\qquad\qquad W_r \ \text{real positive }\approx 3.17748\\
& W_\rho= (W_{\bar \rho})^*\ \text{complex with negative real part.}
\end{align}

We know the following facts about the BPS quiver:
\begin{itemize}
\item should be connected and compatible with $\hat c_\mathrm{uv}=1$. Indeed, were it not connected, the connected components will have at most three nodes, and all such theories are already classified;
\item the numbers of BPS states connecting $\rho$ with $\infty$ (resp.\! $r$) is the same as the number of states connecting $\bar\rho$ with $\infty$ (resp.\! $r$) since they are related by complex conjugation;
\item there are no solitons connecting $r$ and $\infty$. 
\end{itemize}

Then the graph underlying the quiver must have the form
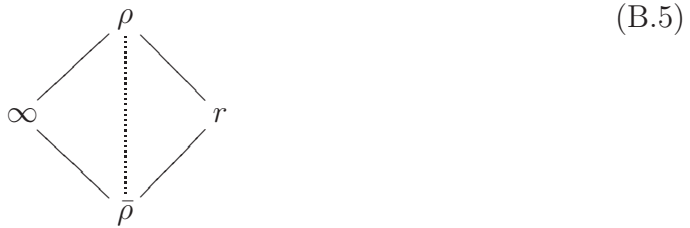
\begin{equation}
\xymatrix{& \rho \ar@{..}[dd] \ar@{-}[dl]\ar@{-}[dr] &\\
\infty & & r\\
& \bar\rho \ar@{-}[ul]\ar@{-}[ur] &}
\end{equation}
where the dashed line means that there may be or not a soliton connecting the two complex vacua. We also know that the \emph{orientation} of the arrows should be invariant under reflection with respect to the horizontal axis (\textit{i.e.}\! under complex conjugation). Finally, we know that the direction of the arrows should be consistent with $\hat c=1$, which requires that any proper sub--quiver should be a minimal model one. This leaves us with three possible BPS quivers
which are all mutation--equivalent to $\widehat{A}_3(2,2)$. 
\subsection{$N_f=3$}

One has
\begin{equation}
W^\prime = 2\, e ^{2X}\, \frac{(e^X-1)^3-1}{(e^X-1)^3}
\end{equation}
so that we have two vacua at $X=-\infty$, and three vacua for
$e^X=1+\varrho$, where $\varrho$ is a primitive third root of $1$. 
 The critical values are $0$ for the vacua at $\infty$, and 
\begin{equation}\begin{split}
W(e^X-1=\varrho)&=(1+\varrho)^2\: \frac{\varrho^2+1}{\varrho^2}=
(1+\varrho)^2(1+\varrho^{-2})=(1+\varrho)^3\\
&= \begin{cases} 8 & \varrho=1\\
(-\varrho^2)^3\equiv -1& \varrho\neq 1.\end{cases}
\end{split}\end{equation}

Thus, all critical values are real (and hence aligned). There are no solitons between the two vacua at infinity, nor between the two vacua at $e^X=1+e^{\pm 2\pi i/3}$. Moreover, complex conjugation exchanges these last two vacua, and hence the number of soliton from each of these two vacua and the other vacua are equal.

\smallskip

Setting $y= e^{-X}$, the equation $W(X)=w$ becomes the quartic equation
\begin{equation}\label{quartic}
y^4-2 y^3 +(1+2/w)y^2+ 2\,y/w-1/w=0\end{equation}
whose discriminant is
\begin{equation}
-16\,(w-8)(w+1)^2/w^5.
\end{equation}

Consider the solitons between infinity and the vacuum $0$. In the $W$--plane they corresponds to the segment $0\leq w\leq 8$ on the \emph{real} axis. For $w\sim 0$ real positive, \eqref{quartic} gives
$y\sim \zeta\,w^{-1/4}$, where $\zeta$ is a fourth--root of $1$. Thus, for $w\sim 0^+$ we have one real positive, one real negative, and a pair of complex conjugate roots. Given that the constant term of \eqref{quartic} never vanishes, this configuration of roots (one positive, one negative, a pair of conjugate ones) will persists as we move $w$ along the real axis until we get at the first zero of the discriminant at $w=8$. Here the two complex roots come together and become \emph{real}. Indeed, at $w=8$ the roots of
\eqref{quartic} are  
\begin{equation}
y=1/2, 1/2, (1-\sqrt{3})/2, (1+\sqrt{3})/2,
\end{equation}
and the two solutions which becomes purely imaginary as $w\rightarrow 0^+$, both have limit $y=1/2$ as $w\rightarrow 8$. The other two roots at $w=8$ corresponds to the two real roots at $w\sim 0$, respectively negative and positive.

$y=1/2$, corresponds to $e^X=2$, that is to the vacuum $0$. Therefore, the two imaginary roots of \eqref{quartic} over the segment $0\leq w\leq 8$ in the $W$--plane are precisely two BPS states connecting vacuum $0$ to, respectively, $\infty_1$ and $\infty_2$, where these two vacua correspond to $e^X= \mp i\, w^{1/4}$, as $w\rightarrow 0$.

In the $W$--plane, the solitons from infinity to $e^X=1+e^{2\pi i/3}$ correspond to the segment $-1,\leq w\leq 0$ on the real axis. For $w\sim 0$ real and \emph{negative} we have from \eqref{quartic} $y\sim \zeta\, |w|^{-1/4}$ where $\zeta$ is a fourth--root of $-1$. Thus for $w\sim 0^-$ we have two pairs of complex conjugate roots with phases $\pm i$ and, respectively, $e^{\pm i\pi/4}$. 

As we decrease $w$ from $0$ to $-1$ these pairs of complex roots will not mix, since the discriminant is not zero, until we reach $w=-1$ where the discriminant has a \emph{double} zero. There the two complex pair --- while remaining complex --- gets together. Indeed, the roots of equation \eqref{quartic} with $t=-1$ are
\begin{equation}
y= e^{\pi i/3}, e^{\pi i/3}, e^{-\pi i/3}, e^{-\pi i/3}.
\end{equation}
One has $e^{\mp \pi i/3}= 1+ e^{\mp 2\pi i/3}$. Hence two of the soliton starting from infinity will reach each complex classical vacua.

Finally, the solitons between $e^X=2$ and $e^X=1+\varrho$ correspond to the segment $-1\leq w\leq 0$ on the real axis. But these all passes through infinity. So no soliton here.

In conclusion, the above results suggest the following form for the quiver  (where the nodes are labelled by the values of $e^X$)
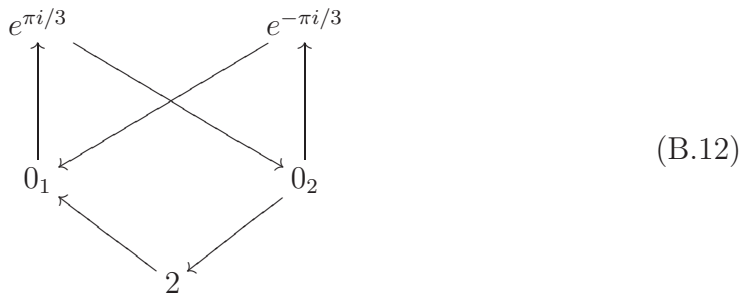
\begin{equation}
\begin{gathered}\begin{xy} 0;<1pt,0pt>:<0pt,-1pt>:: 
(0,0) *+{e^{\pi i/3}} ="0",
(101,0) *+{e^{-\pi i/3}} ="1",
(0,60) *+{0_1} ="2",
(51,99) *+{2} ="3",
(101,60) *+{0_2} ="4",
"2", {\ar"0"},
"0", {\ar"4"},
"1", {\ar"2"},
"4", {\ar"1"},
"3", {\ar"2"},
"4", {\ar"3"},
\end{xy}\end{gathered}
\end{equation}
which is mutation--equivalent to $\widehat{D}_4$.

\subsection{LG with $W(X)=\wp^\prime(X)$}\label{app:weierstrass-prime}

Let $\wp(z)$ be the Weierstrass function
\begin{equation}
(\wp^\prime)^2= 4\, \wp^3-g_2\, \wp-g_3.
\end{equation}
where the cubic polynomial in the \textsc{rhs} has non--vanishing determinant $\Delta\neq 0$.
We consider a LG model with the field $X$ taking value on the corresponding torus and superpotential $W(X)=\wp^\prime(X)$. The vacuum condition is
\begin{equation}
0=W^\prime(X)= 6\, \wp(X)^2- \frac{1}{2}\, g_2.
\end{equation}
The function in the \textsc{rhs} has a pole of order 4 at the origin, and hence four zeros.\medskip

\textbf{Lemma.} \textit{For $g_2\neq 0$, all four classical vacua are massive (and hence distinct). Between any two vacua, the absolute number of BPS solitons is either $1$ or $2$.}\vglue 9pt

\textsc{Proof.} Indeed, $W^{\prime\prime}= 12\, \wp(X)\, \wp^\prime(X)$. At a vacuum $\wp(X)=\pm \sqrt{g_2/12}$, and hence $\wp(X)\neq 0$. is non--zero. Then, in order to have $W^\prime(X)=W^{\prime\prime}(X)=0$, we must have $\wp^\prime(X)=0$ and hence
\begin{equation}
0= 4\wp^4-g_2\,\wp-g_3= \pm \sqrt{g_2/12}\,\big( 4 g_2/12-g_2\big)-g_3
\end{equation}
or
\begin{equation}
0=g_3^2- \frac{1}{12} \cdot \frac{4}{9}\cdot g_2^3=-\frac{\Delta}{432}\neq 0,
\end{equation}
which is absurd. Then the four vacua are $\pm X_\pm$ where
$\wp(X_\pm)=\pm \sqrt{g_2/12}$. 

Let $W_\pm =\wp^\prime(X_\pm)$. Consider the elliptic functions
$F_{\epsilon,\epsilon^\prime}(X)= \wp^\prime(X)-\epsilon\, W_{\epsilon^\prime}$, where $\epsilon,\epsilon^\prime= \pm$.
These meromorphic functions have a pole of order $3$ at the origin, and hence should have three zeros on the torus whose sum must give zero. One the other hand,
\begin{equation}
F_{\epsilon, \epsilon^\prime}(\epsilon X_{\epsilon^\prime})=0,\quad
F^\prime_{\epsilon, \epsilon^\prime}(\epsilon X_{\epsilon^\prime})\equiv \wp^{\prime\prime}(\epsilon X_{\epsilon^\prime})=0,
\end{equation}
and hence $F_{\epsilon, \epsilon^\prime}(X)$ has a \emph{double} zero at $\epsilon X_{\epsilon^\prime}$. 

Consider now the inverse image of the segment in $W$ plane between the points $\epsilon W_{\epsilon^\prime}$ and  $\tilde \epsilon W_{\tilde\epsilon^\prime}$; it may be written as
\begin{equation}
(1-t)\, F_{\epsilon,\epsilon^\prime}+t\, F_{\tilde\epsilon,\tilde\epsilon^\prime}=0\qquad 0\leq t\leq 1.
\end{equation}
For each $t$ in the open interval $0<t<1$, we have three values of $X$ (modulo periods) which satisfy this equation. Moreover, these values are all distinct, except at $t=0,1$, where two of the three values will go to the critical point $\epsilon X_{\epsilon^\prime}$ and, respectively, to $\tilde \epsilon X_{\tilde \epsilon^\prime}$ while the third root approaches at $-2\epsilon X_{\epsilon^\prime}$ and $-2\tilde\epsilon X_{\tilde\epsilon^\prime}$, respectively. Let $X_{(1)}(t)$, $X_{(2)}(t)$ be the two solutions which for $t=0$ go to the classical vacuum $\epsilon X_{\epsilon^\prime}$. Two things may happen: either both $X_{(1)}(t)$, $X_{(2)}(t)$ go to $\tilde \epsilon X_{\tilde\epsilon^\prime}$ as $t\rightarrow 1$, or one of the two go to the third root $-2\tilde \epsilon X_{\tilde\epsilon^\prime}$ while the other one will necessarily go to $\tilde \epsilon X_{\tilde\epsilon^\prime}$.\hfill $\square$

\bigskip

To simplify the analysis, we consider a special case with enhanced symmetry, namely a lemniscatic (square) torus with periods $(1,i)$, corresponding to $g_3=0$, $g_2=\Gamma(1/4)^8/16\pi^2$. Then $\wp^\prime(i X)= i\, \wp^\prime(X)$, and the model has a $\Z_4$ symmetry, $X\rightarrow i\, X$, under which the four (distinct) vacua form an orbit.
The four vacua are at
\begin{equation}
 X_k= i^{k-2}\left(\frac{1}{2}+ i\, \alpha\right),\qquad k=1,2,3,4,\qquad \alpha\approx 0.1988783\in \R.
\end{equation}
 The critical values form a square in $W$--plane with vertices at
\begin{equation}
 W(X_k)=i^{k-1}\, a,\qquad k=1,2,3,4,\qquad a\approx 22.3682\in \R.
\end{equation}

By the $\Z_4$ symmetry, it is enough to determine the number of BPS states along a side and a diagonal of this square. Consider the diagonal corresponding to the segment along the imaginary axis from $-ia$ to $+ia$; a diagonal soliton is a curve on the torus connecting $1/2-i\alpha$ to $1/2+i\alpha$ which maps to this segment in the $W$--plane. Let the $X$--plane be the universal cover of the torus. Along the straight--line $1/2+i\mathbb{R}$ the function $\wp^\prime(X)$ is purely imaginary, so the segment in the $X$--plane connecting $1/2-i\alpha$ to $1/2+i\alpha$ is mapped into the diagonal of the square, and hence it is a soliton. Likewise, the segment in the $X$ plane from 
$1/2-i\alpha$ to $1/2-i(1-\alpha)$ is also a segment betwen the same two vacua on the torus. So there are at least two solitons along each diagonal; since there cannot be more than two by the lemma, we conclude that along the diagonal we have precisely two solitons.   

 It remains to determine the number $\mu$ of solitons along the sides of the square. We have $|\mu|=1, 2$ by the lemma. 
In order to get $\mu$, we may use the general classification of $\Z_4$ symmetric models in \cite{Cecotti:1993rm}.  
 Eqn.(8.5) of ref.\cite{Cecotti:1993rm} implies that
\begin{equation}
\cq(z)\equiv z^4 + \mu\, z^3 \pm 2\, z^2+(-1)^{q+1}\mu\, z+ (-1)^{q+1} 
\end{equation} 
should be a product of cyclotomic polynomials for some choice of signs $\pm$ and $(-1)^q$. The solutions to this condition with $\mu=\pm 1,\pm 2$ are 
\begin{align}
\Phi_3(z)\, \Phi_4(z)&=z^4+z^3+2 z^2+z+1\\
\Phi_6(z)\, \Phi_4(z)&=z^4-z^3+2 z^2-z+1\\
\Phi_4(z)\, \Phi_1(z)^2&=z^4-2 z^3+2 z^2-2 z+1\\
\Phi_4(z)\, \Phi_2(z)^2&=z^4+2 z^3+2 z^2+2 z+1\end{align}
which also implies $(-1)^q=-1$.
Then eqn.(8.4) of ref.\cite{Cecotti:1993rm} gives for the characteristic polynomial of the $2d$ monodromy $M$
\begin{equation}\label{detM}
\det[z-M]=\begin{cases} \Phi_3(-z)\, \Phi_1(-z)^2 & |\mu|=1\\
\Phi_1(-z)^4 & |\mu|=2.
\end{cases}
\end{equation}
The second case corresponds to the four point correlation of the Ising model. The spectrum of $M$ is not compatible with a unitary theory with $\hat c_\mathrm{uv}\leq 1$.

The first case of eqn.\eqref{detM} is perfectly compatible with a AF model with $\hat c_\mathrm{uv}=1$,
and having four chiral primary operators of dimension in the UV $0$, $1/3$, $2/3$ and $1$. Since the two allowed deformations of $W(X)$, namely $\wp(X)$ and $\zeta(X)$, are expected to have UV dimensions $2/3$ and $1/3$, respectively, this solution must correspond to the model $W(X)=\wp^\prime(X)$. 

Then we learn that along the sides of the critical square in the $W$--plane we have just one soliton, $|\mu|=1$.
\bigskip

\textbf{The quiver}\vglue 6pt

We write the elements $S_\theta$ of the Stokes group corresponding to the four BPS rays $e^{i\theta}$ in the lower half plane,  as borrowed from section 8 of ref.\cite{classification} for the relavant $\Z_4$--symmetric model\footnote{\ With respect to that reference, we change the sign to vacua $3$ and $4$, which is natural since the topological metric $\eta$ changes sign as $X\leftrightarrow -X$.}  
\begin{align}
 S_0&= 1-2\, E_{3,1} &  S_{-\pi/4}&= 1- E_{2,1}+E_{3,4}\\
 S_{-\pi/2}&= 1+2\,E_{2,4} & S_{-3\pi/4}& = 1- E_{1,4}-E_{2,3}  
\end{align}
where $(E_{ij})_{kl}$ is the matrix which is $1$ for $k=i, l=j$ and zero otherwise. One has
(the conventions of [classification] correspond to a taking the product of the $S_\theta$ in the \emph{clockwise} order)
\begin{equation}
 S \equiv S_{-3\pi/4}\,S_{-\pi/2}\, S_{-\pi/4}\, S_0 =
\left(
\begin{array}{cccc}
 1 & 0 & 0 & -1 \\
 1 & 1 & -1 & 1 \\
 -2 & 0 & 1 & 1 \\
 0 & 0 & 0 & 1
\end{array}
\right).
\end{equation}
and then 
\begin{equation}
 B\equiv S-S^t = \left(
\begin{array}{cccc}
 0 & -1 & 2 & -1 \\
 1 & 0 & -1 & 1 \\
 -2 & 1 & 0 & 1 \\
 1 & -1 & -1 & 0
\end{array}
\right),
\end{equation}
which corresponds to the quiver
\begin{equation}
 \begin{gathered}
  \xymatrix{1 \ar@<-0.4ex>[dd]\ar@<0.4ex>[dd] && 2\ar[ll]\ar[dd]\\
&&\\
3 \ar[rr]\ar[uurr] &&4\ar[uull]}
 \end{gathered}
\end{equation}
which is the one associated to the unique ideal triangulation of a torus with a boundary having a marked point.
If we change the half--plane used to define $B$, nothing is going to change: in fact by $\Z_4$ symmetry, we have only to check the rotation of the half--plane by $-\pi/4$; this amounts to replacing
\begin{equation}
 S\rightarrow S^\prime =I_3\, (S_0^{-1})^t\, S\, S_0^{-1}\, I_3
\end{equation}
(where $I_3=\mathrm{diag}(1,1,-1,1)$ just a vacuum sign redefinition to reestablish the correct conventions). Then 
$B^\prime=S^\prime- (S^\prime)^t$ gives the quiver 
\begin{equation}
 \begin{gathered}
  \xymatrix{3 \ar@<-0.4ex>[dd]\ar@<0.4ex>[dd] && 2\ar[ll]\ar[dd]\\
&&\\
1 \ar[rr]\ar[uurr] &&4\ar[uull]}
 \end{gathered}
\end{equation}
 which is the same as before, up to a relabeling of the nodes.


\end{document}